\documentclass[
reprint,
amsmath,
amssymb,
floatfix,
superscriptaddress,
twocolumn,
aps
]{revtex4-2}

\usepackage{dcolumn}
\usepackage[normalem]{ulem}
\usepackage{bm}
\usepackage{color}
\usepackage{amssymb}
\usepackage{amsmath}
\usepackage{xfrac}
\usepackage{mathrsfs}
\usepackage{comment}

\usepackage{hyperref}
\usepackage[dvipsnames]{xcolor}
\hypersetup{
    colorlinks=true,
    linkcolor=Blue,
    citecolor=Blue,   
   urlcolor=Blue
   }

\def\la{{\langle}}
\def\ra{{\rangle}}
\def\br{{\textbf{r}}}
\def\tbr{{\tilde{\textbf{r}}}}
\def\Pe{{\rm{Pe}}}
\def\bu{{\hat{\textbf{u}}}}
\def\bR{{\bm{\mathcal{R}}}}

\def\ttt{{\tilde{t}}}

\begin{document}
\title{Exact dynamics and dimensional effects in torque-driven trapped active Brownian particles}

\author{Anweshika Pattanayak}
\email[Equal contribution:~]{anweshika.pattanayak@tifr.res.in}
\affiliation{Department of Physical Sciences, Indian Institute of Science Education and Research Mohali, Sector 81, Knowledge City, S. A. S. Nagar, Manauli PO 140306, India}
\affiliation{Department of Theoretical Physics, Tata Institute of Fundamental Research, Homi Bhabha Road, Mumbai 400005, India}

\author{Amir Shee}
\email[Equal contribution:~]{amir.shee@uvm.edu}
\affiliation{Department of Physics, University of Vermont, Burlington, Vermont 05405, USA}

\author{Abhishek Chaudhuri}
\email[Contact author:~]{abhishek@iisermohali.ac.in}
\affiliation{Department of Physical Sciences, Indian Institute of Science Education and Research Mohali, Sector 81, Knowledge City, S. A. S. Nagar, Manauli PO 140306, India}

\author{Debasish Chaudhuri}
\email[Contact author:~]{debc@iopb.res.in}
\affiliation{Institute of Physics, Sachivalaya Marg, Bhubaneswar 751005, India.}
\affiliation{Homi Bhabha National Institute, Anushakti Nagar, Mumbai 400094, India}

\date{\today}

\begin{abstract} 
The interplay of chirality, self-propulsion, and confinement generates pronounced non-equilibrium fluctuations, with higher-order moments encoding the shape of the position distribution. We present an exact analytical framework, based on the Laplace transform of the Fokker–Planck equation, for the transient dynamics of a chiral active Brownian particle in a harmonic trap in two and three dimensions. Closed-form expressions for time-dependent moments up to fourth order allow a full characterization of the excess kurtosis across transient and steady-state regimes. In two dimensions, the excess kurtosis exhibits damped oscillations with multiple re-entrant crossovers, evolving from negative values, indicative of off-centered ring-like distributions, to small positive values characteristic of near-Gaussian distributions with a weak tail enhancement; increasing trap stiffness progressively suppresses this oscillatory response. In three dimensions, the excess kurtosis of a torque-driven active Brownian particle stays negative, reflecting robust non-Gaussian fluctuations with half-ring– to band-like position distributions in the plane defined by the torque axis. These results reveal how chirality, propulsion, confinement, and dimensionality jointly shape transient dynamics and provide experimentally accessible signatures of confined chiral active motion.
\end{abstract}

\maketitle

\section{Introduction}

Chirality arises across chemical, physical, and biological systems and is a key ingredient in a broad class of active matter models~\cite{van-Teeffelen2008, Chen2017, Kim2018, Oliver2018, Lei2019, Huang2020, Liebchen2022}. At the single-particle level, self-propelled particles with intrinsic rotation exhibit circular and helical trajectories~\cite{Lowen2016, Liebchen2022}, with chirality emerging from asymmetries in propulsion or particle geometry~\cite{Kummel2013, Arora2024}, or through coupling to external fields~\cite{Grzybowski2000, Cruz2024}. Such chiral motion is observed experimentally in biomolecular systems~\cite{Jennings1901}, including proteins~\cite{Loose2014}, microtubules~\cite{Sumino2012}, bacteria~\cite{Brokaw1982, DiLuzio2005, Lauga2006, PerezIpina2019, Otte2021}, and sperm cells~\cite{Riedel2005, Nosrati2015}, as well as in synthetic active systems spanning microscopic~\cite{Yan2015, Massana-Cid2021, Kaur2025} to macroscopic scales~\cite{Farhadi2018, Scholz2021, Vega2022, Lopez2022, Siebers2023, Caprini2024}.

Theoretically, single chiral (circle) swimmers provide a minimal framework to understand how chirality modifies transport~\cite{Van-Teeffelen2009, Mijalkov2013, Volpe2014, Sevilla2016, Markovich2019, Chepizhko2020}, including circular trajectories in \textit{E.~coli}~\cite{Leonardo2011, Araujo2019} and chemically driven circular or helical motion~\cite{Bohmer2005, Taktikos2011}. Chiral active systems display odd viscosity and diffusivity~\cite{Soni2019, Hargus2021}, and more broadly, chirality acts as a control parameter that alters transport, giving rise to phenomena such as spiral patterns, anomalous diffusion, and confinement-induced oscillatory states~\cite{Caprini2019, Li2020, Khatri2022, Pattanayak2024, Olsen2024, Pattanayak2025}. These considerations motivate a focused study of single-particle chiral active motion, with emphasis on transient non-Gaussian fluctuations characterized through higher-order displacement moments.
Deviations from Gaussianity in free chiral active motion have been characterized using the non-Gaussian parameter and higher moments~\cite{Sevilla2016, Pattanayak2024}. Here, we investigate how harmonic confinement reshapes this transient non-Gaussian structure and how this behavior depends on spatial dimension.

Confinement plays a central role in shaping the dynamics of chiral active particles~\cite{Fazli2021, Murali2022, Caprini2023}.
Experimental platforms such as optical traps and microfluidic chambers provide controlled settings that couple particle activity to restoring forces~\cite{Takatori2016, Peng2022}.
In such environments, confinement is known to significantly modify both the mean-squared displacement~\cite{Van-Teeffelen2009, Debets2023, Shee2024, Marconi2025, Barman2025} and the steady-state distributions~\cite{Pattanayak2025}.

In this work, we develop an exact analytical theory for the dynamics of a harmonically trapped chiral active Brownian particle in both two and three dimensions, with particular emphasis on the approach to the steady state. We employ a Fokker–Planck–based method that enables the systematic calculation of arbitrary dynamical moments. Originally introduced for configurational statistics of semiflexible polymers~\cite{Hermans1952, Daniels1952}, this approach has been adapted to active particle dynamics~\cite{Shee2020, Chaudhuri2021, Patel2023} and recently extended to torque-driven motion in free active particles~\cite{Pattanayak2024} and their steady-state properties under confinement~\cite{Pattanayak2025}.

Here, we obtain closed-form expressions for the time-dependent second and fourth moments of particle displacements, providing a complete characterization of the dynamics, including non-Gaussian features arising from the intrinsic nonlinearities of active Brownian motion.
Our analysis reveals pronounced dimensional effects: in two dimensions, the excess kurtosis displays damped oscillations between negative and positive values, indicating crossovers between off-centered and near-Gaussian distributions, whereas in three dimensions, for a torque-driven active Brownian particle, it remains strictly negative, reflecting persistent non-Gaussian fluctuations. These predictions agree with our numerical simulation results and point to experimentally accessible signatures of chirality-induced dynamics.
By elucidating how self-propulsion, chirality, and confinement jointly shape non-equilibrium fluctuations, this work provides a comprehensive, exact theoretical framework for chiral active matter. It also highlights the fundamental role of dimensionality in governing universal features of confined chiral active motion, offering clear pathways for experimental validation.

The remainder of the paper is organized as follows. In Sec.~\ref{sec-2}, we analyze the exact time-dependent moments for a chiral active Brownian particle confined in a harmonic trap in two dimensions. Section~\ref{sec-3} extends the analysis to three dimensions and highlights the qualitative differences arising from dimensionality. In Sec.~\ref{sec-4}, we summarize our main conclusions and discuss the physical implications of our results, including signatures of chirality-induced non-Gaussian fluctuations and their experimental relevance.

\begin{figure}[t]
\centering
\includegraphics[width=\linewidth]{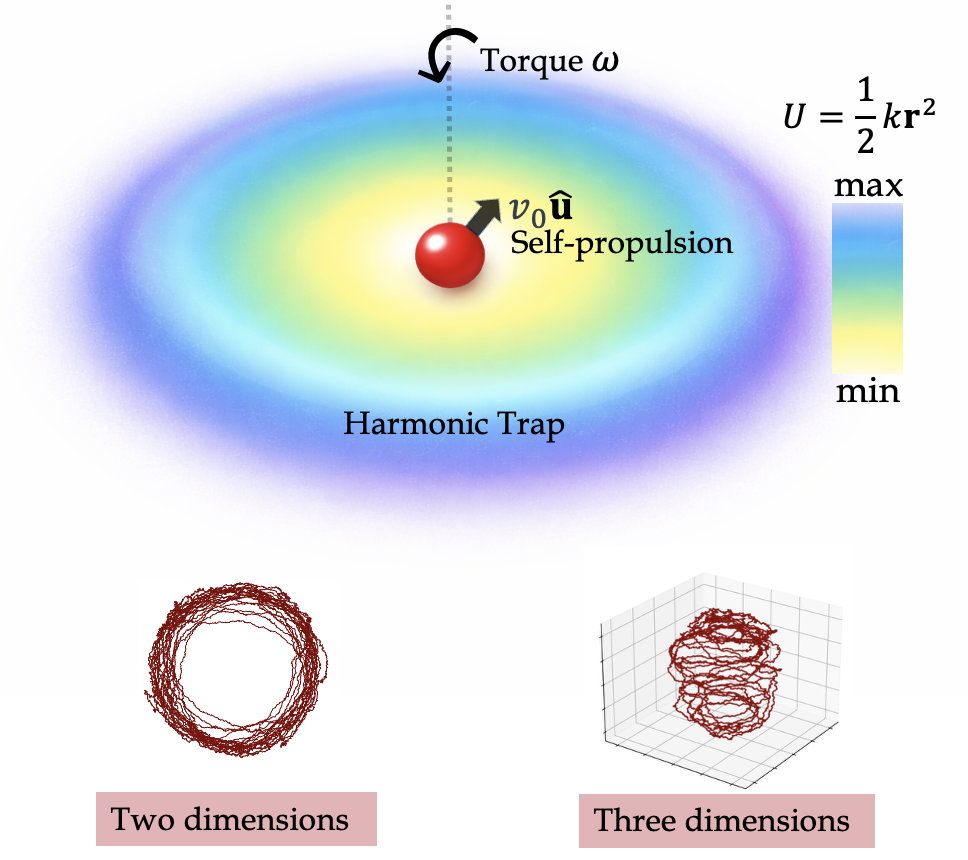}
\caption{Schematic of an active Brownian particle under torque in a harmonic trap.
The particle self-propels with active speed $v_0$ along its orientation vector $\mathbf{\hat{u}}$, possesses intrinsic chirality or experiences an applied torque $\omega$ that rotates its orientation, and is confined by a trap of stiffness $k$. The combined action of torque, propulsion, and confinement generates a characteristic circular trajectory in two dimensions or a spiral trajectory in three dimensions within the trapping potential.
}
\label{fig1}
\end{figure}

\section{Two-dimensions}
\label{sec-2}

\noindent
{\bf Model:~}We consider a single chiral active Brownian particle in two dimensions with position $\br=(x,y)$ and heading direction(or orientation unit vector) $\bu = (\cos\phi , \sin\phi)$. The particle is confined by a harmonic potential $U = k \br^2/2$(see Fig.~\ref{fig1}). Its overdamped dynamics can be written as~\cite{Caprini2019, Liebchen2022, Pattanayak2025}
\begin{align}
d{\br} &= v_0\bu dt +\sqrt{2 D}\,d\bm{B}-\mu \br dt\,,\label{eom1:2d}
\\
d{\phi} &= \omega dt+ \sqrt{2 D_r} \, d W\,,
\label{eom2:2d}
\end{align}
where $v_0$ is the self-propulsion(or active) speed, $D$ and $D_r$ are the translational and rotational diffusivities, respectively, and $\omega$ is the constant chiral (torque) frequency. The parameter $\mu = k/\gamma$ denotes the trap strength scaled by the viscous drag $\gamma$.
The translational and rotational Wiener processes are characterized by variances $\langle dB_i dB_j \rangle = \delta_{ij} dt$ (for $i,j \in {x,y}$) and $\langle dW^2 \rangle = dt$, with all mixed correlations vanishing.
In the absence of confinement ($\mu = 0$), the governing equations reduce to a free chiral active Brownian particle~\cite{Pattanayak2024}.

Time and length are measured in units of the persistence time $\tau_r = D_r^{-1}$ and the diffusive length scale $\ell = \sqrt{D/D_r}$, respectively.
The dimensionless control parameters are the Péclet number $\Pe = v_0/\sqrt{DD_r}$ (activity), chirality $\Omega = \omega/D_r$, and trap strength $\beta = \mu/D_r$.
The  dimensionless position  and time will be  denoted by $\tilde{\br} = \br / \ell$ and $\tilde{t} = t / \tau_r$.
These dimensionless units and parameters are used in both two and three dimensions.

The exact dynamical moments are obtained by formulating the corresponding Fokker–Planck equation, applying a Laplace transform, and systematically solving the resulting moment-generating equation, detailed in Appendix-\ref{app-A}.
We numerically integrate the dimensionless Langevin equations using the Euler–Maruyama scheme and find excellent agreement with our analytical predictions.

\begin{figure*}[t]
\centering
\includegraphics[width=\linewidth]{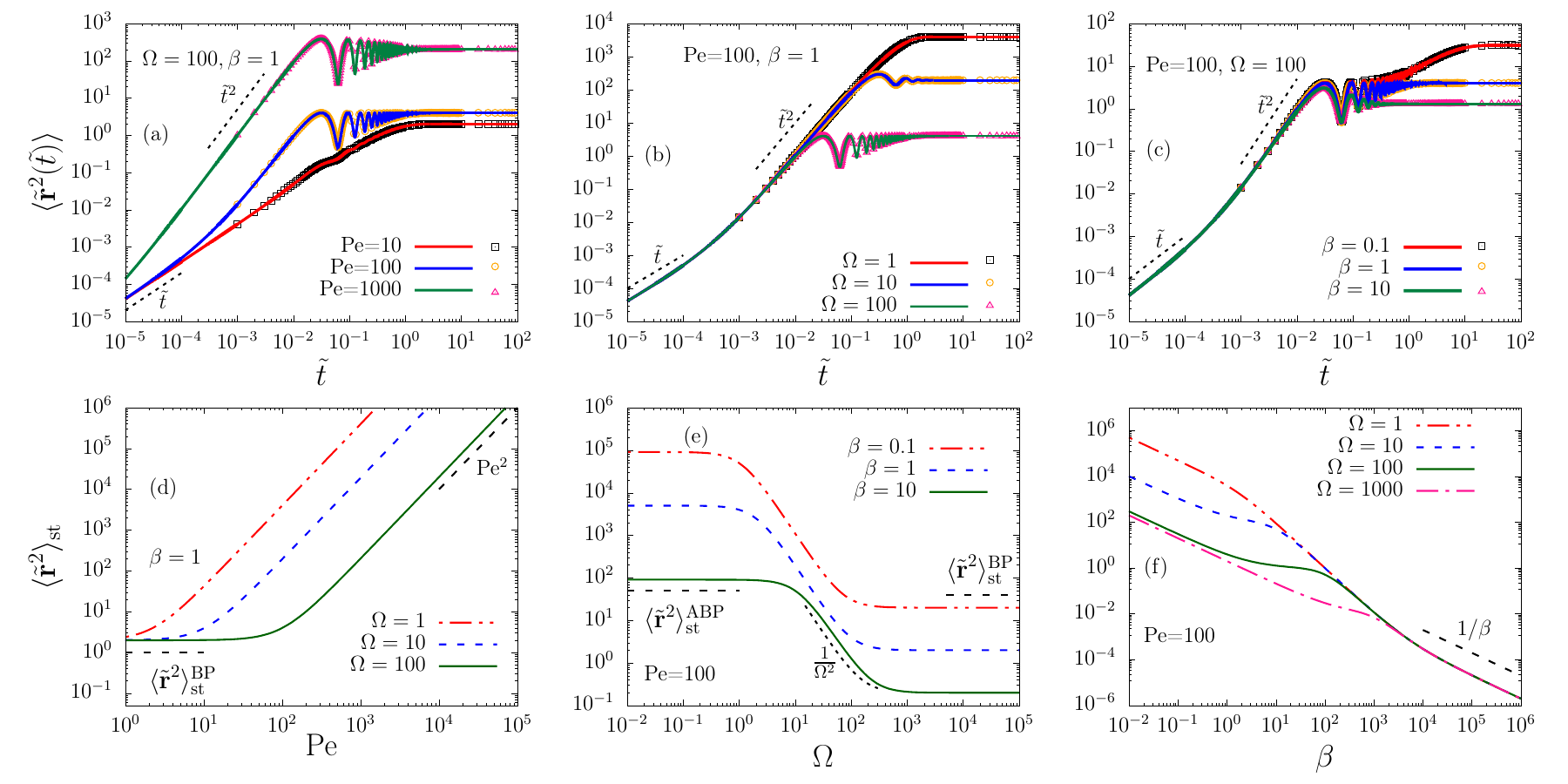}
\caption{
Mean-squared displacement (MSD) of a chiral active Brownian particle (CABP) in a harmonic trap in two dimensions (2d) as a function of time for (a) $\text{Pe}=10, 100, 1000$ with $\Omega=100$ and $\beta=1$, (b) chirality $\Omega=1,10,100$ with activity $\text{Pe}=100$ and harmonic trap stiffness $\beta=1$, and (c) $\beta=0.1,1,10$ with $\Pe=100$ and $\Omega=100$.
Steady state MSD as a function of (d) $\Pe$ for $\Omega=1,10,100$ with $\beta=1$, (e) $\Omega$ for $\beta=0.1,1,10$ with $\Pe=100$, (f) $\beta$ for $\Omega=1,10,100,1000$ with $\Pe=100$.
The lines in (a), (b) and (c) are plots of Eq.~(\ref{eq:msd_2d_dimensionless}) and the points are from simulations. The lines in (d), (e), and (f) are plots of Eq.~\eqref{eq:msd_2d_st}.
}
\label{fig2}
\end{figure*}

\subsection{Mean-squared displacement(MSD)}
In dimensionless form, the mean-squared displacement (MSD) of a chiral active Brownian particle 
in two dimensions is given by~\cite{Lowen2020, Debets2023, Shee2024, Barman2025} (see detailed derivation in Appendix-\ref{app-A})
\begin{align}
\langle \tilde{\mathbf{r}}^2(\tilde{t})\rangle
&= \frac{2}{\beta}\big(1 - e^{-2\beta \tilde{t}}\big) \nonumber\\
&+ \mathrm{Pe}^2\Bigg[
\frac{\beta+1}{\beta\big[(\beta+1)^2 + \Omega^2\big]}
+ \frac{(\beta-1)e^{-2\beta \tilde{t}}}{\beta\big[(\beta-1)^2 + \Omega^2\big]}
\nonumber\\
&
- \frac{2 e^{-(\beta+1)\tilde{t}}
\big[(\beta^2 + \Omega^2 - 1)\cos(\Omega \tilde{t})
+ 2\Omega \sin(\Omega \tilde{t})\big]}
{\big[(\beta-1)^2 + \Omega^2\big]\big[(\beta+1)^2 + \Omega^2\big]}
\Bigg]\,.
\label{eq:msd_2d_dimensionless}
\end{align}
The time-dependent mean-squared displacement (MSD) given in Eq.~\eqref{eq:msd_2d_dimensionless} 
captures the full interplay between activity, chirality, and harmonic confinement in two dimensions.
The achiral (or non-chiral) limit of our model, an active Brownian particle in a trap, has been
studied both theoretically and experimentally~\cite{Takatori2016, Chaudhuri2021, Peng2022},
and Eq.~\eqref{eq:msd_2d_dimensionless} reduces to the corresponding
result~\cite{Chaudhuri2021} as $\Omega\to0$ (see Appendix-\ref{app-A}).

\noindent
\textit{Small time expansion:~} 
In $\ttt \to 0$ limit, expanding Eq.~\eqref{eq:msd_2d_dimensionless}, we get~\cite{Lowen2020, Barman2025}
\begin{align}
\lim_{\ttt \to 0}\langle \tbr^2\rangle& =4 \ttt+\ttt^2 \left(\text{Pe}^2-4 \beta \right)+\frac{\ttt^3}{3}  \left(8 \beta ^2-3\beta  \text{Pe}^2-\text{Pe}^2\right)
\nonumber\\&+\frac{\ttt^4}{12}  \left(-16 \beta ^3+7\beta ^2 \text{Pe}^2+4\beta \text{Pe}^2-\text{Pe}^2 \Omega ^2+\text{Pe}^2\right)\nonumber\\&+O\left(\ttt^5\right)\,.
\label{eq:MSD_2d_small_times}
\end{align}
At very small times the dynamics is dominated by thermal fluctuations $\langle \tbr^2\rangle=4\ttt$, which crosses over to ballistic behavior $\langle \tbr^2\rangle\sim \ttt^2$ at $\ttt_{I}=4/(\Pe^2-4\beta)$ with condition $\Pe>>2\sqrt{\beta}$. Notably, the $\mathcal{O}(\tilde{t}^{2})$ carries an explicit dependence on the trap strength $\beta$, indicating that confinement affects the short-time growth once self-propulsion becomes relevant. In contrast, chirality contributes only at $\mathcal{O}(\tilde{t}^{4})$, demonstrating its limited relevance during the early-time behavior.

\medskip 

\noindent
In the long time limit, the MSD in Eq.~(\ref{eq:msd_2d_dimensionless}) reaches steady state $\langle \tbr^2\rangle_{\rm st}=  \lim_{\ttt \to \infty} \langle \tbr^2\rangle$~\cite{Lowen2020, Debets2023, Shee2024, Marconi2025, Pattanayak2025}
\begin{align}
\langle \tilde{\mathbf{r}}^2\rangle_{\mathrm{st}}
= \frac{2}{\beta} + \frac{(\beta + 1)\,\mathrm{Pe}^2}{\beta\left[(\beta + 1)^2 + \Omega^2\right]}\,.
\label{eq:msd_2d_st}
\end{align}
The approach to the steady state(Eq.~\eqref{eq:msd_2d_st}) is not algebraic but instead governed by exponential relaxation. Eq.~\eqref{eq:msd_2d_dimensionless} around the steady state gives
\begin{align}
\langle \tilde{\mathbf{r}}^2\rangle = \langle\tilde{\mathbf{r}}^2\rangle_{\mathrm{st}} + \mathcal{O}(e^{-\lambda \ttt})\,,\nonumber
\end{align}
with $\lambda=\mathrm{min}(2\beta, \beta+1)$(see Eq.~\eqref{eq:msd_2d_dimensionless}), demonstrating exponential convergence. Accordingly, $\lambda=2\beta$ for $\beta<1$ and $\lambda=\beta+1$ for $\beta>1$ (degenerate at $\beta=1$). The characteristic relaxation time scales as $\sim\lambda^{-1}$, and the steady state is reached for $\lambda \ttt\gg1$(see Fig.~\ref{fig2}(a–c)). Thus, for weak trapping ($\beta<1$) the long-time relaxation is governed by thermal fluctuation, whereas for strong trapping ($\beta>1$) it is dominated by activity–trap coupling.

Figure~\ref{fig2} shows the time evolution and steady-state behavior of the dimensionless mean-squared displacement (MSD), 
$\langle \tilde{\mathbf{r}}^2(\tilde{t})\rangle$, for a chiral active Brownian particle confined in a harmonic potential.
Figure~\ref{fig2}(a)--(c) presents the analytic prediction from Eq.~\eqref{eq:msd_2d_dimensionless} (solid lines) 
compared with simulation results (points), showing excellent agreement across all parameter regimes.
The MSD evolution reveals four distinct dynamical regimes that appear consistently across parameters: an initial diffusive regime($\langle \tilde{\mathbf{r}}^2\rangle \sim \tilde{t}$) dominated by translational noise, ballistic regime($\langle \tilde{\mathbf{r}}^2\rangle \sim \tilde{t}^2$) dominated by persistent propulsion, an intermediate crossover regime with damped oscillatory behavior, and a final steady-state plateau where confinement balances chiral activity and diffusion. Such oscillatory dynamics due to chiral motion has been reported before~\cite{van-Teeffelen2008, Caprini2019, Otte2021, PerezIpina2019, Pattanayak2024, Pattanayak2025}. We include it here as an essential reference point for comparing with our time-dependent excess kurtosis results, as well as with the corresponding three-dimensional behavior.

For fixed $\Omega = 100$ and $\beta = 1$ [Fig.~\ref{fig2}(a)], increasing ${\rm Pe}$ enhances both the amplitude and the duration of the ballistic regime.
At fixed ${\rm Pe} = 100$ and $\beta = 1$ [Fig.~\ref{fig2}(b)], increasing chirality $\Omega$ suppresses the MSD at longer times, as rapid rotational motion diminishes directed propulsion.
For fixed ${\rm Pe} = 100$ and $\Omega = 100$ [Fig.~\ref{fig2}(c)], stronger confinement (larger $\beta$) enhances relaxation and lowers the steady-state value.
In the weak-trap limit ($\beta \ll 1$), the MSD remains unbounded over long times, with steady-state saturation occurring only at very late times.

The steady-state MSD(Eq.~\eqref{eq:msd_2d_st}) plotted in Figs.~\ref{fig2}(d)--(f), quantifies its dependence on activity ${\rm Pe}$, chirality $\Omega$, and trap strength $\beta$.
Figure~\ref{fig2}(d) highlights the passive Brownian limit, $\langle \tilde{\mathbf{r}}^{2}\rangle_{\rm st}^{\rm BP} = 2/\beta$, at low activity ($\Pe \to 0$), and the strong quadratic scaling $\langle \tilde{\mathbf{r}}^{2}\rangle_{\rm st} \propto \Pe^{2}$ that emerges at high activity ($\Pe \to \infty$). This transition occurs at critical activity $\Pe^{*}=\sqrt{2[(\beta+1)^2+\Omega^2]/(\beta+1)}$.
Figure~\ref{fig2}(e) shows that $\langle \tilde{\mathbf{r}}^{2}\rangle_{\rm st}$ decreases monotonically with increasing chirality, transitioning from the non-chiral active Brownian limit, $\langle \tilde{\mathbf{r}}^{2}\rangle_{\rm st}^{\rm ABP}=\langle \tilde{\mathbf{r}}^{2}\rangle_{\rm st}(\Omega=0)$~\cite{Chaudhuri2021}, in the low-chirality regime ($\Omega\to 0$), and saturating toward the passive Brownian value $\langle \tilde{\mathbf{r}}^{2}\rangle_{\rm st}^{\rm BP}$ in the high-chirality limit ($\Omega\to\infty$). The initial decay below the achiral active Brownian value begins at the critical chirality $\Omega^{*}_{1}=(\beta+1)$. A second critical chirality, defined as the point where the active correction in the $\Omega^{-2}$ tail becomes comparable to the Brownian contribution, is given by $\Omega^{*}_{2}=\Pe\sqrt{(\beta+1)/2}$. Beyond this scale, the MSD is dominated by the Brownian term and effectively saturates to $\langle \tilde{\mathbf{r}}^{2}\rangle_{\rm st}^{\rm BP}$.
Figure~\ref{fig2}(f) demonstrates the characteristic inverse dependence on trap strength, $\langle \tilde{\mathbf{r}}^{2}\rangle_{\rm st} \sim 1/\beta$, reflecting how strong confinement suppresses spatial fluctuations.

The steady-state fluctuations can also be expressed, per Cartesian component, through
the product $\sigma_{F_x}\sigma_x$ of the standard deviations of the trap force and the
displacement, which for a passive Brownian particle equals $k_BT$ by equipartition.
Since $\mathbf{F}=-k\br$ is proportional to the position, $\sigma_{F_x}=k\sigma_x$
exactly, independently of the mean, and since $\langle\br\rangle_{\rm st}=0$
(Appendix-\ref{app-A}), the product is
$\sigma_{F_x}\sigma_x=k\langle x^2\rangle_{\rm st}$.
Using $D=k_BT/\gamma$, this is $\beta\langle\tilde x^2\rangle_{\rm st}$ in units of
$k_BT$, giving
\begin{align}
\frac{\sigma_{F_x} \sigma_x}{k_BT}= 1 + \frac{\Pe^2}{2}
\frac{\beta+1}{(\beta+1)^2+\Omega^2}\,.
\label{eq:force_disp_2d}
\end{align}
The active contribution is the second term, which depends non-monotonically on trap
strength whenever $\Omega>1$: the product attains a maximum $1+\Pe^2/(4\Omega)$ at
$\beta^{*}=\Omega-1$, in contrast to the achiral case $\Omega=0$, where the active
term decays monotonically as $\Pe^2/[2(\beta+1)]$.
The maximum occurs where the decay rate $\beta+1$ of the position--orientation
correlation $\langle\tilde{\mathbf r}\cdot\bu\rangle$
(Eq.~\eqref{eq:ur_cross_corr_dimensionless_2d}), which carries activity into the
position fluctuations, matches the chiral frequency $\Omega$. In a weak trap the
torque winds the trajectory into circular orbits whose radius is set by $\Pe/\Omega$
rather than by the trap, so the small restoring force $k\sigma_x$ keeps the product
low; in a strong trap the displacement itself is suppressed. The chiral active
particle therefore departs furthest from the equilibrium value at intermediate
confinement.

Although the mean-squared displacement (MSD) provides a quantitative view of the different dynamical regimes, it does not fully capture how the position distribution behaves, especially within the oscillatory regime. To characterize these deviations from Gaussian behavior, we compute the excess kurtosis as a measure of non-Gaussianity.

\begin{figure*}[t]
\centering
\includegraphics[width=\linewidth]{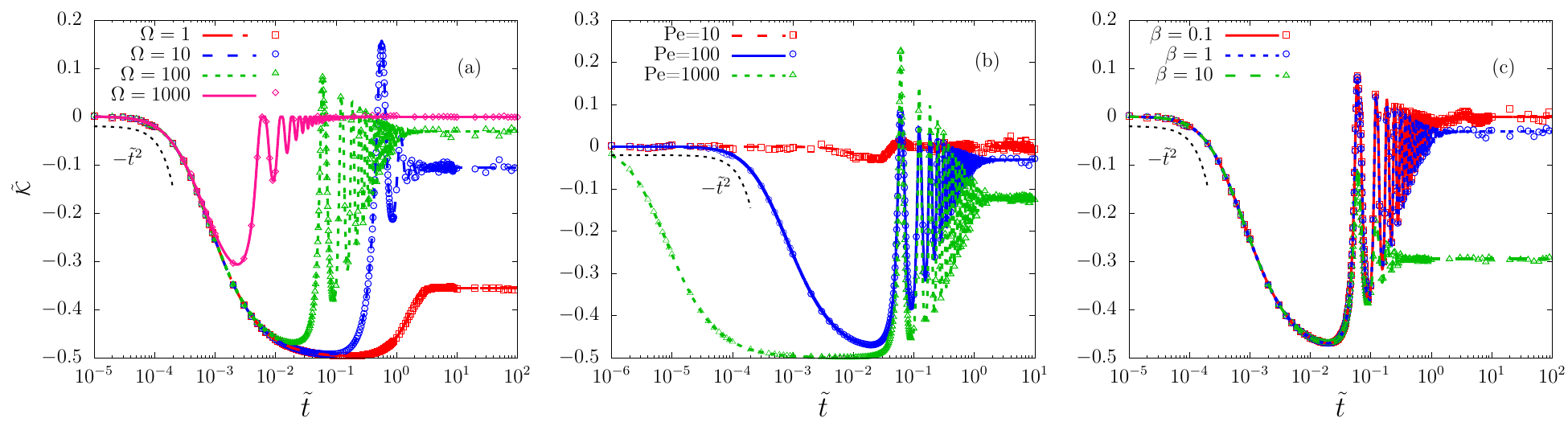}
\includegraphics[width=0.27\linewidth]{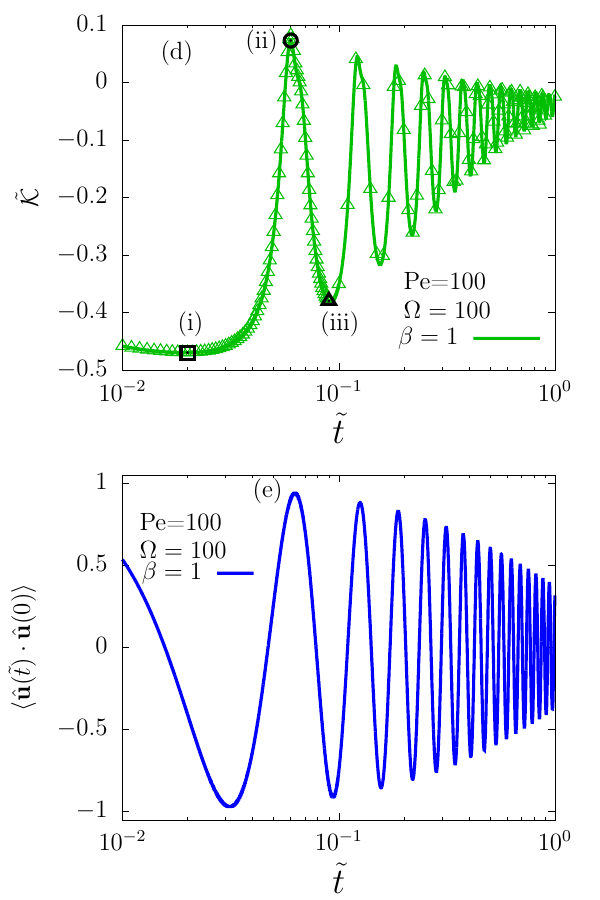}
\includegraphics[width=0.71\linewidth]{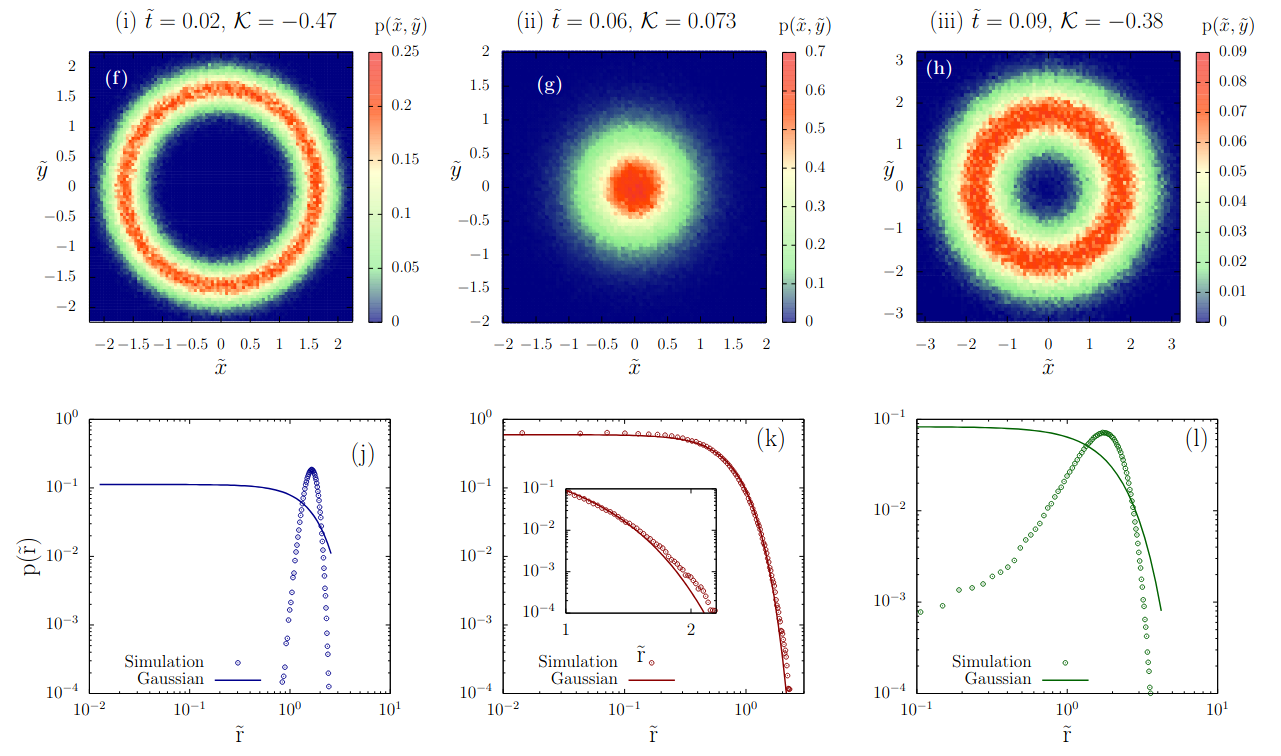}
\caption{
Excess kurtosis of a chiral active Brownian particle (CABP) in a harmonic trap in two dimensions (2d) as a function of time for (a) chirality $\Omega=1,10,100, 1000$ with activity $\text{Pe}=100$ and harmonic trap stiffness $\beta=1$, (b) $\text{Pe}=10, 100, 1000$ with $\Omega=100$ and $\beta=1$, and (c) $\beta=0.1,1,10$ with $\Pe=100$ and $\Omega=100$.
The lines in (a), (b) and (c) are plots of Eq.~(\ref{ex_kurt_2d}) and the points are from simulations.
(d) Three representative points with negative, positive, and negative excess kurtosis are selected to examine the corresponding position distributions in the $x-y$ plane (f-h) and in the radial direction (j-l).
(e) Analytic orientation autocorrelation function $\langle\bu\cdot\bu_0\rangle=e^{- \tilde{t}}\cos (\Omega \tilde{t})$(see detailed derivation in Appendix-\ref{app-A}) showing oscillations that occur in phase with the excess kurtosis. 
(f,h) Bimodal position distributions associated with the negative excess kurtosis active states (see (i) and (iii) in (d)), with the corresponding radial distributions in (j) and (l) exhibiting clear deviations from the Gaussian profile obtained from the MSD(Eq.~\eqref{eq:msd_2d_dimensionless}).
(g) Unimodal position distribution associated with the positive excess kurtosis state (see (ii) in (d)), with the corresponding radial distribution in (k) displaying a weak tail enhancement relative to the Gaussian(see inset of (k)).
}
\label{fig3}
\end{figure*}
\subsection{Excess kurtosis: Deviation from Gaussian process}
The deviation from the Gaussian process is measured by the excess kurtosis which can be written as~\cite{Pattanayak2024, Pattanayak2025},
\begin{equation}
    \tilde{\cal{K}}(\ttt) = \frac{\langle{\tbr^4}(\ttt)\rangle}{\mu_4(\ttt)} - 1\,,
    \label{ex_kurt_2d}
\end{equation}
where $\tilde{\mu}_4=\langle \delta \tbr^2\rangle^2 + 2\langle \delta \tilde{r}_i \delta \tilde{r}_j \rangle^2 + 2\langle \delta \tbr^2\rangle\langle \tilde{r}\rangle^2
+ 4\langle \tilde{r}_i\rangle\langle \tilde{r}_j \rangle\langle \delta \tilde{r}_i\delta \tilde{r}_j \rangle + \langle \tilde{r}\rangle ^4$. Averaging over all initial orientations, $\mu_4$ reduces to,
\begin{align}
    \tilde{\mu}_4=\langle  \tbr^2\rangle^2 + 2\langle  \tilde{r}_i  \tilde{r}_j \rangle^2=2 \langle  \tbr^2\rangle^2\,. 
    \label{mu4_avg}
\end{align}
Using the expressions of $\langle{\tbr^4}\rangle$ (see Appendix-\ref{app-A}; Eq.~\eqref{eq:r4avg_2d_dimensionless} and the figure in Appendix Fig.~\ref{app_fig1}) and $\langle  \tbr^2\rangle$(Eq.~\eqref{eq:msd_2d_dimensionless}), we get the exact time-dependent expression of the excess kurtosis.

\noindent
The small-time expansion of the excess kurtosis,
\begin{align}
    &\lim_{\tilde{t}\to 0} \tilde{\cal{K}} = -\frac{\text{Pe}^4 }{32}\ttt^2+\frac{1}{192} \text{Pe}^4 \left(3 \text{Pe}^2+4\right) \ttt^3\nonumber\\&-\frac{ \left(135 \text{Pe}^8+360 \text{Pe}^6-120 \beta ^2 \text{Pe}^4-120 \text{Pe}^4 \Omega ^2+136 \text{Pe}^4\right)}{23040}\ttt^4\nonumber\\&+ O\left(\ttt^5\right)\,.
    \label{eq:excess_kurtosis_2d_small_times}
\end{align}
The small-time expansion of the excess kurtosis shows that non-Gaussianity emerges through a sequence of well-separated processes. At very short times ($\tilde{t} \to 0$), the dynamics is dominated by translational diffusion, and $\tilde{\mathcal{K}} \simeq 0$, indicating nearly Gaussian behavior. The first leading term, $\tilde{\cal{K}}\simeq -\text{Pe}^4 \ttt^2/32$, indicates that the distribution initially becomes off-centered bimodal, reflecting the short-time regime where motion is dominated by self-propulsion. On the next timescale, $\ttt \simeq 6/(3\Pe^2+4) \sim \Pe^{-2}$, the $\ttt^3$ term partly offsets this trend, marking the onset of coherent active displacements. Confinement($\beta$), chirality($\Omega$) appear only at order $\ttt^4$, demonstrating that their influence enters more slowly. Thus, the early-time behavior reflects a clear hierarchy: activity controls the initial non-Gaussianity, while confinement and chirality modify it only at higher order. It is important to note that confinement first influences the excess kurtosis at $\mathcal{O}(\tilde{t}^{4})$, whereas its effect on the MSD already appears at $\mathcal{O}(\tilde{t}^{2})$(Eq.~\eqref{eq:MSD_2d_small_times}) and for fourth order moment of displacement at $\mathcal{O}(\tilde{t}^{3})$(see Appendix-\ref{app-A}; Eq.~\eqref{eq:r4avg_2d_small_time_dimensionless}).

In Fig.~\ref{fig3}(a)-(c), we show the time evolution of the excess kurtosis 
$\tilde{\mathcal{K}}(\ttt)$ of the displacement distribution for varying chirality ($\Omega$), activity ($\mathrm{Pe}$), and 
confinement strength ($\beta$). The analytical predictions (lines) show excellent agreement with the simulation results (points). At short times ($\tilde{t} \ll 1$), $\tilde{\mathcal{K}} \simeq 0$, indicating that the displacement 
statistics are nearly Gaussian and dominated by translational diffusion. 
As time increases, $\tilde{\mathcal{K}}$ becomes negative, signifying an activity-dominated state. 
This transition reflects dominance of persistent active motion as predicted in Eq.~\eqref{eq:excess_kurtosis_2d_small_times}. 

In the short to intermediate time regime, the effects of chirality and confinement on non-Gaussian fluctuations are minimal(see Eq.~\eqref{eq:excess_kurtosis_2d_small_times} and Fig.~\ref{fig3}(a),(c)).
At longer timescales ($\ttt > \Omega^{-1}$) and for sufficiently strong chirality ($\Omega > 1$), the influence of chirality becomes pronounced. In this regime, which corresponds to the onset of damped oscillatory behavior in the orientation autocorrelation (see Appendix-\ref{app-A} and Fig.~\ref{fig3}(e)), the excess kurtosis develops damped oscillations (Fig.~\ref{fig3}(a),(d)).

Damped oscillations in the higher-order statistics of free chiral active motion have been
reported previously~\cite{Sevilla2016, Pattanayak2024}.
They arise from the coherent rotation of the propulsion direction, which also causes the mean displacement to oscillate.
From Eq.~\eqref{eq:mean_disp_2d_dimensionless}, $\tilde{\rm r}_{\rm mean}$ is maximal after half a revolution, when $\langle\bu\cdot\bu_0\rangle=e^{-\ttt}\cos(\Omega\ttt)$ is most negative.
The distribution is then off-centered and ring-like, giving $\tilde{\cal K}<0$.
Half a period later, the orientational memory becomes positive, the mean offset decreases, and the distribution becomes centrally peaked with slightly enhanced tails, giving a weakly positive $\tilde{\cal K}$.
Thus, the excess kurtosis oscillates with the orientational memory and opposite to $\tilde{\rm r}_{\rm mean}$ [Figs.~\ref{fig3}(d),(e) and \ref{app_fig2}(a)].
Confinement leaves the frequency $\Omega$ unchanged but increases the damping from $e^{-\ttt}$ to $e^{-(\beta+1)\ttt}$ because the trap acts on position, not orientation.
The excess kurtosis therefore inherits an oscillatory memory damped jointly by rotational diffusion and confinement.

In the damped oscillatory regime, the excess kurtosis crosses zero and exhibits smooth transitions between negative and positive values as chirality increases.
The magnitude of these oscillations is most pronounced at intermediate chirality, reflecting a non-monotonic evolution of the active displacement distribution from off-centered configurations to near-Gaussian fluctuations. In contrast, the number of zero crossings increases with increasing chirality, although the positive excess kurtosis remains very close to zero. Furthermore, at sufficiently large chirality, the damped oscillatory regime is progressively suppressed and eventually disappears, with the excess kurtosis approaching zero without further oscillations.

With increasing activity ($\Pe$), the magnitude of the negative excess kurtosis increases, in agreement with the prediction of Eq.~\eqref{eq:excess_kurtosis_2d_small_times}. Concurrently, its minimum shifts to earlier times, scaling as $\tilde{t} \sim \Pe^{-2}$. This indicates that stronger self-propulsion advances the crossover from diffusive to non-Gaussian active dynamics to progressively earlier times (Fig.~\ref{fig3}(b)). Moreover, enhanced activity increases the amplitude of the damped oscillations in the excess kurtosis and leads to a greater number of zero crossings. In contrast, stronger confinement ($\beta$) suppresses the damped oscillatory regime and reduces the number of zero crossings, as the harmonic trap limits large spatial fluctuations (Fig.~\ref{fig3}(c)).

Figure~\ref{fig3}(d)–(l) illustrate how the sign of the excess kurtosis governs the time-dependent structure of the displacement distribution.
Notably, the oscillatory behavior of the excess kurtosis is out of phase with both the MSD and the fourth-order displacement moment(see Fig.~\ref{fig2}(a) and Appendix-\ref{app-A}; Fig.~\ref{app_fig1}(a) and Fig.~\ref{app_fig2}(c)), yet appears in phase with the oscillatory orientation autocorrelation shown in Fig.~\ref{fig3}(e). This out-of-phase relationship between the MSD or fourth-order moment and the orientation autocorrelation originates from the underlying position–orientation cross-correlation(see Appendix-\ref{app-A}; Fig.~\ref{app_fig2}(b)).

We identify three representative parameter points in Fig.~\ref{fig3}(d), corresponding to negative, positive, and again negative excess kurtosis, capturing the multiple re-entrant behavior observed over time.  These points are used to examine the structure of the position distribution in the $x-y$ plane(Fig.~\ref{fig3}(f),(g),(h)) and their radial projection(Fig.~\ref{fig3}(j),(k),(l)).
For the two points with negative excess kurtosis, shown in (f) and (h), the displacement distribution becomes bimodal and off-centered. This behavior arises when active circular motion dominates, producing ring-like structures rather than a single central peak. The corresponding radial distributions in (j) and (l) deviate clearly from the Gaussian reference obtained from the MSD(Eq.~\eqref{eq:msd_2d_dimensionless}), reflecting the off-centered nature of these states.
In contrast, the point with positive excess kurtosis, shown in (g), exhibits a unimodal, centrally peaked distribution that develops a weak tail enhancement. This is evident in the radial distribution in (k), which shows a slight enhancement relative to the Gaussian tail.
We note that the positive excursions are small, $\tilde{\cal{K}}\lesssim0.1$, so these
states are near-Gaussian; the physically significant feature is the sign change itself rather
than the magnitude of the positive value.
Together, (d)–(l) demonstrate that the sign and magnitude of the excess kurtosis provide a clear diagnostic of the spatial statistics: negative values correspond to off-centered bimodal distributions driven by persistent circular motion, while small positive values signal near-Gaussian fluctuations with a slight tail enhancement associated with a more diffusive, centrally peaked state.
The smooth crossovers between these behaviors underlie the multiple re-entrant structure identified in the excess kurtosis.

Finally, the steady-state excess kurtosis $\tilde{\mathcal{K}}(t\to\infty)$ ranges from negative values, corresponding to active states with off-centered bimodal position distributions, to small positive values that reflect near-Gaussian fluctuations. These behaviors are discussed in detail by Pattanayak {\it et al.}~\cite{Pattanayak2025}. A wide range of experiments exhibiting chiral rotation are described by theoretical models of chiral active Brownian particles at the level of second-order moments~\cite{Kummel2013, Lowen2016, Lopez2022, Kaur2025}. These systems therefore provide promising platforms for direct experimental verification of our excess kurtosis findings.

\section{Three-dimensions}
\label{sec-3}

\noindent
\textbf{Model:~}
In three dimensions, the heading direction (or orientation unit vector) is expressed as $\bu = (\sin\theta \cos\phi, \sin\theta \sin\phi, \cos\theta)$, where $\theta$ and $\phi$ denote the polar and azimuthal angles, respectively.
Within the It\^o interpretation~\cite{Ito1975, VandenBerg1985, Raible2004, Mijatovic2020, Sevilla2016}, the Langevin dynamics of the position expressed as~\cite{Pattanayak2025}:
\begin{align}
d{\br}(t) &= v_0\bu dt+\sqrt{2 D}\, d \bm{B}(t)-\mu \br dt\,.\label{eom1:3d}
\end{align}
We consider a torque $\bm{\omega} = \omega_0 (\sin\theta_\omega \cos\phi_\omega, \sin\theta_\omega \sin\phi_\omega, \cos\theta_\omega)$, characterized by its magnitude $\omega_0$ and orientation specified by the angles $\theta_\omega$ and $\phi_\omega$(see Fig.~\ref{fig1}).
The evolution of the heading direction $\bu$ is governed by the generalized torque ${\bm\omega} \times \bu$ together with orientational noise. The heading  direction angles evolves as~\cite{Pattanayak2024, Pattanayak2025}
\begin{align}
d\theta(t) &= \omega_{\theta} dt+ \frac{D_r}{\tan\theta} dt+\sqrt{2 D_r}\, d W_\theta(t)\,,\label{eom2:3d}
\\
d\phi(t) &=  \omega_{\phi} dt+\frac{\sqrt{2 D_r}\, d W_\phi(t)}{\sin\theta}\,,
\label{eom3:3d}
\end{align}
where the generalized forces are $\omega_{\theta} = {\bf \hat u}_\theta \cdot ({\bm\omega} \times {\bf \hat u}) =  \omega_0 \sin\theta_\omega \sin(\phi_\omega-\phi)$ and
$\omega_{\phi}
= {\bf \hat u}_\phi \cdot ({\bm\omega} \times {\bf \hat u})=\omega_0\left(\cos\theta_\omega-\cot\theta\sin\theta_\omega \cos(\phi_\omega-\phi)\right)$~\footnote{The unit vectors ${\bf \hat u}_\theta = \partial \hat{\bm u}/\partial \theta$ and ${\bf \hat u}_\phi =(1/\sin\theta) (\partial \hat{\bm u}/\partial \phi)$.}. 
Considering a constant torque, ${\bm{\omega}} =  \omega_0 \hat{z}$ which implies $\theta_\omega=0$, the equations reduce to:
\begin{align}
 d\theta(t) = \frac{D_r}{\tan\theta} dt+\sqrt{2 D_r}d W _\theta ; \,\,\, d\phi(t) =  \omega_0 dt+\frac{\sqrt{2 D_r}d W_\phi }{\sin\theta}.
 \end{align}
The translational and rotational Gaussian noise are characterized by $\langle d\bm{B}_{i}d\bm{B}_{j}\rangle=\delta_{ij}dt$, $\langle dW_\theta^2\rangle=dt$, $\langle dW_\phi^2\rangle=dt$ and $\langle d\bm{B}_{i}\rangle=\langle dW_\theta\rangle=\langle dW_\phi\rangle=0$.
As before, we formulate the corresponding Fokker–Planck equation, apply the Laplace transform, and carry out the subsequent analysis to obtain the dynamical moments exactly; the full derivations are provided in Appendix-\ref{app-B}.
We perform numerical integration of the above Langevin equations using the Euler–Maruyama scheme and verify our analytical predictions against the simulation results.

\begin{figure*}[t]
\centering
\includegraphics[width=\linewidth]{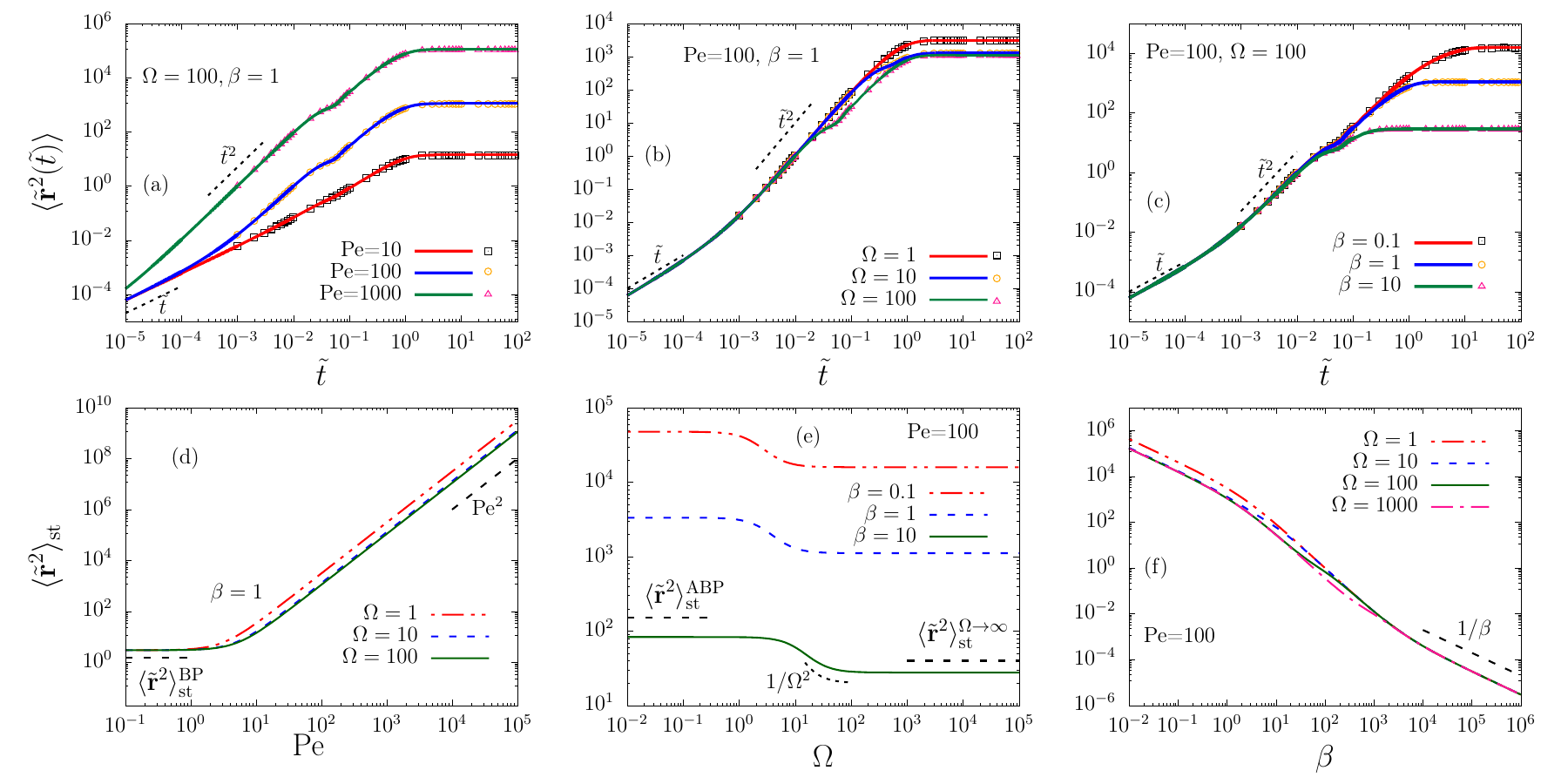}
\caption{
Mean-squared displacement (MSD) of an active Brownian particle under torque in a harmonic trap in three dimensions (3d) as a function of time for (a) $\text{Pe}=10, 100, 1000$ with $\Omega=100$ and $\beta=1$, (b) chirality $\Omega=1,10,100$ with activity $\text{Pe}=100$ and harmonic trap stiffness $\beta=1$, and (c) $\beta=0.1,1,10$ with $\Pe=100$ and $\Omega=100$.
Steady state MSD as a function of (d) $\Pe$ for $\Omega=1,10,100$ with $\beta=1$, (e) $\Omega$ for $\beta=0.1,1,10$ with $\Pe=100$, and (f) $\beta$ for $\Omega=1,10,100,1000$ with $\Pe=100$.
The lines in (a), (b) and (c) are plots of Eq.~(\ref{eq:msd_3d_dimensionless}) and the points are from simulations. The lines in (d), (e), and (f) are plots of Eq.~(\ref{eq:msd_3d_st_dimensionless}).
}
\label{fig4}
\end{figure*}

\subsection{Mean-squared displacement(MSD)}

In dimensionless form, the mean-squared displacement
(MSD) of a chiral active Brownian particle in three-dimensions is given by (see detailed derivation in Appendix-\ref{app-B})
\begin{widetext}
\begin{align}
&\langle \tbr^2 (\ttt)\rangle = \frac{3}{\beta} \left(1-e^{-2 \beta  \ttt}\right) +\frac{\text{Pe}^2 \left[3 (\beta +2)^2+\Omega ^2\right]}{3\beta(\beta +2)\left[(\beta +2)^2+\Omega ^2\right]}+\frac{\text{Pe}^2 e^{-6 \ttt} \left(3 \text{u}_{z0}^2-1\right) \Omega ^2}{3(\beta -4) (\beta -3) \left[(\beta -4)^2+\Omega ^2\right]}-\frac{2 \text{Pe}^2 e^{-(\beta +2) \ttt} \left[(\beta +2) \text{u}_{z0}^2-2\right]}{(\beta -4) (\beta -2) (\beta +2)} \nonumber\\
&+\frac{\text{Pe}^2 e^{-2 \beta  \ttt} \left[(\beta -3) (\beta -2)^2+\Omega ^2 \left(\beta  \text{u}_{z0}^2-1\right)\right]}{\beta(\beta -3) (\beta -2) \left[(\beta -2)^2+\Omega ^2\right]}+2 \text{Pe}^2 e^{-\ttt(\beta +2) } \cos (\Omega\ttt )\left[ \frac{ -2 ((\beta -3) \beta +6) \Omega ^2-(\beta -4) (\beta -2)^2 (\beta +2)-\Omega^4}{\left((\beta -4)^2+\Omega ^2\right) \left((\beta -2)^2+\Omega ^2\right) \left((\beta +2)^2+\Omega ^2\right)}\right.\nonumber\\&  \left.+\frac{\text{u}_{z0}^2 \left[(\beta -6) \beta +\Omega ^2+8\right] }{\left[(\beta -4)^2+\Omega ^2\right] \left[(\beta -2)^2+\Omega ^2\right] }\right] -\frac{4 \text{Pe}^2 \Omega  e^{-\ttt(\beta +2)} \sin ( \Omega \ttt) \left[(\beta -12) \beta +\text{u}_{z0}^2 \left[(\beta +2)^2+\Omega ^2\right]+\Omega ^2+20\right]}{\left[(\beta -4)^2+\Omega ^2\right] \left[(\beta -2)^2+\Omega ^2\right] \left[(\beta +2)^2+\Omega ^2\right]}\,.
\label{eq:msd_3d_dimensionless}
\end{align}
\end{widetext}
The total mean-squared displacement in three dimensions naturally decomposes into perpendicular and longitudinal contributions relative to the torque axis, $\langle \tbr^2 (\ttt)\rangle = \langle \tbr^2_\perp(\ttt)\rangle + \langle \tilde z^2(\ttt)\rangle$
where $\tilde{\mathbf r}_{\perp}=(\tilde x,\tilde y)$ denotes motion in the plane transverse to the torque axis and $\tilde z$ denotes motion along the torque axis. The closed form expressions of $\langle \tbr^2_\perp(\ttt)\rangle$ and $\langle \tilde z^2(\ttt)\rangle$  are given by (see detailed derivation in Appendix-\ref{app-B})
\begin{widetext}
    \begin{align}
    \nonumber
&\langle \tbr^2_\perp(\ttt)\rangle =  \frac{2}{\beta}\left(1 -e^{-2\beta\ttt}\right)+\frac{2 (\beta +2) \text{Pe}^2}{3 \beta  \left((\beta +2)^2+\Omega ^2\right)}-\frac{\text{Pe}^2 e^{-6 \ttt}(\beta -4)  \left(3 u_{z0}^2-1\right)}{3 (\beta -3) \left((\beta -4)^2+\Omega ^2\right)}-\frac{ \text{Pe}^2\,e^{-2\beta\ttt} (\beta -2)\left(\beta  \left(u_{z0}^2-1\right)+2\right)}{\beta(\beta -3) \left((\beta -2)^2+\Omega ^2\right)}+
\\\nonumber
&\frac{2 \text{Pe}^2 e^{-(\beta +2) \ttt} \cos ( \Omega \ttt) \left(-2 ((\beta -3) \beta +6) \Omega ^2-(\beta -4) (\beta -2)^2 (\beta +2)+u_{z0}^2 \left((\beta -6) \beta +\Omega ^2+8\right) \left((\beta +2)^2+\Omega ^2\right)-\Omega ^4\right)}{\left((\beta -4)^2+\Omega ^2\right) \left((\beta -2)^2+\Omega ^2\right) \left((\beta +2)^2+\Omega ^2\right)}
\\&
-\frac{4 \text{Pe}^2 \Omega  e^{-(\beta +2) \ttt} \sin ( \Omega \ttt) \left((\beta -12) \beta +u_{z0}^2 \left((\beta +2)^2+\Omega ^2\right)+\Omega ^2+20\right)}{\left((\beta -4)^2+\Omega ^2\right) \left((\beta -2)^2+\Omega ^2\right) \left((\beta +2)^2+\Omega ^2\right)}\,,
\label{eq:msd_3d_rperp}
\\
&\langle \tilde z^2(\ttt)\rangle = \frac{1}{\beta}\left(1 -e^{-2\beta\ttt}\right)+\frac{{\rm Pe}^2}{3\beta(\beta+2)}+\frac{\Pe^2\,e^{-6 \ttt} \left(3 u_{z0}^2-1\right) }{3 \left(\beta ^2-7 \beta +12\right)}-\frac{2\, \text{Pe}^2 e^{-(\beta +2) \ttt} \left((\beta +2) u_{z0}^2-2\right)}{(\beta -4) (\beta -2) (\beta +2)}-\frac{\text{Pe}^2\,e^{-2 \beta  \ttt} \left(1 -\beta\, u_{z0}^2\right)}{\beta  \left(\beta ^2-5 \beta +6\right)}\,.
\label{eq:msd_3d_z}
    \end{align}
\end{widetext}
The explicit expressions given above for $\langle \tilde{\mathbf{r}}_{\perp}^2\rangle$(Eq.~\eqref{eq:msd_3d_rperp}) and $\langle \tilde z^2\rangle$(Eq.~\eqref{eq:msd_3d_z}) together constitute Eq.~(\ref{eq:msd_3d_dimensionless}).
The first terms of right hand side of Eq.~\eqref{eq:msd_3d_rperp} and Eq.~\eqref{eq:msd_3d_z} are the passive translational contributions: they grow linearly for $\tilde t\ll \beta^{-1}$ and saturate for $\tilde t\gg \beta^{-1}$, indicating loss of memory beyond the translational relaxation time $\beta^{-1}$. The activity enters at order ${\rm Pe}^2$ and splits into qualitatively different transverse and longitudinal behaviors. Along the torque axis, the propulsion component parallel to $\hat{\mathbf z}$ is not rotated by the torque; consequently $\langle \tilde z^{2}\rangle$ contains no oscillatory $\cos(\Omega \tilde t)$ or $\sin(\Omega \tilde t)$ terms and its steady part is independent of $\Omega$, with $\langle \tilde z^{2}\rangle_{\rm st}\sim {\rm Pe}^2/[3\beta(\beta+2)]$. In contrast, the perpendicular propulsion is continuously rotated, producing helical/circular motion in the transverse plane; this gives the damped oscillatory pieces $\propto e^{-(\beta+2)\tilde t}\cos(\Omega\tilde t)$ and $\propto e^{-(\beta+2)\tilde t}\sin(\Omega\tilde t)$ in $\langle \tilde{\mathbf r}_{\perp}^{2}\rangle$, and an explicit torque dependence of its steady part through the factor $\left[(\beta+2)^2+\Omega^2\right]^{-1}$, i.e., increasing $\Omega$ suppresses transverse exploration. Finally, the dependence on $u_{z0}^2$ encodes the anisotropic short-time (ballistic) bias set by the initial orientation (more longitudinal growth for $u_{z0}^2 \to 1$, more transverse for $u_{z0}^2 \to 0$), while all such memory terms decay exponentially so that the long-time MSD becomes independent of initial conditions.

Note that $\langle \tilde{\mathbf r}_\perp^2\rangle$ describes motion in a plane but is not equivalent to the strictly 2d MSD in Eq.~\eqref{eq:msd_2d_dimensionless}, because it arises from projecting the full 3d dynamics, which has a different orientational relaxation spectrum. This difference is reflected not only in the replacement $(\beta+1)\to(\beta+2)$ in the active and oscillatory terms, but also in additional relaxation channels and anisotropic memory contributions proportional to $(3u_{z0}^2-1)$, which are absent in two dimensions. For isotropic initial orientation, $u_{z0}^2=\langle u_{z0}^2\rangle=1/3$, the anisotropic memory terms vanish, but $\langle \tilde{\mathbf r}_\perp^2\rangle$ still differs from the 2d MSD due to the distinct 3d orientational relaxation and additional relaxation channels.

Eq.~\eqref{eq:msd_3d_dimensionless} represents the full time-dependent mean-squared displacement (MSD) of an active Brownian particle under torque in a harmonic trap, expressed in dimensionless form.
It consists of a passive thermal relaxation(first term), a constant active contribution that sets the steady-state plateau(second term), and transient exponential terms that encode the crossover from short-time diffusion to an intermediate ballistic regime before confinement enforces saturation.
For isotropic initial orientations, corresponding to a uniform distribution over 
the unit sphere, the higher-harmonic transient contributions proportional to 
$e^{-6\tilde{t}}$ vanish exactly upon orientational averaging, since their 
prefactors contain $(3u_{z0}^2-1)\big|_{u_{z0}^2=1/3} = 0$. The oscillatory 
terms proportional to $e^{-(\beta+2)\tilde{t}}\cos(\Omega\tilde{t})$ and 
$e^{-(\beta+2)\tilde{t}}\sin(\Omega\tilde{t})$ survive but their amplitudes are 
strongly suppressed relative to the two-dimensional case, owing to the additional 
$\Omega$-dependent poles in the three-dimensional orientational relaxation 
spectrum and the dominant non-oscillatory longitudinal contribution $\langle 
\tilde{z}^2\rangle$, which together result in a smooth, effectively 
non-oscillatory MSD evolution toward the steady state (Fig.~\ref{fig4}).
In the short time limit,
\begin{align}
&\lim_{\ttt \to 0} \langle \tbr^2\rangle = 6 \ttt+\ttt^2 \left(\text{Pe}^2-6 \beta \right)+\frac{\ttt^3}{3} \left[12 \beta ^2-(3 \beta +2) \text{Pe}^2\right]\nonumber\\
&+ \frac{\ttt^4}{12} \left[-24 \beta ^3+\text{Pe}^2 \left(7 \beta ^2+8 \beta -\Omega ^2+4\right)+\text{Pe}^2 \Omega ^2 \text{u}_{z0}^2\right]\nonumber\\&+\mathcal{O}\left(t^5\right)\,.
\label{eq:MSD_3d_small_times}
\end{align}
At very small times the dynamics is dominated by thermal fluctuations $\langle \tbr^2\rangle=6\ttt$, which crosses over to ballistic behavior $\langle \tbr^2\rangle\sim \ttt^2$ at $\ttt_{I}=6/(\Pe^2-6\beta)$ with condition $\Pe>>\sqrt{6\beta}$. Notably, the $\mathcal{O}(\tilde{t}^{2})$ carries an explicit dependence on the trap strength $\beta$, indicating that confinement affects the short-time growth once self-propulsion becomes relevant similar to two-dimensions.
The effect of torque appears only at $\mathcal{O}(\tilde{t}^{4})$, demonstrating its limited relevance in the early-time dynamics.
We note that the torque-dependent contribution at $\mathcal{O}(\tilde{t}^{4})$ vanishes when the initial propulsion direction is aligned with the torque axis, i.e., $u_{z0}=\pm1$. In this configuration, the propulsion vector is parallel to the axis of rotation and therefore remains unchanged by the torque, resulting in no modification of the particle's displacement to leading order. In contrast, when the initial orientation has a transverse component, torque continuously rotates the propulsion direction in the perpendicular plane, reducing directional persistence and thereby suppressing the MSD. This behavior is fundamentally different from the two-dimensional case, where torque always rotates the propulsion direction and thus affects the dynamics regardless of the initial orientation.
To disentangle the perpendicular and longitudinal components relative to the torque axis
\begin{align}
&\langle \tilde{\mathbf r}_\perp^2(\ttt)\rangle
=
4\,\ttt +\left[{\rm Pe}^2(1-u_{z0}^2)
-4\beta\right]\ttt^2
\nonumber\\
&+\left[
\frac{8}{3}\beta^2
+{\rm Pe}^2\left(
\beta(u_{z0}^2-1)+\frac{8}{3}u_{z0}^2-\frac{4}{3}
\right)
\right]\ttt^3
\nonumber\\
&+\bigg[
-\frac{4}{3}\beta^3
+\frac{{\rm Pe}^2}{12}
\left(
\Omega^2 u_{z0}^2-\Omega^2
-7\beta^2 u_{z0}^2+7\beta^2
\right.\nonumber\\
&\left.-26\beta u_{z0}^2+14\beta
-52u_{z0}^2+20
\right)
\bigg]\ttt^4 +\mathcal O(\ttt^5)\,,
\label{eq:MSD_perp_3d_small_times}
\\
&\langle \tilde z^2(\ttt)\rangle =
2\,\ttt +\left[{\rm Pe}^2 u_{z0}^2
-2\beta\right]\ttt^2
\nonumber\\
& +\left[
\frac{4}{3}\beta^2
+{\rm Pe}^2\left(
-\beta u_{z0}^2-\frac{8}{3}u_{z0}^2+\frac{2}{3}
\right)
\right]\ttt^3
\nonumber\\
&+\left[
-\frac{2}{3}\beta^3
+\frac{{\rm Pe}^2}{12}
\left(
7\beta^2 u_{z0}^2
+26\beta u_{z0}^2
-6\beta
+52u_{z0}^2
-16
\right)
\right]\ttt^4
\nonumber\\
&+\mathcal O(\ttt^5)\,.
\label{eq:MSD_z_3d_small_times}
\end{align}
Equations~\eqref{eq:MSD_perp_3d_small_times} and \eqref{eq:MSD_z_3d_small_times} are obtained by performing a short-time expansion of the closed-form expressions given in Eqs.~\eqref{eq:msd_3d_rperp} and \eqref{eq:msd_3d_z}, respectively.
The leading terms, $4\ttt$ and $2\ttt$, simply reflect thermal diffusion in three dimensions, with two transverse and one longitudinal degrees of freedom contributing additively.
At order $\ttt^2$, activity enters anisotropically through the weights $(1-u_{z0}^2)$ and $u_{z0}^2$, showing that self-propulsion initially enhances fluctuations primarily along the instantaneous propulsion direction: an initially transverse orientation ($u_{z0}=0$) maximizes the active growth of $\langle \tilde{\mathbf r}_\perp^2\rangle$ while leaving the leading active contribution to $\langle \tilde z^2\rangle$ zero, whereas an initially axial orientation ($u_{z0}=\pm1$) does the opposite.
Confinement produces the negative terms $-4\beta\ttt^2$ and $-2\beta\ttt^2$ in the transverse and longitudinal components, respectively, indicating that the harmonic restoring force suppresses spreading already at the earliest nonlinear order.
Finally, torque affects the short-time dynamics only at $\mathcal O(\ttt^4)$ and only through the transverse MSD, since rotation about the $z$ axis changes the propulsion direction in the perpendicular plane but does not modify motion along the axis. Indeed, the torque-dependent part can be written as
\begin{align}
\left.\langle \tilde{\mathbf r}_\perp^2(\ttt)\rangle\right|_{\Omega}
=
-\frac{\Pe^2\Omega^2}{12}\,(1-u_{z0}^2)\,\ttt^4+\mathcal O(\ttt^5)\,,\nonumber
\end{align}
which vanishes for $u_{z0}=\pm1$ (initial alignment with the torque axis) and is maximal for $u_{z0}=0$ (purely transverse initial orientation), reflecting that torque reduces directional persistence only when the propulsion has a transverse component. Notably, upon averaging over isotropic initial orientations (uniform distribution of $\bu_0$ on the unit sphere, so that $u_{z0}^2=\langle u_{z0}^2\rangle = 1/3$), the short-time expansions up to $\mathcal{O}(\ttt^3)$ satisfy $\langle \tilde z^2(\ttt)\rangle= \langle \tilde{\mathbf r}_\perp^2(\ttt)\rangle/2$.

\noindent
Steady state MSD:
In the long time limit, the MSD reaches its steady state $\la\tbr^2\ra_{\rm st}=  \lim_{t\to \infty} \langle \tbr^2\rangle$~\cite{Pattanayak2025},
\begin{align}
  \la\tbr^2\ra_{\rm st} &= \frac{3}{\beta} + \frac{\text{Pe}^2 \left[3 (\beta +2)^2+\Omega ^2\right]}{3 \beta (\beta +2) [(\beta +2)^2 + \Omega ^2]}\,.
  \label{eq:msd_3d_st_dimensionless}
\end{align}

Note that the MSD(Eq.~\eqref{eq:msd_3d_dimensionless}) approaches to its steady-state value(Eq.~\eqref{eq:msd_3d_st_dimensionless}) by exponential relaxation i.e.,
\begin{align}
    \la \tbr^2\ra = \la\tbr^2\ra_{\rm st}+ \mathcal{O}(e^{-\lambda \ttt})\,,\nonumber
\end{align}
with $\lambda=\rm{min}(2\beta,\beta+2,6)$.
Thus, as in two dimensions, the relaxation toward the steady state is exponential, with characteristic time $\lambda^{-1}$ and steady state attained for $\lambda \ttt \gg 1$. The $e^{-6\ttt}$ term, originating from higher angular harmonics, vanishes upon averaging over initial orientations, reducing the effective rate to $\lambda=\min(2\beta,\beta+2)$. Hence, the long-time relaxation is governed by $\lambda=2\beta$ for $\beta<2$ and by $\lambda=\beta+2$ for $\beta>2$ (with degeneracy at $\beta=2$)(see Fig.~\ref{fig4}(a-c)). Physically, for weak trapping ($\beta<2$) the relaxation is governed by thermal fluctuations, whereas for strong trapping ($\beta>2$) it is controlled by the coupling between activity and confinement.

The first term in Eq.~\eqref{eq:msd_3d_st_dimensionless} represents the thermal equilibrium contribution from passive Brownian motion, while the second term corresponds to active contribution, which decreases with increasing confinement strength $\beta$. 
The active term grows quadratically with ${\rm Pe}$, indicating that stronger propulsion enhances fluctuations.
The steady-state MSD can be decomposed into perpendicular and longitudinal components relative to the torque axis,
\begin{align}
  \la\tbr_{\perp}^2\ra_{\rm st} &= \frac{2}{\beta} + \frac{2(\beta+2)\text{Pe}^2}{3 \beta [(\beta +2)^2 + \Omega ^2]}\,,
  \label{eq:msd_rperp_3d_st_dimensionless}\\
\la \tilde{z}^2\ra_{\rm st} &= \frac{1}{\beta} + \frac{\text{Pe}^2}{3 \beta (\beta +2)}\,.
  \label{eq:msd_z_3d_st_dimensionless}
\end{align}
These expressions reveal a fundamental anisotropy induced by torque: (i) The parallel component $\la \tilde{z}^2\ra_{\rm st}$ is independent of $\Omega$. This reflects that torque rotates the propulsion direction only in the plane perpendicular to its axis, leaving motion along the axis unaffected. (ii) The perpendicular component $\la\tbr_{\perp}^2\ra_{\rm st}$, decreases with increasing $\Omega$, since torque continuously redirects propulsion within the perpendicular plane, reducing persistence length and limiting spatial spread. Thus, torque selectively suppresses fluctuations transverse to its axis while leaving longitudinal fluctuations unchanged, producing anisotropic confinement even in an isotropic harmonic trap.

The corresponding trap-force--displacement products follow, per Cartesian
component, from Eqs.~\eqref{eq:msd_rperp_3d_st_dimensionless} and
\eqref{eq:msd_z_3d_st_dimensionless}, with
$\sigma_{F_x}\sigma_x/k_BT = 1 + \Pe^2(\beta+2)/\left\{3\left[(\beta+2)^2+\Omega^2
\right]\right\}$ transverse to the torque axis and
$\sigma_{F_z}\sigma_z/k_BT = 1+\Pe^2/\left[3(\beta+2)\right]$ along it. The transverse
product is again non-monotonic whenever $\Omega>2$, peaking at $1+\Pe^2/(6\Omega)$ at
$\beta^{*}=\Omega-2$, whereas the longitudinal one is independent of $\Omega$ and
decreases monotonically, consistent with the absence of oscillatory terms in
$\langle\tilde z^2\rangle$ (Eq.~\eqref{eq:msd_3d_z}).

Figure~\ref{fig4} shows the time evolution and steady-state behavior of the dimensionless mean-squared displacement (MSD), 
$\langle \tilde{\mathbf{r}}^2(\tilde{t})\rangle$, for an active Brownian particle under torque confined in a harmonic potential. Here, we consider isotropic initial orientations, corresponding to a uniform distribution of $\bu_0$ on the unit sphere, for which ensemble averaging $u_{z0}=\langle u_{z0}\rangle=0$ and $u_{z0}^2=\langle u_{z0}^2\rangle=1/3$.
The MSD evolution reveals four distinct dynamical regimes that appear consistently across parameters: an initial diffusive regime($\langle \tilde{\mathbf{r}}^2\rangle \sim \tilde{t}$) dominated by translational noise, ballistic regime($\langle \tilde{\mathbf{r}}^2\rangle \sim \tilde{t}^2$) dominated by persistent propulsion, an intermediate crossover regime without oscillatory behavior in MSD characterized by rotational decorrelation, and a final steady-state plateau where confinement balances chiral activity and diffusion.
Figure~\ref{fig4}(a)--(c) presents the analytic prediction from Eq.~\eqref{eq:msd_3d_dimensionless} (solid lines) 
compared with simulation results (points), showing excellent agreement across all parameter regimes.

For fixed $\Omega=100$ and $\beta=1$ [Fig.~\ref{fig4}(a)], increasing ${\rm Pe}$ enhances both the amplitude and duration of the ballistic regime, as stronger self-propulsion drives larger excursions before confinement becomes dominant. 
The steady-state value grows approximately as $\langle \tilde{\mathbf{r}}^2\rangle_{\rm st} \sim {\rm Pe}^2/\beta$, consistent with the analytical expression in Eq.~\eqref{eq:msd_3d_st_dimensionless}. 
At fixed ${\rm Pe}=100$ and $\beta=1$ [Fig.~\ref{fig4}(b)], increasing torque $\Omega$ suppresses MSD in the intermediate regime. 
Larger $\Omega$ values suppress the overall MSD amplitude, since rapid precession averages out directed motion and confines the particle more tightly near the trap center. 
For fixed ${\rm Pe}=100$ and $\Omega=100$ [Fig.~\ref{fig4}(c)], stronger confinement ($\beta$) accelerates relaxation and reduces the saturation value. 
In the weak-trap limit ($\beta \ll 1$), the MSD remains unbounded for long times, approaching steady-state at late times.

The steady-state results in Figs.~\ref{fig4}(d)--(f) quantify these dependencies. 
Figure~\ref{fig4}(d) shows the steady-state MSD increases with activity, transitioning from the passive Brownian value $\langle \tilde{\mathbf{r}}^2\rangle^{\rm BP}_{\rm st}=3/\beta$ at low $\Pe$ to a quadratic scaling $\langle \tilde{\mathbf{r}}^2\rangle_{\rm st} \propto \Pe^{2}$ at high activity, indicating that steady-state fluctuations are predominantly controlled by the strength of self-propulsion. This transition occurs at critical activity \begin{align}
\Pe^{*}=\left[\frac{9(\beta+2)[(\beta+2)^2+\Omega^2]}{3(\beta+2)^2+\Omega^2}\right]^{1/2}\,.
\end{align}
In Fig.~\ref{fig4}(e) shows that $\langle \tilde{\mathbf{r}}^2\rangle_{\rm st}$ decreases monotonically from a constant value($\la\tbr^2\ra_{\rm st}\simeq\Pe^2/\beta(\beta+2)+3/\beta$ at $\Omega \to 0$; which is MSD of ABPs without torque $\la\tbr^2\ra^{\rm ABP}_{\rm st}$) with increasing torque $\Omega$, approaching a constant($\la\tbr^2\ra_{\rm st}\simeq\Pe^2/3\beta(\beta+2)+3/\beta$; marked as $\la\tbr^2\ra^{\Omega\to\infty}_{\rm st}$) at large torque($\Omega\to\infty$) where fast rotational dynamics average out propulsion. The crossover between these two limits is governed by a single frequency scale $\Omega\sim(\beta+2)$ which corresponds to the point where the MSD has decreased halfway between the two plateaus. The onset of decay is marked by the inflection point $\Omega^{*}_{1}=(\beta+2)/\sqrt{3}$ obtained from $\mathrm{d}^2\la\tbr^2\ra_{\rm st}/\mathrm{d}^2\Omega=0$. A second characteristic torque can be defined by considering $95\%$  saturation toward the high-$\Omega$ plateau, giving $\Omega^{*}_{2}\sim (\beta+2)\sqrt{19}$
beyond which the MSD is effectively indistinguishable from $\la\tbr^2\ra^{\Omega\to\infty}_{\rm st}$. The decay at intermediate torque scales as $\la\tbr^2\ra_{\rm st}\sim\Omega^{-2}$.
Finally, Fig.~\ref{fig4}(f) demonstrates an inverse dependence $\langle \tilde{\mathbf{r}}^2\rangle_{\rm st} \sim 1/\beta$, highlighting that confinement directly suppresses the spatial extent of active fluctuations. 
In the strong-trap regime, all curves collapse, showing that torque becomes irrelevant once the trap dominates the dynamics.

In summary, it illustrates the complete dynamical crossover from diffusive to ballistic to an intermediate crossover regime without oscillation to localized motion.
Notably, the time-dependent non-oscillatory MSD exhibits a clear qualitative difference from its two-dimensional, oscillatory counterpart(Fig.~\ref{fig2}(a)--(c)), whereas the steady-state MSD shows the same qualitative scaling behavior as in two dimensions(Fig.~\ref{fig2}(d)--(f)).

Similar to the two-dimensional case, although the exact mean-squared displacement (MSD) offers a quantitative view of the dynamical regimes, it does not fully capture the behavior of the position distribution. To assess deviations from Gaussianity, we therefore compute the excess kurtosis as a measure of non-Gaussian fluctuations.

\begin{figure*}[t]
\centering
\includegraphics[width=\linewidth]{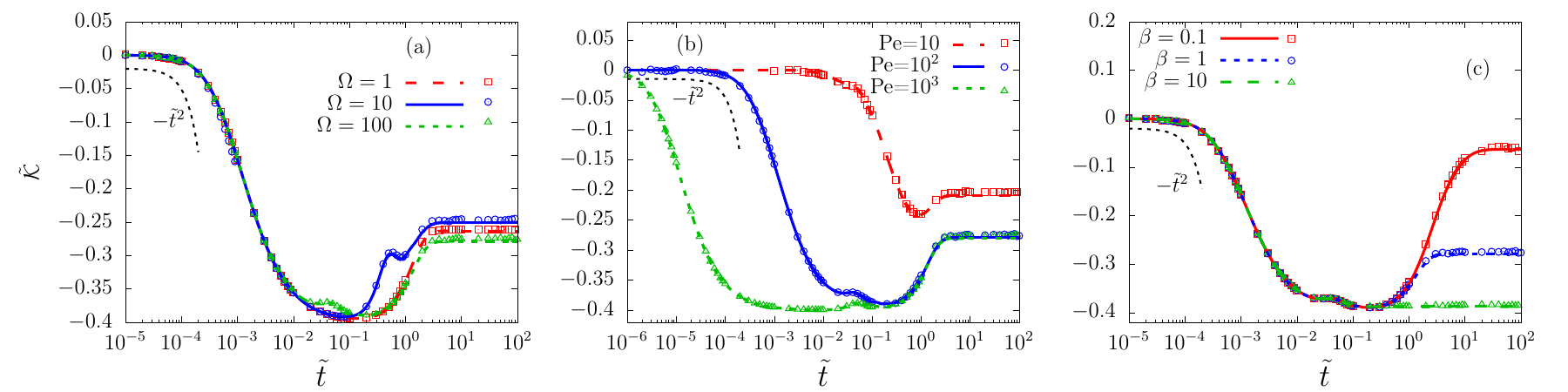}
\includegraphics[width=\linewidth]{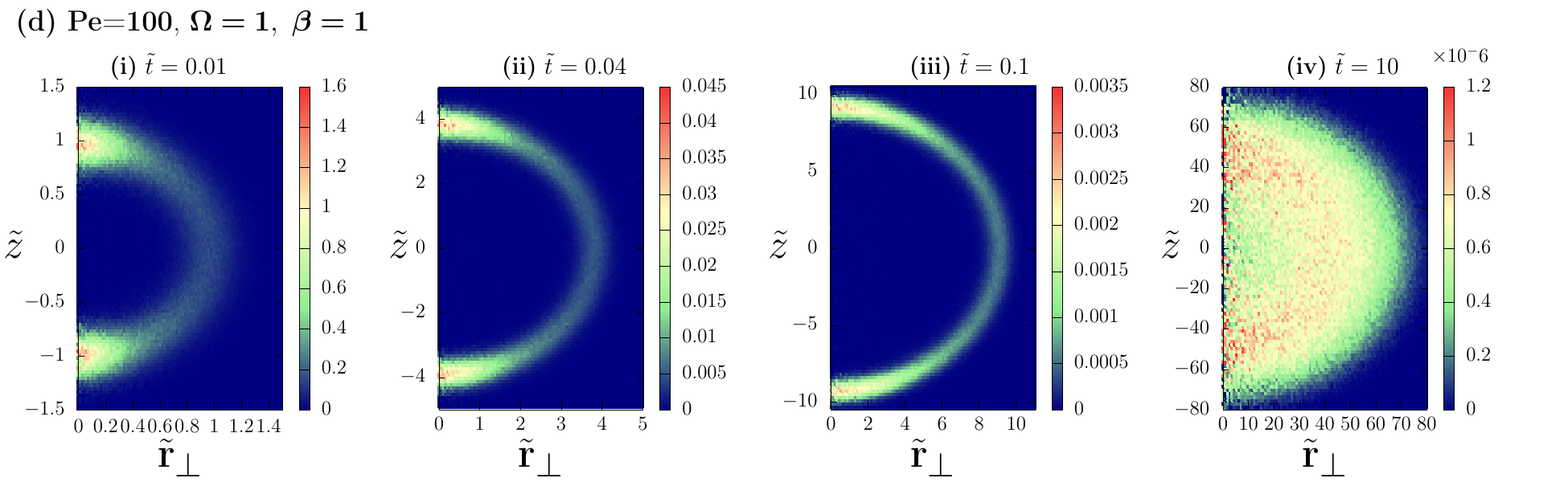}
\includegraphics[width=\linewidth]{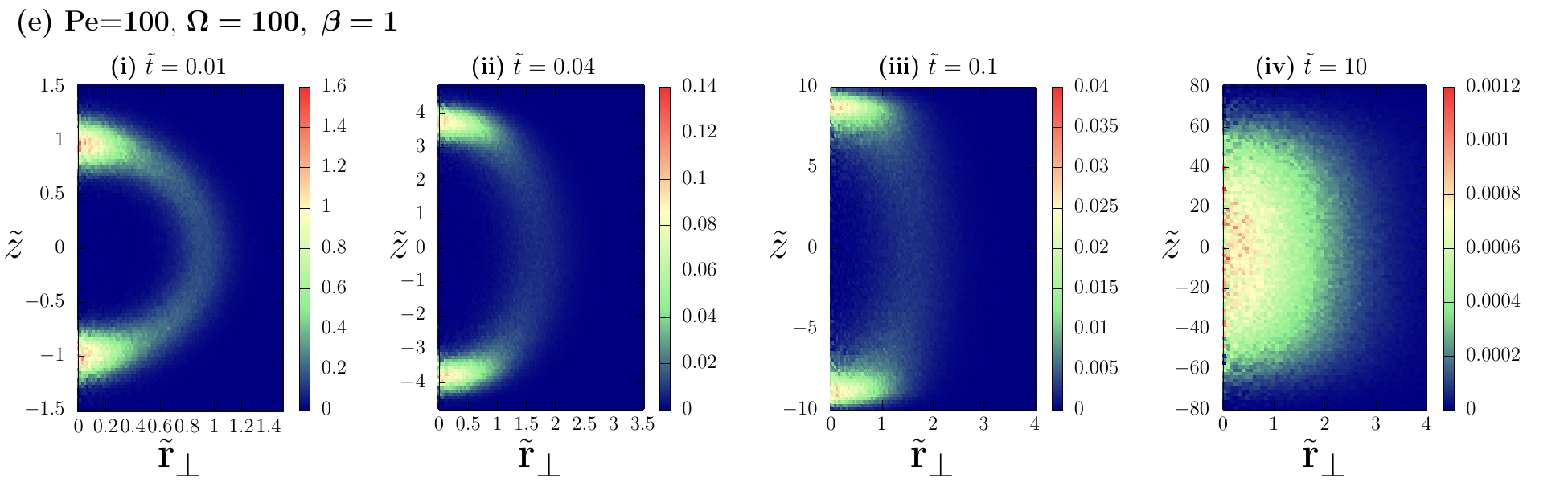}
\caption{
Excess kurtosis of an active Brownian particle under torque in a harmonic trap in three dimensions (3d) as a function of time for (a) torque $\Omega=1,10,100$ with activity $\text{Pe}=100$ and harmonic trap stiffness $\beta=1$, (b) $\text{Pe}=10, 100, 1000$ with $\Omega=100$ and $\beta=1$, and (c) $\beta=0.1,1,10$ with $\Pe=100$ and $\Omega=100$. Lines represent analytic predictions, while points denote simulation results.
(d,e) Two-dimensional projections of the position distribution in the $(\tilde r_{\perp}, \tilde z)$ plane(where 
$\tilde r_{\perp} = \sqrt{\tilde x^{2} + \tilde y^{2}}$) at four different time snapshots in the active regimes($\tilde{\cal{K}}<0$) for weak torque $\Omega = 1$ (d) and strong torque $\Omega = 100$ (e) with $\mathrm{Pe} = 100$ and $\beta = 1$. In the active regime, the distribution evolves from an initial half-ring–like structure to a long-time off-centered half ring at weak torque, and to band-like structures at strong torque, reflecting how increasing rotational precession suppresses radial spreading and enhances anisotropic active motion.
}
\label{fig5}
\end{figure*}

\subsection{Excess kurtosis: Deviation from Gaussian process}
We can now determine the excess kurtosis as a function of time to capture the deviation from Gaussian-like behavior. 
To evaluate it, we first express the fourth central moment $\mu_4$ in terms of the second moments. 
After averaging over all possible initial orientations, Eq.~\eqref{mu4_avg} reduces to~\cite{Pattanayak2025}
\begin{align}
&\mu_4  = \langle\br^2\rangle^2+2\Big( \langle  x^2 \rangle^2 +\langle  y^2 \rangle^2+ \langle  z^2 \rangle^2\nonumber\\
&+\langle  x y \rangle^2+\langle  yz \rangle^2+\langle  z x \rangle^2\Big)\,.
\label{eq:mu4avg_3d}
\end{align}
We can calculate the expression of $\mu_4$ using the expressions derived in previous section. Similar method of two dimensional excess kurtosis in Eq.~\eqref{ex_kurt_2d}, we calculate exact excess kurtosis in three dimensions using $\langle{\tbr^4}\rangle(\ttt)$(see Appendix-\ref{app-B}; Eq.~\eqref{eq:r4_3d_dimensionless} and Fig.~\ref{app_fig3}) and Eq.~\eqref{eq:mu4avg_3d}. In the short-time regime, the excess kurtosis $\tilde{\mathcal{K}}$ approaches zero, 
indicating that the displacement distribution remains nearly Gaussian. 
This behavior can be quantified by performing a series expansion around $\tilde{t} = 0$, 
which gives
\begin{align}
\lim_{\ttt \to 0}\tilde {\cal{K}} &= -\frac{\text{Pe}^4 \ttt^2}{90}+\frac{ \text{Pe}^4 \left(\text{Pe}^2+4\right) \ttt^3}{270}\nonumber\\
&+\frac{\text{Pe}^4 \ttt^4 \left(30 \beta ^2-15 \left(\text{Pe}^2+8\right) \text{Pe}^2+20 \Omega ^2-152\right)}{16200}\nonumber\\
&+O\left(\ttt^5\right)\,.
\label{eq:excess_kurtosis_3d_small_times}
\end{align}
Its leading behavior, $\tilde{\mathcal{K}} \simeq -\mathrm{Pe}^4\tilde{t}^2/90 + \mathcal{O}(\tilde{t}^3)$, indicates activity-induced non-Gaussianity and is consistent with the behavior of a free active Brownian particle under torque reported in~\cite{Pattanayak2024}. Higher-order corrections become significant as $\tilde{t}$ approaches the intrinsic dynamical timescales associated with confinement and torque.
It is important to note that confinement first influences the excess kurtosis at $\mathcal{O}(\tilde{t}^{4})$, whereas its effect on the MSD already appears at $\mathcal{O}(\tilde{t}^{2})$(Eq.~\eqref{eq:MSD_3d_small_times}) and for fourth order moment of displacement at $\mathcal{O}(\tilde{t}^{3})$(see Appendix-\ref{app-B}; Eq.~\eqref{eq:r4avg_3d_small_time_dimensionless}).
The steady-state excess kurtosis remains negative, as explicitly discussed in our recent work in Pattanayak {\it et al.}~\cite{Pattanayak2025}. Here, we focus on the exact temporal behavior of the excess kurtosis, which serves as a key indicator of dynamical crossovers between Gaussian-like and non-Gaussian regimes.

In Fig.~\ref{fig5}, we show the time evolution of the excess kurtosis 
$\tilde{\mathcal{K}}$ of the displacement distribution for varying activity ($\mathrm{Pe}$), 
confinement strength ($\beta$), and torque ($\Omega$). 
At short times ($\tilde{t} \ll 1$), $\tilde{\mathcal{K}} \approx 0$, indicating that the displacement 
statistics are nearly Gaussian and dominated by translational diffusion. 
As time increases, $\tilde{\mathcal{K}}$ becomes negative, signifying an activity-dominated state. 
This transition reflects the competition between persistent active motion and the restoring 
effect of the harmonic trap.
Torque ($\Omega$), activity ($\mathrm{Pe}$), and trap stiffness ($\beta$) each exert a non-monotonic influence as a function of time on the degree of non-Gaussianity.
In the small to intermediate time regime, the effect of torque on non-Gaussian fluctuations remains minimal, consistent with the prediction of Eq.~\eqref{eq:excess_kurtosis_3d_small_times}. At larger timescales, however, the influence of torque becomes significant in Fig.~\ref{fig5}(a). For increasing activity ($\mathrm{Pe}$), the magnitude of the negative excess kurtosis grows and its minimum 
shifts to shorter times, showing that strong self-propulsion accelerates the crossover from diffusive 
to non-Gaussian dynamics in Fig.~\ref{fig5}(b). 
In contrast, stronger confinement ($\beta$) suppresses these deviations, as the trap restricts large 
fluctuations and restores Gaussian-like behavior at earlier times in Fig.~\ref{fig5}(c).

We characterize further the active regimes(negative excess kurtosis) in Fig.~\ref{fig5}(a)–(c) by calculating spatial distribution of particle displacements.
In Fig.~\ref{fig5}(d), for weak torque ($\Omega = 1$), the distribution in the $(\tilde r_{\perp},\tilde z)$ plane shows a pronounced half–ring–like structure at intermediate times, reflecting the emergence of directed active motion. At longer times, this structure becomes increasingly off-centered, consistent with the negative excess kurtosis and the particle’s persistent escape from the trap center.
In contrast, Fig.~\ref{fig5}(e) demonstrates that strong torque ($\Omega = 100$) dramatically alters this evolution: the distribution rapidly develops into narrow, band-like patterns aligned with the rotation axis, indicating that fast rotational precession suppresses radial spreading and enhances anisotropic active motion.
These visual signatures in Fig.~\ref{fig5}(d) and Fig.~\ref{fig5}(e) mirror the temporal evolution of $\tilde{\mathcal{K}}$ and highlight how torque controls both the magnitude of non-Gaussian fluctuations and the geometry of displacement statistics.

The absence of re-entrant positive excess kurtosis in three dimensions arises because the torque rotates only the propulsion component perpendicular to its axis.
In two dimensions, the entire orientation rotates, so the orientational memory periodically reverses sign and the ring-like distribution collapses into a centrally peaked one, producing transient positive $\tilde{\cal K}$ (Fig.~\ref{fig3}).
In three dimensions, however, the longitudinal component $u_z$ remains unaffected.
Consequently, motion along the torque axis broadens the distribution monotonically while the transverse motion rotates.
The distribution therefore remains half-ring-like or band-like [Fig.~\ref{fig5}(d),(e)], keeping weight away from the origin and $\tilde{\cal K}<0$.
The faster orientational relaxation in three dimensions, $2D_r$ rather than $D_r$, further suppresses the oscillations.

\section{Conclusions}
\label{sec-4}
In this work, we developed an exact analytical framework for the transient dynamics of a torque-driven (chiral) active Brownian particle in a harmonic trap in both two and three dimensions. Using a Laplace-transform approach to the Fokker–Planck equation, we obtained closed-form expressions for time-dependent displacement moments up to fourth order, enabling a complete characterization of excess kurtosis across transient and steady-state regimes.

Our analysis reveals a striking dimensional contrast. In two dimensions, the excess kurtosis exhibits damped oscillations with multiple re-entrant sign changes, reflecting crossovers between off-centered, ring-like distributions and transient near-Gaussian fluctuations with weakly enhanced tails. These oscillations are controlled by the interplay of activity, chirality, and confinement, and are directly linked to the rotational dynamics of particle orientation. In three dimensions, by contrast, the excess kurtosis remains strictly negative throughout, indicating robust non-Gaussian fluctuations without oscillatory crossovers.
This qualitative difference originates in the fact that a torque rotates only the propulsion
component transverse to its axis, so that in three dimensions the longitudinal motion
broadens the position distribution monotonically and prevents the periodic collapse to a
centrally peaked form that generates positive excess kurtosis in two dimensions.

The theoretical predictions show excellent agreement with numerical simulations and identify clear experimental signatures, including oscillatory versus monotonic kurtosis evolution and its dependence on activity, chirality, and confinement strength.
The closed-form expressions rely on the moment hierarchy closing, which requires a
linear restoring force, so they are specific to harmonic confinement. The mechanism above
does not use the form of the potential: the axial coordinate broadens monotonically under
any confining potential, and the absence of re-entrant positive kurtosis in three
dimensions should therefore persist beyond the harmonic case.
Fourth-order statistics are sensitive to measurement error, but additive localization noise
attenuates the signal rather than creating spurious structure. If the recorded position
carries an independent, approximately Gaussian error of standard deviation
$\sigma_\epsilon$ (in units of $\ell$), the fourth cumulant is unchanged while the variance
increases, so the measured excess kurtosis is
$\tilde{\cal{K}}_{\rm meas}=\tilde{\cal{K}}\,[1+\sigma_\epsilon^2/\langle\tilde
x^2\rangle]^{-2}$ (two-dimensions). The prefactor is strictly positive, so the sign of $\tilde{\cal{K}}$ and
the times of its zero crossings are unaffected, and the attenuation is correctable once
$\sigma_\epsilon$ is calibrated from immobilized particles.
These features are most readily accessed in synthetic systems where large ensembles can
be accumulated with high localization accuracy, including circle
microswimmers~\cite{Kummel2013}, chiral granular rotors~\cite{Lopez2022}, and air-driven
spinners~\cite{Farhadi2018}; the excess kurtosis itself has already been measured in
self-propelled colloidal systems~\cite{Zheng2013}. In bacteria and sperm cells the same
signatures should be present, but smaller ensembles and cell-to-cell variability in propulsion
speed and chiral frequency make quantitative validation a longer-term prospect.
Overall, our results establish a rigorous framework for quantifying higher-order fluctuations in confined chiral active systems and highlight the fundamental role of dimensionality in shaping non-equilibrium dynamics.

\section*{Acknowledgments}
AP and AC thank the computing facility at IISER Mohali. AS and DC acknowledge SAMKHYA, the high-performance computing facility at the Institute of Physics, Bhubaneswar. AC acknowledges support from the Indo-German grant (IC-12025(22)/1/2023-ICD-DBT).  D.C. acknowledges the Department of Atomic Energy (India) for grant 1603/2/2020/IoP/R\&D-II/150288 and CY Cergy Paris Universit{\'e} for a Visiting Professorship.

\section*{Author contributions}
DC and AC conceived and designed the research. AP and AS performed the analytical and numerical calculations. All authors contributed equally to the preparation of the manuscript.

\section*{Data availability} 
The data that support the findings of this study are available within the article.

\section*{Code availability}
Code supporting this study is available from the corresponding author upon reasonable request.

\section*{Conflict of Interest}
The authors declare no conflicts of interest.

\bibliography{reference}

\begin{thebibliography}{70}%
\makeatletter
\providecommand \@ifxundefined [1]{%
 \@ifx{#1\undefined}
}%
\providecommand \@ifnum [1]{%
 \ifnum #1\expandafter \@firstoftwo
 \else \expandafter \@secondoftwo
 \fi
}%
\providecommand \@ifx [1]{%
 \ifx #1\expandafter \@firstoftwo
 \else \expandafter \@secondoftwo
 \fi
}%
\providecommand \natexlab [1]{#1}%
\providecommand \enquote  [1]{``#1''}%
\providecommand \bibnamefont  [1]{#1}%
\providecommand \bibfnamefont [1]{#1}%
\providecommand \citenamefont [1]{#1}%
\providecommand \href@noop [0]{\@secondoftwo}%
\providecommand \href [0]{\begingroup \@sanitize@url \@href}%
\providecommand \@href[1]{\@@startlink{#1}\@@href}%
\providecommand \@@href[1]{\endgroup#1\@@endlink}%
\providecommand \@sanitize@url [0]{\catcode `\\12\catcode `\$12\catcode
  `\&12\catcode `\#12\catcode `\^12\catcode `\_12\catcode `\%12\relax}%
\providecommand \@@startlink[1]{}%
\providecommand \@@endlink[0]{}%
\providecommand \url  [0]{\begingroup\@sanitize@url \@url }%
\providecommand \@url [1]{\endgroup\@href {#1}{\urlprefix }}%
\providecommand \urlprefix  [0]{URL }%
\providecommand \Eprint [0]{\href }%
\providecommand \doibase [0]{https://doi.org/}%
\providecommand \selectlanguage [0]{\@gobble}%
\providecommand \bibinfo  [0]{\@secondoftwo}%
\providecommand \bibfield  [0]{\@secondoftwo}%
\providecommand \translation [1]{[#1]}%
\providecommand \BibitemOpen [0]{}%
\providecommand \bibitemStop [0]{}%
\providecommand \bibitemNoStop [0]{.\EOS\space}%
\providecommand \EOS [0]{\spacefactor3000\relax}%
\providecommand \BibitemShut  [1]{\csname bibitem#1\endcsname}%
\let\auto@bib@innerbib\@empty
\bibitem [{\citenamefont {van Teeffelen}\ and\ \citenamefont
  {L\"owen}(2008)}]{van-Teeffelen2008}%
  \BibitemOpen
  \bibfield  {author} {\bibinfo {author} {\bibfnamefont {S.}~\bibnamefont {van
  Teeffelen}}\ and\ \bibinfo {author} {\bibfnamefont {H.}~\bibnamefont
  {L\"owen}},\ }\bibfield  {title} {\bibinfo {title} {Dynamics of a brownian
  circle swimmer},\ }\href {https://doi.org/10.1103/PhysRevE.78.020101}
  {\bibfield  {journal} {\bibinfo  {journal} {Phys. Rev. E}\ }\textbf {\bibinfo
  {volume} {78}},\ \bibinfo {pages} {020101} (\bibinfo {year}
  {2008})}\BibitemShut {NoStop}%
\bibitem [{\citenamefont {Chen}\ \emph {et~al.}(2017)\citenamefont {Chen},
  \citenamefont {Liu}, \citenamefont {qing Shi}, \citenamefont {Chaté},\ and\
  \citenamefont {Wu}}]{Chen2017}%
  \BibitemOpen
  \bibfield  {author} {\bibinfo {author} {\bibfnamefont {C.}~\bibnamefont
  {Chen}}, \bibinfo {author} {\bibfnamefont {S.}~\bibnamefont {Liu}}, \bibinfo
  {author} {\bibfnamefont {X.}~\bibnamefont {qing Shi}}, \bibinfo {author}
  {\bibfnamefont {H.}~\bibnamefont {Chaté}},\ and\ \bibinfo {author}
  {\bibfnamefont {Y.}~\bibnamefont {Wu}},\ }\bibfield  {title} {\bibinfo
  {title} {Weak synchronization and large-scale collective oscillation in dense
  bacterial suspensions},\ }\href {https://doi.org/10.1038/nature20817}
  {\bibfield  {journal} {\bibinfo  {journal} {Nature}\ }\textbf {\bibinfo
  {volume} {542}},\ \bibinfo {pages} {210} (\bibinfo {year}
  {2017})}\BibitemShut {NoStop}%
\bibitem [{\citenamefont {Kim}\ \emph {et~al.}(2018)\citenamefont {Kim},
  \citenamefont {Yoshinaga}, \citenamefont {Bhattacharyya}, \citenamefont
  {Nakazawa}, \citenamefont {Umetsu},\ and\ \citenamefont {Teizer}}]{Kim2018}%
  \BibitemOpen
  \bibfield  {author} {\bibinfo {author} {\bibfnamefont {K.}~\bibnamefont
  {Kim}}, \bibinfo {author} {\bibfnamefont {N.}~\bibnamefont {Yoshinaga}},
  \bibinfo {author} {\bibfnamefont {S.}~\bibnamefont {Bhattacharyya}}, \bibinfo
  {author} {\bibfnamefont {H.}~\bibnamefont {Nakazawa}}, \bibinfo {author}
  {\bibfnamefont {M.}~\bibnamefont {Umetsu}},\ and\ \bibinfo {author}
  {\bibfnamefont {W.}~\bibnamefont {Teizer}},\ }\bibfield  {title} {\bibinfo
  {title} {Large-scale chirality in an active layer of microtubules and kinesin
  motor proteins},\ }\href {https://doi.org/10.1039/c7sm02298k} {\bibfield
  {journal} {\bibinfo  {journal} {Soft Matter}\ }\textbf {\bibinfo {volume}
  {14}},\ \bibinfo {pages} {3221} (\bibinfo {year} {2018})}\BibitemShut
  {NoStop}%
\bibitem [{\citenamefont {Oliver}\ \emph {et~al.}(2018)\citenamefont {Oliver},
  \citenamefont {Alpmann}, \citenamefont {Álvaro Barroso}, \citenamefont
  {Dewenter}, \citenamefont {Woerdemann},\ and\ \citenamefont
  {Denz}}]{Oliver2018}%
  \BibitemOpen
  \bibfield  {author} {\bibinfo {author} {\bibfnamefont {N.}~\bibnamefont
  {Oliver}}, \bibinfo {author} {\bibfnamefont {C.}~\bibnamefont {Alpmann}},
  \bibinfo {author} {\bibnamefont {Álvaro Barroso}}, \bibinfo {author}
  {\bibfnamefont {L.}~\bibnamefont {Dewenter}}, \bibinfo {author}
  {\bibfnamefont {M.}~\bibnamefont {Woerdemann}},\ and\ \bibinfo {author}
  {\bibfnamefont {C.}~\bibnamefont {Denz}},\ }\bibfield  {title} {\bibinfo
  {title} {Synchronization in pairs of rotating active biomotors},\ }\href
  {https://doi.org/10.1039/C8SM00022K} {\bibfield  {journal} {\bibinfo
  {journal} {Soft Matter}\ }\textbf {\bibinfo {volume} {14}},\ \bibinfo {pages}
  {3073} (\bibinfo {year} {2018})}\BibitemShut {NoStop}%
\bibitem [{\citenamefont {Lei}\ \emph {et~al.}(2019)\citenamefont {Lei},
  \citenamefont {Ciamarra},\ and\ \citenamefont {Ni}}]{Lei2019}%
  \BibitemOpen
  \bibfield  {author} {\bibinfo {author} {\bibfnamefont {Q.-L.}\ \bibnamefont
  {Lei}}, \bibinfo {author} {\bibfnamefont {M.~P.}\ \bibnamefont {Ciamarra}},\
  and\ \bibinfo {author} {\bibfnamefont {R.}~\bibnamefont {Ni}},\ }\bibfield
  {title} {\bibinfo {title} {{Nonequilibrium strongly hyperuniform fluids of
  circle active particles with large local density fluctuations}},\ }\href
  {https://doi.org/10.1126/sciadv.aau7423} {\bibfield  {journal} {\bibinfo
  {journal} {Science Advances}\ }\textbf {\bibinfo {volume} {5}},\ \bibinfo
  {pages} {eaau7423} (\bibinfo {year} {2019})}\BibitemShut {NoStop}%
\bibitem [{\citenamefont {Huang}\ \emph {et~al.}(2020)\citenamefont {Huang},
  \citenamefont {Menzel},\ and\ \citenamefont {L\"owen}}]{Huang2020}%
  \BibitemOpen
  \bibfield  {author} {\bibinfo {author} {\bibfnamefont {Z.-F.}\ \bibnamefont
  {Huang}}, \bibinfo {author} {\bibfnamefont {A.~M.}\ \bibnamefont {Menzel}},\
  and\ \bibinfo {author} {\bibfnamefont {H.}~\bibnamefont {L\"owen}},\
  }\bibfield  {title} {\bibinfo {title} {Dynamical crystallites of active
  chiral particles},\ }\href {https://doi.org/10.1103/PhysRevLett.125.218002}
  {\bibfield  {journal} {\bibinfo  {journal} {Phys. Rev. Lett.}\ }\textbf
  {\bibinfo {volume} {125}},\ \bibinfo {pages} {218002} (\bibinfo {year}
  {2020})}\BibitemShut {NoStop}%
\bibitem [{\citenamefont {Liebchen}\ and\ \citenamefont
  {Levis}(2022)}]{Liebchen2022}%
  \BibitemOpen
  \bibfield  {author} {\bibinfo {author} {\bibfnamefont {B.}~\bibnamefont
  {Liebchen}}\ and\ \bibinfo {author} {\bibfnamefont {D.}~\bibnamefont
  {Levis}},\ }\bibfield  {title} {\bibinfo {title} {{Chiral active matter}},\
  }\href {https://doi.org/10.1209/0295-5075/ac8f69} {\bibfield  {journal}
  {\bibinfo  {journal} {EPL}\ }\textbf {\bibinfo {volume} {139}},\ \bibinfo
  {pages} {67001} (\bibinfo {year} {2022})}\BibitemShut {NoStop}%
\bibitem [{\citenamefont {L{\"{o}}wen}(2016)}]{Lowen2016}%
  \BibitemOpen
  \bibfield  {author} {\bibinfo {author} {\bibfnamefont {H.}~\bibnamefont
  {L{\"{o}}wen}},\ }\bibfield  {title} {\bibinfo {title} {{Chirality in
  microswimmer motion: From circle swimmers to active turbulence}},\ }\href
  {https://doi.org/10.1140/epjst/e2016-60054-6} {\bibfield  {journal} {\bibinfo
   {journal} {The European Physical Journal Special Topics}\ }\textbf {\bibinfo
  {volume} {225}},\ \bibinfo {pages} {2319} (\bibinfo {year}
  {2016})}\BibitemShut {NoStop}%
\bibitem [{\citenamefont {K\"ummel}\ \emph {et~al.}(2013)\citenamefont
  {K\"ummel}, \citenamefont {ten Hagen}, \citenamefont {Wittkowski},
  \citenamefont {Buttinoni}, \citenamefont {Eichhorn}, \citenamefont {Volpe},
  \citenamefont {L\"owen},\ and\ \citenamefont {Bechinger}}]{Kummel2013}%
  \BibitemOpen
  \bibfield  {author} {\bibinfo {author} {\bibfnamefont {F.}~\bibnamefont
  {K\"ummel}}, \bibinfo {author} {\bibfnamefont {B.}~\bibnamefont {ten Hagen}},
  \bibinfo {author} {\bibfnamefont {R.}~\bibnamefont {Wittkowski}}, \bibinfo
  {author} {\bibfnamefont {I.}~\bibnamefont {Buttinoni}}, \bibinfo {author}
  {\bibfnamefont {R.}~\bibnamefont {Eichhorn}}, \bibinfo {author}
  {\bibfnamefont {G.}~\bibnamefont {Volpe}}, \bibinfo {author} {\bibfnamefont
  {H.}~\bibnamefont {L\"owen}},\ and\ \bibinfo {author} {\bibfnamefont
  {C.}~\bibnamefont {Bechinger}},\ }\bibfield  {title} {\bibinfo {title}
  {Circular motion of asymmetric self-propelling particles},\ }\href
  {https://doi.org/10.1103/PhysRevLett.110.198302} {\bibfield  {journal}
  {\bibinfo  {journal} {Phys. Rev. Lett.}\ }\textbf {\bibinfo {volume} {110}},\
  \bibinfo {pages} {198302} (\bibinfo {year} {2013})}\BibitemShut {NoStop}%
\bibitem [{\citenamefont {Arora}\ \emph {et~al.}(2024)\citenamefont {Arora},
  \citenamefont {Sadhukhan}, \citenamefont {Nandi}, \citenamefont {Bi},
  \citenamefont {Sood},\ and\ \citenamefont {Ganapathy}}]{Arora2024}%
  \BibitemOpen
  \bibfield  {author} {\bibinfo {author} {\bibfnamefont {P.}~\bibnamefont
  {Arora}}, \bibinfo {author} {\bibfnamefont {S.}~\bibnamefont {Sadhukhan}},
  \bibinfo {author} {\bibfnamefont {S.~K.}\ \bibnamefont {Nandi}}, \bibinfo
  {author} {\bibfnamefont {D.}~\bibnamefont {Bi}}, \bibinfo {author}
  {\bibfnamefont {A.~K.}\ \bibnamefont {Sood}},\ and\ \bibinfo {author}
  {\bibfnamefont {R.}~\bibnamefont {Ganapathy}},\ }\bibfield  {title} {\bibinfo
  {title} {A shape-driven reentrant jamming transition in confluent monolayers
  of synthetic cell-mimics},\ }\href
  {https://doi.org/10.1038/s41467-024-49044-z} {\bibfield  {journal} {\bibinfo
  {journal} {Nature Communications}\ }\textbf {\bibinfo {volume} {15}},\
  \bibinfo {pages} {5645} (\bibinfo {year} {2024})}\BibitemShut {NoStop}%
\bibitem [{\citenamefont {Grzybowski}\ \emph {et~al.}(2000)\citenamefont
  {Grzybowski}, \citenamefont {Stone},\ and\ \citenamefont
  {Whitesides}}]{Grzybowski2000}%
  \BibitemOpen
  \bibfield  {author} {\bibinfo {author} {\bibfnamefont {B.~A.}\ \bibnamefont
  {Grzybowski}}, \bibinfo {author} {\bibfnamefont {H.~A.}\ \bibnamefont
  {Stone}},\ and\ \bibinfo {author} {\bibfnamefont {G.~M.}\ \bibnamefont
  {Whitesides}},\ }\bibfield  {title} {\bibinfo {title} {{Dynamic self-assembly
  of magnetized, millimetre-sized objects rotating at a liquid–air
  interface}},\ }\href {https://doi.org/10.1038/35016528} {\bibfield  {journal}
  {\bibinfo  {journal} {Nature}\ }\textbf {\bibinfo {volume} {405}},\ \bibinfo
  {pages} {1033} (\bibinfo {year} {2000})}\BibitemShut {NoStop}%
\bibitem [{\citenamefont {Cruz}\ \emph {et~al.}(2024)\citenamefont {Cruz},
  \citenamefont {Díaz-Hernández}, \citenamefont {Castañeda-Jonapá},
  \citenamefont {Morales-Padrón}, \citenamefont {Estudillo},\ and\
  \citenamefont {Salgado-García}}]{Cruz2024}%
  \BibitemOpen
  \bibfield  {author} {\bibinfo {author} {\bibfnamefont {J.-M.}\ \bibnamefont
  {Cruz}}, \bibinfo {author} {\bibfnamefont {O.}~\bibnamefont
  {Díaz-Hernández}}, \bibinfo {author} {\bibfnamefont {A.}~\bibnamefont
  {Castañeda-Jonapá}}, \bibinfo {author} {\bibfnamefont {G.}~\bibnamefont
  {Morales-Padrón}}, \bibinfo {author} {\bibfnamefont {A.}~\bibnamefont
  {Estudillo}},\ and\ \bibinfo {author} {\bibfnamefont {R.}~\bibnamefont
  {Salgado-García}},\ }\bibfield  {title} {\bibinfo {title} {Active chiral
  dynamics and boundary accumulation phenomenon in confined camphor
  particles},\ }\href {https://doi.org/10.1039/D3SM01407J} {\bibfield
  {journal} {\bibinfo  {journal} {Soft Matter}\ }\textbf {\bibinfo {volume}
  {20}},\ \bibinfo {pages} {1199} (\bibinfo {year} {2024})}\BibitemShut
  {NoStop}%
\bibitem [{\citenamefont {Jennings}(1901)}]{Jennings1901}%
  \BibitemOpen
  \bibfield  {author} {\bibinfo {author} {\bibfnamefont {H.~S.}\ \bibnamefont
  {Jennings}},\ }\bibfield  {title} {\bibinfo {title} {On the significance of
  the spiral swimming of organisms},\ }\href {https://doi.org/10.1086/277922}
  {\bibfield  {journal} {\bibinfo  {journal} {The American Naturalist}\
  }\textbf {\bibinfo {volume} {35}},\ \bibinfo {pages} {369} (\bibinfo {year}
  {1901})}\BibitemShut {NoStop}%
\bibitem [{\citenamefont {Loose}\ and\ \citenamefont
  {Mitchison}(2014)}]{Loose2014}%
  \BibitemOpen
  \bibfield  {author} {\bibinfo {author} {\bibfnamefont {M.}~\bibnamefont
  {Loose}}\ and\ \bibinfo {author} {\bibfnamefont {T.~J.}\ \bibnamefont
  {Mitchison}},\ }\bibfield  {title} {\bibinfo {title} {{The bacterial cell
  division proteins FtsA and FtsZ self-organize into dynamic cytoskeletal
  patterns}},\ }\href {https://doi.org/10.1038/ncb2885} {\bibfield  {journal}
  {\bibinfo  {journal} {Nature Cell Biology}\ }\textbf {\bibinfo {volume}
  {16}},\ \bibinfo {pages} {38} (\bibinfo {year} {2014})}\BibitemShut {NoStop}%
\bibitem [{\citenamefont {Sumino}\ \emph {et~al.}(2012)\citenamefont {Sumino},
  \citenamefont {Nagai}, \citenamefont {Shitaka}, \citenamefont {Tanaka},
  \citenamefont {Yoshikawa}, \citenamefont {Chat{\'{e}}},\ and\ \citenamefont
  {Oiwa}}]{Sumino2012}%
  \BibitemOpen
  \bibfield  {author} {\bibinfo {author} {\bibfnamefont {Y.}~\bibnamefont
  {Sumino}}, \bibinfo {author} {\bibfnamefont {K.~H.}\ \bibnamefont {Nagai}},
  \bibinfo {author} {\bibfnamefont {Y.}~\bibnamefont {Shitaka}}, \bibinfo
  {author} {\bibfnamefont {D.}~\bibnamefont {Tanaka}}, \bibinfo {author}
  {\bibfnamefont {K.}~\bibnamefont {Yoshikawa}}, \bibinfo {author}
  {\bibfnamefont {H.}~\bibnamefont {Chat{\'{e}}}},\ and\ \bibinfo {author}
  {\bibfnamefont {K.}~\bibnamefont {Oiwa}},\ }\bibfield  {title} {\bibinfo
  {title} {{Large-scale vortex lattice emerging from collectively moving
  microtubules}},\ }\href {https://doi.org/10.1038/nature10874} {\bibfield
  {journal} {\bibinfo  {journal} {Nature}\ }\textbf {\bibinfo {volume} {483}},\
  \bibinfo {pages} {448} (\bibinfo {year} {2012})}\BibitemShut {NoStop}%
\bibitem [{\citenamefont {Brokaw}\ \emph {et~al.}(1982)\citenamefont {Brokaw},
  \citenamefont {Luck},\ and\ \citenamefont {Huang}}]{Brokaw1982}%
  \BibitemOpen
  \bibfield  {author} {\bibinfo {author} {\bibfnamefont {C.~J.}\ \bibnamefont
  {Brokaw}}, \bibinfo {author} {\bibfnamefont {D.~J.}\ \bibnamefont {Luck}},\
  and\ \bibinfo {author} {\bibfnamefont {B.}~\bibnamefont {Huang}},\ }\bibfield
   {title} {\bibinfo {title} {{Analysis of the movement of Chlamydomonas
  flagella: the function of the radial-spoke system is revealed by comparison
  of wild-type and mutant flagella.}},\ }\href
  {https://doi.org/10.1083/JCB.92.3.722} {\bibfield  {journal} {\bibinfo
  {journal} {Journal of Cell Biology}\ }\textbf {\bibinfo {volume} {92}},\
  \bibinfo {pages} {722} (\bibinfo {year} {1982})}\BibitemShut {NoStop}%
\bibitem [{\citenamefont {DiLuzio}\ \emph {et~al.}(2005)\citenamefont
  {DiLuzio}, \citenamefont {Turner}, \citenamefont {Mayer}, \citenamefont
  {Garstecki}, \citenamefont {Weibel}, \citenamefont {Berg},\ and\
  \citenamefont {Whitesides}}]{DiLuzio2005}%
  \BibitemOpen
  \bibfield  {author} {\bibinfo {author} {\bibfnamefont {W.~R.}\ \bibnamefont
  {DiLuzio}}, \bibinfo {author} {\bibfnamefont {L.}~\bibnamefont {Turner}},
  \bibinfo {author} {\bibfnamefont {M.}~\bibnamefont {Mayer}}, \bibinfo
  {author} {\bibfnamefont {P.}~\bibnamefont {Garstecki}}, \bibinfo {author}
  {\bibfnamefont {D.~B.}\ \bibnamefont {Weibel}}, \bibinfo {author}
  {\bibfnamefont {H.~C.}\ \bibnamefont {Berg}},\ and\ \bibinfo {author}
  {\bibfnamefont {G.~M.}\ \bibnamefont {Whitesides}},\ }\bibfield  {title}
  {\bibinfo {title} {{Escherichia coli swim on the right-hand side}},\ }\href
  {https://doi.org/10.1038/nature03660} {\bibfield  {journal} {\bibinfo
  {journal} {Nature}\ }\textbf {\bibinfo {volume} {435}},\ \bibinfo {pages}
  {1271} (\bibinfo {year} {2005})}\BibitemShut {NoStop}%
\bibitem [{\citenamefont {Lauga}\ \emph {et~al.}(2006)\citenamefont {Lauga},
  \citenamefont {DiLuzio}, \citenamefont {Whitesides},\ and\ \citenamefont
  {Stone}}]{Lauga2006}%
  \BibitemOpen
  \bibfield  {author} {\bibinfo {author} {\bibfnamefont {E.}~\bibnamefont
  {Lauga}}, \bibinfo {author} {\bibfnamefont {W.~R.}\ \bibnamefont {DiLuzio}},
  \bibinfo {author} {\bibfnamefont {G.~M.}\ \bibnamefont {Whitesides}},\ and\
  \bibinfo {author} {\bibfnamefont {H.~A.}\ \bibnamefont {Stone}},\ }\bibfield
  {title} {\bibinfo {title} {{Swimming in Circles: Motion of Bacteria near
  Solid Boundaries}},\ }\href {https://doi.org/10.1529/biophysj.105.069401}
  {\bibfield  {journal} {\bibinfo  {journal} {Biophysical Journal}\ }\textbf
  {\bibinfo {volume} {90}},\ \bibinfo {pages} {400} (\bibinfo {year}
  {2006})}\BibitemShut {NoStop}%
\bibitem [{\citenamefont {{Perez Ipi{\~{n}}a}}\ \emph
  {et~al.}(2019)\citenamefont {{Perez Ipi{\~{n}}a}}, \citenamefont {Otte},
  \citenamefont {Pontier-Bres}, \citenamefont {Czerucka},\ and\ \citenamefont
  {Peruani}}]{PerezIpina2019}%
  \BibitemOpen
  \bibfield  {author} {\bibinfo {author} {\bibfnamefont {E.}~\bibnamefont
  {{Perez Ipi{\~{n}}a}}}, \bibinfo {author} {\bibfnamefont {S.}~\bibnamefont
  {Otte}}, \bibinfo {author} {\bibfnamefont {R.}~\bibnamefont {Pontier-Bres}},
  \bibinfo {author} {\bibfnamefont {D.}~\bibnamefont {Czerucka}},\ and\
  \bibinfo {author} {\bibfnamefont {F.}~\bibnamefont {Peruani}},\ }\bibfield
  {title} {\bibinfo {title} {{Bacteria display optimal transport near
  surfaces}},\ }\href {https://doi.org/10.1038/s41567-019-0460-5} {\bibfield
  {journal} {\bibinfo  {journal} {Nature Physics}\ }\textbf {\bibinfo {volume}
  {15}},\ \bibinfo {pages} {610} (\bibinfo {year} {2019})}\BibitemShut
  {NoStop}%
\bibitem [{\citenamefont {Otte}\ \emph {et~al.}(2021)\citenamefont {Otte},
  \citenamefont {Ipi{\~n}a}, \citenamefont {Pontier-Bres}, \citenamefont
  {Czerucka},\ and\ \citenamefont {Peruani}}]{Otte2021}%
  \BibitemOpen
  \bibfield  {author} {\bibinfo {author} {\bibfnamefont {S.}~\bibnamefont
  {Otte}}, \bibinfo {author} {\bibfnamefont {E.~P.}\ \bibnamefont {Ipi{\~n}a}},
  \bibinfo {author} {\bibfnamefont {R.}~\bibnamefont {Pontier-Bres}}, \bibinfo
  {author} {\bibfnamefont {D.}~\bibnamefont {Czerucka}},\ and\ \bibinfo
  {author} {\bibfnamefont {F.}~\bibnamefont {Peruani}},\ }\bibfield  {title}
  {\bibinfo {title} {Statistics of pathogenic bacteria in the search of host
  cells},\ }\href {https://doi.org/10.1038/s41467-021-22156-6} {\bibfield
  {journal} {\bibinfo  {journal} {Nature Communications}\ }\textbf {\bibinfo
  {volume} {12}},\ \bibinfo {pages} {1990} (\bibinfo {year}
  {2021})}\BibitemShut {NoStop}%
\bibitem [{\citenamefont {Riedel}\ \emph {et~al.}(2005)\citenamefont {Riedel},
  \citenamefont {Kruse},\ and\ \citenamefont {Howard}}]{Riedel2005}%
  \BibitemOpen
  \bibfield  {author} {\bibinfo {author} {\bibfnamefont {I.~H.}\ \bibnamefont
  {Riedel}}, \bibinfo {author} {\bibfnamefont {K.}~\bibnamefont {Kruse}},\ and\
  \bibinfo {author} {\bibfnamefont {J.}~\bibnamefont {Howard}},\ }\bibfield
  {title} {\bibinfo {title} {{A self-organized vortex array of hydrodynamically
  entrained sperm cells}},\ }\href {https://doi.org/10.1126/science.1110329}
  {\bibfield  {journal} {\bibinfo  {journal} {Science}\ }\textbf {\bibinfo
  {volume} {309}},\ \bibinfo {pages} {300} (\bibinfo {year}
  {2005})}\BibitemShut {NoStop}%
\bibitem [{\citenamefont {Nosrati}\ \emph {et~al.}(2015)\citenamefont
  {Nosrati}, \citenamefont {Driouchi}, \citenamefont {Yip},\ and\ \citenamefont
  {Sinton}}]{Nosrati2015}%
  \BibitemOpen
  \bibfield  {author} {\bibinfo {author} {\bibfnamefont {R.}~\bibnamefont
  {Nosrati}}, \bibinfo {author} {\bibfnamefont {A.}~\bibnamefont {Driouchi}},
  \bibinfo {author} {\bibfnamefont {C.~M.}\ \bibnamefont {Yip}},\ and\ \bibinfo
  {author} {\bibfnamefont {D.}~\bibnamefont {Sinton}},\ }\bibfield  {title}
  {\bibinfo {title} {{Two-dimensional slither swimming of sperm within a
  micrometre of a surface}},\ }\href {https://doi.org/10.1038/ncomms9703}
  {\bibfield  {journal} {\bibinfo  {journal} {Nature Communications}\ }\textbf
  {\bibinfo {volume} {6}},\ \bibinfo {pages} {8703} (\bibinfo {year}
  {2015})}\BibitemShut {NoStop}%
\bibitem [{\citenamefont {Yan}\ \emph {et~al.}(2015)\citenamefont {Yan},
  \citenamefont {Bae},\ and\ \citenamefont {Granick}}]{Yan2015}%
  \BibitemOpen
  \bibfield  {author} {\bibinfo {author} {\bibfnamefont {J.}~\bibnamefont
  {Yan}}, \bibinfo {author} {\bibfnamefont {S.~C.}\ \bibnamefont {Bae}},\ and\
  \bibinfo {author} {\bibfnamefont {S.}~\bibnamefont {Granick}},\ }\bibfield
  {title} {\bibinfo {title} {Rotating crystals of magnetic janus colloids},\
  }\href {https://doi.org/10.1039/C4SM01962H} {\bibfield  {journal} {\bibinfo
  {journal} {Soft Matter}\ }\textbf {\bibinfo {volume} {11}},\ \bibinfo {pages}
  {147} (\bibinfo {year} {2015})}\BibitemShut {NoStop}%
\bibitem [{\citenamefont {Massana-Cid}\ \emph {et~al.}(2021)\citenamefont
  {Massana-Cid}, \citenamefont {Levis}, \citenamefont {Hern\'andez},
  \citenamefont {Pagonabarraga},\ and\ \citenamefont
  {Tierno}}]{Massana-Cid2021}%
  \BibitemOpen
  \bibfield  {author} {\bibinfo {author} {\bibfnamefont {H.}~\bibnamefont
  {Massana-Cid}}, \bibinfo {author} {\bibfnamefont {D.}~\bibnamefont {Levis}},
  \bibinfo {author} {\bibfnamefont {R.~J.~H.}\ \bibnamefont {Hern\'andez}},
  \bibinfo {author} {\bibfnamefont {I.}~\bibnamefont {Pagonabarraga}},\ and\
  \bibinfo {author} {\bibfnamefont {P.}~\bibnamefont {Tierno}},\ }\bibfield
  {title} {\bibinfo {title} {Arrested phase separation in chiral fluids of
  colloidal spinners},\ }\href
  {https://doi.org/10.1103/PhysRevResearch.3.L042021} {\bibfield  {journal}
  {\bibinfo  {journal} {Phys. Rev. Res.}\ }\textbf {\bibinfo {volume} {3}},\
  \bibinfo {pages} {L042021} (\bibinfo {year} {2021})}\BibitemShut {NoStop}%
\bibitem [{\citenamefont {Kaur}\ \emph {et~al.}(2025)\citenamefont {Kaur},
  \citenamefont {Khuntia}, \citenamefont {Taneja}, \citenamefont {Chaudhuri},
  \citenamefont {Yogendran},\ and\ \citenamefont {Rakshit}}]{Kaur2025}%
  \BibitemOpen
  \bibfield  {author} {\bibinfo {author} {\bibfnamefont {V.}~\bibnamefont
  {Kaur}}, \bibinfo {author} {\bibfnamefont {S.~S.}\ \bibnamefont {Khuntia}},
  \bibinfo {author} {\bibfnamefont {C.}~\bibnamefont {Taneja}}, \bibinfo
  {author} {\bibfnamefont {A.}~\bibnamefont {Chaudhuri}}, \bibinfo {author}
  {\bibfnamefont {K.~P.}\ \bibnamefont {Yogendran}},\ and\ \bibinfo {author}
  {\bibfnamefont {S.}~\bibnamefont {Rakshit}},\ }\bibfield  {title} {\bibinfo
  {title} {De-novo design of actively spinning and gyrating spherical
  micro-vesicles},\ }\href {https://doi.org/10.1002/adma.202419716} {\bibfield
  {journal} {\bibinfo  {journal} {Advanced Materials}\ }\textbf {\bibinfo
  {volume} {37}},\ \bibinfo {pages} {e2419716} (\bibinfo {year}
  {2025})}\BibitemShut {NoStop}%
\bibitem [{\citenamefont {Farhadi}\ \emph {et~al.}(2018)\citenamefont
  {Farhadi}, \citenamefont {Machaca}, \citenamefont {Aird}, \citenamefont
  {Maldonado}, \citenamefont {Davis}, \citenamefont {Arratia},\ and\
  \citenamefont {Durian}}]{Farhadi2018}%
  \BibitemOpen
  \bibfield  {author} {\bibinfo {author} {\bibfnamefont {S.}~\bibnamefont
  {Farhadi}}, \bibinfo {author} {\bibfnamefont {S.}~\bibnamefont {Machaca}},
  \bibinfo {author} {\bibfnamefont {J.}~\bibnamefont {Aird}}, \bibinfo {author}
  {\bibfnamefont {B.~O.~T.}\ \bibnamefont {Maldonado}}, \bibinfo {author}
  {\bibfnamefont {S.}~\bibnamefont {Davis}}, \bibinfo {author} {\bibfnamefont
  {P.~E.}\ \bibnamefont {Arratia}},\ and\ \bibinfo {author} {\bibfnamefont
  {D.~J.}\ \bibnamefont {Durian}},\ }\bibfield  {title} {\bibinfo {title}
  {Dynamics and thermodynamics of air-driven active spinners},\ }\href
  {https://doi.org/10.1039/C8SM00403J} {\bibfield  {journal} {\bibinfo
  {journal} {Soft Matter}\ }\textbf {\bibinfo {volume} {14}},\ \bibinfo {pages}
  {5588} (\bibinfo {year} {2018})}\BibitemShut {NoStop}%
\bibitem [{\citenamefont {Scholz}\ \emph {et~al.}(2021)\citenamefont {Scholz},
  \citenamefont {Liu}, \citenamefont {Schatz}, \citenamefont {Wu},
  \citenamefont {Rückerl}, \citenamefont {Stark},\ and\ \citenamefont
  {Groszek}}]{Scholz2021}%
  \BibitemOpen
  \bibfield  {author} {\bibinfo {author} {\bibfnamefont {C.}~\bibnamefont
  {Scholz}}, \bibinfo {author} {\bibfnamefont {Q.-Y.}\ \bibnamefont {Liu}},
  \bibinfo {author} {\bibfnamefont {C.}~\bibnamefont {Schatz}}, \bibinfo
  {author} {\bibfnamefont {Q.-J.}\ \bibnamefont {Wu}}, \bibinfo {author}
  {\bibfnamefont {R.}~\bibnamefont {Rückerl}}, \bibinfo {author}
  {\bibfnamefont {H.}~\bibnamefont {Stark}},\ and\ \bibinfo {author}
  {\bibfnamefont {A.~A.}\ \bibnamefont {Groszek}},\ }\bibfield  {title}
  {\bibinfo {title} {Surfactants and rotelles in active chiral fluids},\ }\href
  {https://doi.org/10.1126/sciadv.abf8998} {\bibfield  {journal} {\bibinfo
  {journal} {Science Advances}\ }\textbf {\bibinfo {volume} {7}},\ \bibinfo
  {pages} {eabf8998} (\bibinfo {year} {2021})}\BibitemShut {NoStop}%
\bibitem [{\citenamefont {Vega~Reyes}\ \emph {et~al.}(2022)\citenamefont
  {Vega~Reyes}, \citenamefont {L{\'o}pez-Casta{\~n}o},\ and\ \citenamefont
  {Rodr{\'\i}guez-Rivas}}]{Vega2022}%
  \BibitemOpen
  \bibfield  {author} {\bibinfo {author} {\bibfnamefont {F.}~\bibnamefont
  {Vega~Reyes}}, \bibinfo {author} {\bibfnamefont {M.~A.}\ \bibnamefont
  {L{\'o}pez-Casta{\~n}o}},\ and\ \bibinfo {author} {\bibfnamefont
  {{\'A}.}~\bibnamefont {Rodr{\'\i}guez-Rivas}},\ }\bibfield  {title} {\bibinfo
  {title} {Diffusive regimes in a two-dimensional chiral fluid},\ }\href
  {https://doi.org/10.1038/s42005-022-01032-9} {\bibfield  {journal} {\bibinfo
  {journal} {Commun. Phys.}\ }\textbf {\bibinfo {volume} {5}},\ \bibinfo
  {pages} {256} (\bibinfo {year} {2022})}\BibitemShut {NoStop}%
\bibitem [{\citenamefont {L\'opez-Casta\~no}\ \emph {et~al.}(2022)\citenamefont
  {L\'opez-Casta\~no}, \citenamefont {M\'arquez~Seco}, \citenamefont
  {M\'arquez~Seco}, \citenamefont {Rodr\'{\i}guez-Rivas},\ and\ \citenamefont
  {Reyes}}]{Lopez2022}%
  \BibitemOpen
  \bibfield  {author} {\bibinfo {author} {\bibfnamefont {M.~A.}\ \bibnamefont
  {L\'opez-Casta\~no}}, \bibinfo {author} {\bibfnamefont {A.}~\bibnamefont
  {M\'arquez~Seco}}, \bibinfo {author} {\bibfnamefont {A.}~\bibnamefont
  {M\'arquez~Seco}}, \bibinfo {author} {\bibfnamefont {A.}~\bibnamefont
  {Rodr\'{\i}guez-Rivas}},\ and\ \bibinfo {author} {\bibfnamefont {F.~V.}\
  \bibnamefont {Reyes}},\ }\bibfield  {title} {\bibinfo {title} {Chirality
  transitions in a system of active flat spinners},\ }\href
  {https://doi.org/10.1103/PhysRevResearch.4.033230} {\bibfield  {journal}
  {\bibinfo  {journal} {Phys. Rev. Res.}\ }\textbf {\bibinfo {volume} {4}},\
  \bibinfo {pages} {033230} (\bibinfo {year} {2022})}\BibitemShut {NoStop}%
\bibitem [{\citenamefont {Siebers}\ \emph {et~al.}(2023)\citenamefont
  {Siebers}, \citenamefont {Jayaram}, \citenamefont {Bl{\"u}mler},\ and\
  \citenamefont {Speck}}]{Siebers2023}%
  \BibitemOpen
  \bibfield  {author} {\bibinfo {author} {\bibfnamefont {F.}~\bibnamefont
  {Siebers}}, \bibinfo {author} {\bibfnamefont {A.}~\bibnamefont {Jayaram}},
  \bibinfo {author} {\bibfnamefont {P.}~\bibnamefont {Bl{\"u}mler}},\ and\
  \bibinfo {author} {\bibfnamefont {T.}~\bibnamefont {Speck}},\ }\bibfield
  {title} {\bibinfo {title} {Exploiting compositional disorder in collectives
  of light-driven circle walkers},\ }\href
  {https://doi.org/10.1126/sciadv.adf5443} {\bibfield  {journal} {\bibinfo
  {journal} {Sci. Adv.}\ }\textbf {\bibinfo {volume} {9}},\ \bibinfo {pages}
  {eadf5443} (\bibinfo {year} {2023})}\BibitemShut {NoStop}%
\bibitem [{\citenamefont {Caprini}\ \emph {et~al.}(2024)\citenamefont
  {Caprini}, \citenamefont {Liebchen},\ and\ \citenamefont
  {L{\"{o}}wen}}]{Caprini2024}%
  \BibitemOpen
  \bibfield  {author} {\bibinfo {author} {\bibfnamefont {L.}~\bibnamefont
  {Caprini}}, \bibinfo {author} {\bibfnamefont {B.}~\bibnamefont {Liebchen}},\
  and\ \bibinfo {author} {\bibfnamefont {H.}~\bibnamefont {L{\"{o}}wen}},\
  }\bibfield  {title} {\bibinfo {title} {{Self-reverting vortices in chiral
  active matter}},\ }\href {https://doi.org/10.1038/s42005-024-01637-2}
  {\bibfield  {journal} {\bibinfo  {journal} {Communications Physics}\ }\textbf
  {\bibinfo {volume} {7}},\ \bibinfo {pages} {153} (\bibinfo {year}
  {2024})}\BibitemShut {NoStop}%
\bibitem [{\citenamefont {Van~Teeffelen}\ \emph {et~al.}(2009)\citenamefont
  {Van~Teeffelen}, \citenamefont {Zimmermann},\ and\ \citenamefont
  {L{\"{o}}wen}}]{Van-Teeffelen2009}%
  \BibitemOpen
  \bibfield  {author} {\bibinfo {author} {\bibfnamefont {S.}~\bibnamefont
  {Van~Teeffelen}}, \bibinfo {author} {\bibfnamefont {U.}~\bibnamefont
  {Zimmermann}},\ and\ \bibinfo {author} {\bibfnamefont {H.}~\bibnamefont
  {L{\"{o}}wen}},\ }\bibfield  {title} {\bibinfo {title}
  {{Clockwise-directional circle swimmer moves counter-clockwise in Petri dish-
  and ring-like confinements}},\ }\href {https://doi.org/10.1039/B911365G}
  {\bibfield  {journal} {\bibinfo  {journal} {Soft Matter}\ }\textbf {\bibinfo
  {volume} {5}},\ \bibinfo {pages} {4510} (\bibinfo {year} {2009})}\BibitemShut
  {NoStop}%
\bibitem [{\citenamefont {Mijalkov}\ and\ \citenamefont
  {Volpe}(2013)}]{Mijalkov2013}%
  \BibitemOpen
  \bibfield  {author} {\bibinfo {author} {\bibfnamefont {M.}~\bibnamefont
  {Mijalkov}}\ and\ \bibinfo {author} {\bibfnamefont {G.}~\bibnamefont
  {Volpe}},\ }\bibfield  {title} {\bibinfo {title} {Sorting of chiral
  microswimmers},\ }\href {https://doi.org/10.1039/C3SM27923E} {\bibfield
  {journal} {\bibinfo  {journal} {Soft Matter}\ }\textbf {\bibinfo {volume}
  {9}},\ \bibinfo {pages} {6376} (\bibinfo {year} {2013})}\BibitemShut
  {NoStop}%
\bibitem [{\citenamefont {Volpe}\ \emph {et~al.}(2014)\citenamefont {Volpe},
  \citenamefont {Gigan},\ and\ \citenamefont {Volpe}}]{Volpe2014}%
  \BibitemOpen
  \bibfield  {author} {\bibinfo {author} {\bibfnamefont {G.}~\bibnamefont
  {Volpe}}, \bibinfo {author} {\bibfnamefont {S.}~\bibnamefont {Gigan}},\ and\
  \bibinfo {author} {\bibfnamefont {G.}~\bibnamefont {Volpe}},\ }\bibfield
  {title} {\bibinfo {title} {{Simulation of the active Brownian motion of a
  microswimmer}},\ }\href {https://doi.org/10.1119/1.4870398} {\bibfield
  {journal} {\bibinfo  {journal} {American Journal of Physics}\ }\textbf
  {\bibinfo {volume} {82}},\ \bibinfo {pages} {659} (\bibinfo {year}
  {2014})}\BibitemShut {NoStop}%
\bibitem [{\citenamefont {Sevilla}(2016)}]{Sevilla2016}%
  \BibitemOpen
  \bibfield  {author} {\bibinfo {author} {\bibfnamefont {F.~J.}\ \bibnamefont
  {Sevilla}},\ }\bibfield  {title} {\bibinfo {title} {Diffusion of active
  chiral particles},\ }\href {https://doi.org/10.1103/PhysRevE.94.062120}
  {\bibfield  {journal} {\bibinfo  {journal} {Phys. Rev. E}\ }\textbf {\bibinfo
  {volume} {94}},\ \bibinfo {pages} {062120} (\bibinfo {year}
  {2016})}\BibitemShut {NoStop}%
\bibitem [{\citenamefont {Markovich}\ \emph {et~al.}(2019)\citenamefont
  {Markovich}, \citenamefont {Tjhung},\ and\ \citenamefont
  {Cates}}]{Markovich2019}%
  \BibitemOpen
  \bibfield  {author} {\bibinfo {author} {\bibfnamefont {T.}~\bibnamefont
  {Markovich}}, \bibinfo {author} {\bibfnamefont {E.}~\bibnamefont {Tjhung}},\
  and\ \bibinfo {author} {\bibfnamefont {M.~E.}\ \bibnamefont {Cates}},\
  }\bibfield  {title} {\bibinfo {title} {{Chiral active matter: microscopic
  ‘torque dipoles’ have more than one hydrodynamic description}},\ }\href
  {https://doi.org/10.1088/1367-2630/AB54AF} {\bibfield  {journal} {\bibinfo
  {journal} {New Journal of Physics}\ }\textbf {\bibinfo {volume} {21}},\
  \bibinfo {pages} {112001} (\bibinfo {year} {2019})}\BibitemShut {NoStop}%
\bibitem [{\citenamefont {Chepizhko}\ and\ \citenamefont
  {Franosch}(2020)}]{Chepizhko2020}%
  \BibitemOpen
  \bibfield  {author} {\bibinfo {author} {\bibfnamefont {O.}~\bibnamefont
  {Chepizhko}}\ and\ \bibinfo {author} {\bibfnamefont {T.}~\bibnamefont
  {Franosch}},\ }\bibfield  {title} {\bibinfo {title} {Random motion of a
  circle microswimmer in a random environment},\ }\href
  {https://doi.org/10.1088/1367-2630/ab9708} {\bibfield  {journal} {\bibinfo
  {journal} {New Journal of Physics}\ }\textbf {\bibinfo {volume} {22}},\
  \bibinfo {pages} {073022} (\bibinfo {year} {2020})}\BibitemShut {NoStop}%
\bibitem [{\citenamefont {Di~Leonardo}\ \emph {et~al.}(2011)\citenamefont
  {Di~Leonardo}, \citenamefont {Dell’Arciprete}, \citenamefont {Angelani},\
  and\ \citenamefont {Iebba}}]{Leonardo2011}%
  \BibitemOpen
  \bibfield  {author} {\bibinfo {author} {\bibfnamefont {R.}~\bibnamefont
  {Di~Leonardo}}, \bibinfo {author} {\bibfnamefont {D.}~\bibnamefont
  {Dell’Arciprete}}, \bibinfo {author} {\bibfnamefont {L.}~\bibnamefont
  {Angelani}},\ and\ \bibinfo {author} {\bibfnamefont {V.}~\bibnamefont
  {Iebba}},\ }\bibfield  {title} {\bibinfo {title} {{Swimming with an Image}},\
  }\href {https://doi.org/10.1103/PhysRevLett.106.038101} {\bibfield  {journal}
  {\bibinfo  {journal} {Physical Review Letters}\ }\textbf {\bibinfo {volume}
  {106}},\ \bibinfo {pages} {038101} (\bibinfo {year} {2011})}\BibitemShut
  {NoStop}%
\bibitem [{\citenamefont {Araujo}\ \emph {et~al.}(2019)\citenamefont {Araujo},
  \citenamefont {Chen}, \citenamefont {Mani},\ and\ \citenamefont
  {Tang}}]{Araujo2019}%
  \BibitemOpen
  \bibfield  {author} {\bibinfo {author} {\bibfnamefont {G.}~\bibnamefont
  {Araujo}}, \bibinfo {author} {\bibfnamefont {W.}~\bibnamefont {Chen}},
  \bibinfo {author} {\bibfnamefont {S.}~\bibnamefont {Mani}},\ and\ \bibinfo
  {author} {\bibfnamefont {J.~X.}\ \bibnamefont {Tang}},\ }\bibfield  {title}
  {\bibinfo {title} {{Orbiting of Flagellated Bacteria within a Thin Fluid Film
  around Micrometer-Sized Particles}},\ }\href
  {https://doi.org/10.1016/j.bpj.2019.06.005} {\bibfield  {journal} {\bibinfo
  {journal} {Biophysical Journal}\ }\textbf {\bibinfo {volume} {117}},\
  \bibinfo {pages} {346} (\bibinfo {year} {2019})}\BibitemShut {NoStop}%
\bibitem [{\citenamefont {B{\"{o}}hmer}\ \emph {et~al.}(2005)\citenamefont
  {B{\"{o}}hmer}, \citenamefont {Van}, \citenamefont {Weyand}, \citenamefont
  {Hagen}, \citenamefont {Beyermann}, \citenamefont {Matsumoto}, \citenamefont
  {Hoshi}, \citenamefont {Hildebrand},\ and\ \citenamefont
  {Kaupp}}]{Bohmer2005}%
  \BibitemOpen
  \bibfield  {author} {\bibinfo {author} {\bibfnamefont {M.}~\bibnamefont
  {B{\"{o}}hmer}}, \bibinfo {author} {\bibfnamefont {Q.}~\bibnamefont {Van}},
  \bibinfo {author} {\bibfnamefont {I.}~\bibnamefont {Weyand}}, \bibinfo
  {author} {\bibfnamefont {V.}~\bibnamefont {Hagen}}, \bibinfo {author}
  {\bibfnamefont {M.}~\bibnamefont {Beyermann}}, \bibinfo {author}
  {\bibfnamefont {M.}~\bibnamefont {Matsumoto}}, \bibinfo {author}
  {\bibfnamefont {M.}~\bibnamefont {Hoshi}}, \bibinfo {author} {\bibfnamefont
  {E.}~\bibnamefont {Hildebrand}},\ and\ \bibinfo {author} {\bibfnamefont
  {U.~B.}\ \bibnamefont {Kaupp}},\ }\bibfield  {title} {\bibinfo {title} {{Ca2+
  spikes in the flagellum control chemotactic behavior of sperm}},\ }\href
  {https://doi.org/10.1038/sj.emboj.7600744} {\bibfield  {journal} {\bibinfo
  {journal} {The EMBO Journal}\ }\textbf {\bibinfo {volume} {24}},\ \bibinfo
  {pages} {2741} (\bibinfo {year} {2005})}\BibitemShut {NoStop}%
\bibitem [{\citenamefont {Taktikos}\ \emph {et~al.}(2011)\citenamefont
  {Taktikos}, \citenamefont {Zaburdaev},\ and\ \citenamefont
  {Stark}}]{Taktikos2011}%
  \BibitemOpen
  \bibfield  {author} {\bibinfo {author} {\bibfnamefont {J.}~\bibnamefont
  {Taktikos}}, \bibinfo {author} {\bibfnamefont {V.}~\bibnamefont
  {Zaburdaev}},\ and\ \bibinfo {author} {\bibfnamefont {H.}~\bibnamefont
  {Stark}},\ }\bibfield  {title} {\bibinfo {title} {{Modeling a self-propelled
  autochemotactic walker}},\ }\href
  {https://doi.org/10.1103/PhysRevE.84.041924} {\bibfield  {journal} {\bibinfo
  {journal} {Physical Review E}\ }\textbf {\bibinfo {volume} {84}},\ \bibinfo
  {pages} {41924} (\bibinfo {year} {2011})}\BibitemShut {NoStop}%
\bibitem [{\citenamefont {Soni}\ \emph {et~al.}(2019)\citenamefont {Soni},
  \citenamefont {Bililign}, \citenamefont {Magkiriadou}, \citenamefont
  {Sacanna}, \citenamefont {Bartolo}, \citenamefont {Shelley},\ and\
  \citenamefont {Irvine}}]{Soni2019}%
  \BibitemOpen
  \bibfield  {author} {\bibinfo {author} {\bibfnamefont {V.}~\bibnamefont
  {Soni}}, \bibinfo {author} {\bibfnamefont {E.~S.}\ \bibnamefont {Bililign}},
  \bibinfo {author} {\bibfnamefont {S.}~\bibnamefont {Magkiriadou}}, \bibinfo
  {author} {\bibfnamefont {S.}~\bibnamefont {Sacanna}}, \bibinfo {author}
  {\bibfnamefont {D.}~\bibnamefont {Bartolo}}, \bibinfo {author} {\bibfnamefont
  {M.~J.}\ \bibnamefont {Shelley}},\ and\ \bibinfo {author} {\bibfnamefont
  {W.~T.~M.}\ \bibnamefont {Irvine}},\ }\bibfield  {title} {\bibinfo {title}
  {{The odd free surface flows of a colloidal chiral fluid}},\ }\href
  {https://doi.org/10.1038/s41567-019-0603-8} {\bibfield  {journal} {\bibinfo
  {journal} {Nature Physics}\ }\textbf {\bibinfo {volume} {15}},\ \bibinfo
  {pages} {1188} (\bibinfo {year} {2019})}\BibitemShut {NoStop}%
\bibitem [{\citenamefont {Hargus}\ \emph {et~al.}(2021)\citenamefont {Hargus},
  \citenamefont {Epstein},\ and\ \citenamefont {Mandadapu}}]{Hargus2021}%
  \BibitemOpen
  \bibfield  {author} {\bibinfo {author} {\bibfnamefont {C.}~\bibnamefont
  {Hargus}}, \bibinfo {author} {\bibfnamefont {J.~M.}\ \bibnamefont
  {Epstein}},\ and\ \bibinfo {author} {\bibfnamefont {K.~K.}\ \bibnamefont
  {Mandadapu}},\ }\bibfield  {title} {\bibinfo {title} {Odd diffusivity of
  chiral random motion},\ }\href
  {https://doi.org/10.1103/PhysRevLett.127.178001} {\bibfield  {journal}
  {\bibinfo  {journal} {Phys. Rev. Lett.}\ }\textbf {\bibinfo {volume} {127}},\
  \bibinfo {pages} {178001} (\bibinfo {year} {2021})}\BibitemShut {NoStop}%
\bibitem [{\citenamefont {Caprini}\ and\ \citenamefont {Marini
  Bettolo~Marconi}(2019)}]{Caprini2019}%
  \BibitemOpen
  \bibfield  {author} {\bibinfo {author} {\bibfnamefont {L.}~\bibnamefont
  {Caprini}}\ and\ \bibinfo {author} {\bibfnamefont {U.}~\bibnamefont {Marini
  Bettolo~Marconi}},\ }\bibfield  {title} {\bibinfo {title} {{Active chiral
  particles under confinement: surface currents and bulk accumulation
  phenomena}},\ }\href {https://doi.org/10.1039/C8SM02492H} {\bibfield
  {journal} {\bibinfo  {journal} {Soft Matter}\ }\textbf {\bibinfo {volume}
  {15}},\ \bibinfo {pages} {2627} (\bibinfo {year} {2019})}\BibitemShut
  {NoStop}%
\bibitem [{\citenamefont {Li}\ \emph {et~al.}(2020)\citenamefont {Li},
  \citenamefont {Li}, \citenamefont {Marchesoni}, \citenamefont {Debnath},\
  and\ \citenamefont {Ghosh}}]{Li2020}%
  \BibitemOpen
  \bibfield  {author} {\bibinfo {author} {\bibfnamefont {Y.}~\bibnamefont
  {Li}}, \bibinfo {author} {\bibfnamefont {L.}~\bibnamefont {Li}}, \bibinfo
  {author} {\bibfnamefont {F.}~\bibnamefont {Marchesoni}}, \bibinfo {author}
  {\bibfnamefont {D.}~\bibnamefont {Debnath}},\ and\ \bibinfo {author}
  {\bibfnamefont {P.~K.}\ \bibnamefont {Ghosh}},\ }\bibfield  {title} {\bibinfo
  {title} {Diffusion of chiral janus particles in convection rolls},\ }\href
  {https://doi.org/10.1103/PhysRevResearch.2.013250} {\bibfield  {journal}
  {\bibinfo  {journal} {Phys. Rev. Res.}\ }\textbf {\bibinfo {volume} {2}},\
  \bibinfo {pages} {013250} (\bibinfo {year} {2020})}\BibitemShut {NoStop}%
\bibitem [{\citenamefont {Khatri}\ and\ \citenamefont
  {Burada}(2022)}]{Khatri2022}%
  \BibitemOpen
  \bibfield  {author} {\bibinfo {author} {\bibfnamefont {N.}~\bibnamefont
  {Khatri}}\ and\ \bibinfo {author} {\bibfnamefont {P.~S.}\ \bibnamefont
  {Burada}},\ }\bibfield  {title} {\bibinfo {title} {Diffusion of chiral active
  particles in a poiseuille flow},\ }\href
  {https://doi.org/10.1103/PhysRevE.105.024604} {\bibfield  {journal} {\bibinfo
   {journal} {Phys. Rev. E}\ }\textbf {\bibinfo {volume} {105}},\ \bibinfo
  {pages} {024604} (\bibinfo {year} {2022})}\BibitemShut {NoStop}%
\bibitem [{\citenamefont {Pattanayak}\ \emph {et~al.}(2024)\citenamefont
  {Pattanayak}, \citenamefont {Shee}, \citenamefont {Chaudhuri},\ and\
  \citenamefont {Chaudhuri}}]{Pattanayak2024}%
  \BibitemOpen
  \bibfield  {author} {\bibinfo {author} {\bibfnamefont {A.}~\bibnamefont
  {Pattanayak}}, \bibinfo {author} {\bibfnamefont {A.}~\bibnamefont {Shee}},
  \bibinfo {author} {\bibfnamefont {D.}~\bibnamefont {Chaudhuri}},\ and\
  \bibinfo {author} {\bibfnamefont {A.}~\bibnamefont {Chaudhuri}},\ }\bibfield
  {title} {\bibinfo {title} {Impact of torque on active brownian particle:
  exact moments in two and three dimensions},\ }\href
  {https://doi.org/10.1088/1367-2630/ad6a32} {\bibfield  {journal} {\bibinfo
  {journal} {New Journal of Physics}\ }\textbf {\bibinfo {volume} {26}},\
  \bibinfo {pages} {083024} (\bibinfo {year} {2024})}\BibitemShut {NoStop}%
\bibitem [{\citenamefont {Olsen}\ and\ \citenamefont
  {L\"owen}(2024)}]{Olsen2024}%
  \BibitemOpen
  \bibfield  {author} {\bibinfo {author} {\bibfnamefont {K.~S.}\ \bibnamefont
  {Olsen}}\ and\ \bibinfo {author} {\bibfnamefont {H.}~\bibnamefont
  {L\"owen}},\ }\bibfield  {title} {\bibinfo {title} {Optimal diffusion of
  chiral active particles with strategic reorientations},\ }\href
  {https://doi.org/10.1103/PhysRevE.110.064606} {\bibfield  {journal} {\bibinfo
   {journal} {Phys. Rev. E}\ }\textbf {\bibinfo {volume} {110}},\ \bibinfo
  {pages} {064606} (\bibinfo {year} {2024})}\BibitemShut {NoStop}%
\bibitem [{\citenamefont {Pattanayak}\ \emph {et~al.}(2025)\citenamefont
  {Pattanayak}, \citenamefont {Shee}, \citenamefont {Chaudhuri},\ and\
  \citenamefont {Chaudhuri}}]{Pattanayak2025}%
  \BibitemOpen
  \bibfield  {author} {\bibinfo {author} {\bibfnamefont {A.}~\bibnamefont
  {Pattanayak}}, \bibinfo {author} {\bibfnamefont {A.}~\bibnamefont {Shee}},
  \bibinfo {author} {\bibfnamefont {D.}~\bibnamefont {Chaudhuri}},\ and\
  \bibinfo {author} {\bibfnamefont {A.}~\bibnamefont {Chaudhuri}},\ }\bibfield
  {title} {\bibinfo {title} {Chirality, confinement and dimensionality govern
  re-entrant transitions in active matter},\ }\href
  {https://doi.org/10.1063/5.0301938} {\bibfield  {journal} {\bibinfo
  {journal} {Journal of Chemical Physics}\ }\textbf {\bibinfo {volume} {163}},\
  \bibinfo {pages} {244902} (\bibinfo {year} {2025})}\BibitemShut {NoStop}%
\bibitem [{\citenamefont {Fazli}\ and\ \citenamefont {Naji}(2021)}]{Fazli2021}%
  \BibitemOpen
  \bibfield  {author} {\bibinfo {author} {\bibfnamefont {Z.}~\bibnamefont
  {Fazli}}\ and\ \bibinfo {author} {\bibfnamefont {A.}~\bibnamefont {Naji}},\
  }\bibfield  {title} {\bibinfo {title} {Active particles with polar alignment
  in ring-shaped confinement},\ }\href
  {https://doi.org/10.1103/PhysRevE.103.022601} {\bibfield  {journal} {\bibinfo
   {journal} {Phys. Rev. E}\ }\textbf {\bibinfo {volume} {103}},\ \bibinfo
  {pages} {022601} (\bibinfo {year} {2021})}\BibitemShut {NoStop}%
\bibitem [{\citenamefont {Murali}\ \emph {et~al.}(2022)\citenamefont {Murali},
  \citenamefont {Dolai}, \citenamefont {Krishna}, \citenamefont {Kumar},\ and\
  \citenamefont {Thutupalli}}]{Murali2022}%
  \BibitemOpen
  \bibfield  {author} {\bibinfo {author} {\bibfnamefont {A.}~\bibnamefont
  {Murali}}, \bibinfo {author} {\bibfnamefont {P.}~\bibnamefont {Dolai}},
  \bibinfo {author} {\bibfnamefont {A.}~\bibnamefont {Krishna}}, \bibinfo
  {author} {\bibfnamefont {K.~V.}\ \bibnamefont {Kumar}},\ and\ \bibinfo
  {author} {\bibfnamefont {S.}~\bibnamefont {Thutupalli}},\ }\bibfield  {title}
  {\bibinfo {title} {Geometric constraints alter the emergent dynamics of an
  active particle},\ }\href {https://doi.org/10.1103/PhysRevResearch.4.013136}
  {\bibfield  {journal} {\bibinfo  {journal} {Phys. Rev. Res.}\ }\textbf
  {\bibinfo {volume} {4}},\ \bibinfo {pages} {013136} (\bibinfo {year}
  {2022})}\BibitemShut {NoStop}%
\bibitem [{\citenamefont {Caprini}\ \emph {et~al.}(2023)\citenamefont
  {Caprini}, \citenamefont {L{\"{o}}wen},\ and\ \citenamefont {Marini
  Bettolo~Marconi}}]{Caprini2023}%
  \BibitemOpen
  \bibfield  {author} {\bibinfo {author} {\bibfnamefont {L.}~\bibnamefont
  {Caprini}}, \bibinfo {author} {\bibfnamefont {H.}~\bibnamefont
  {L{\"{o}}wen}},\ and\ \bibinfo {author} {\bibfnamefont {U.}~\bibnamefont
  {Marini Bettolo~Marconi}},\ }\bibfield  {title} {\bibinfo {title} {{Chiral
  active matter in external potentials}},\ }\href
  {https://doi.org/10.1039/d3sm00793f} {\bibfield  {journal} {\bibinfo
  {journal} {Soft Matter}\ }\textbf {\bibinfo {volume} {19}},\ \bibinfo {pages}
  {6234} (\bibinfo {year} {2023})}\BibitemShut {NoStop}%
\bibitem [{\citenamefont {Takatori}\ \emph {et~al.}(2016)\citenamefont
  {Takatori}, \citenamefont {De~Dier}, \citenamefont {Vermant},\ and\
  \citenamefont {Brady}}]{Takatori2016}%
  \BibitemOpen
  \bibfield  {author} {\bibinfo {author} {\bibfnamefont {S.~C.}\ \bibnamefont
  {Takatori}}, \bibinfo {author} {\bibfnamefont {R.}~\bibnamefont {De~Dier}},
  \bibinfo {author} {\bibfnamefont {J.}~\bibnamefont {Vermant}},\ and\ \bibinfo
  {author} {\bibfnamefont {J.~F.}\ \bibnamefont {Brady}},\ }\bibfield  {title}
  {\bibinfo {title} {Acoustic trapping of active matter},\ }\href
  {https://doi.org/10.1038/ncomms10694} {\bibfield  {journal} {\bibinfo
  {journal} {Nature Communications}\ }\textbf {\bibinfo {volume} {7}},\
  \bibinfo {pages} {10694} (\bibinfo {year} {2016})}\BibitemShut {NoStop}%
\bibitem [{\citenamefont {Peng}\ and\ \citenamefont {Brady}(2022)}]{Peng2022}%
  \BibitemOpen
  \bibfield  {author} {\bibinfo {author} {\bibfnamefont {Z.}~\bibnamefont
  {Peng}}\ and\ \bibinfo {author} {\bibfnamefont {J.~F.}\ \bibnamefont
  {Brady}},\ }\bibfield  {title} {\bibinfo {title} {Trapped-particle
  microrheology of active suspensions},\ }\href
  {https://doi.org/10.1063/5.0108014} {\bibfield  {journal} {\bibinfo
  {journal} {The Journal of Chemical Physics}\ }\textbf {\bibinfo {volume}
  {157}},\ \bibinfo {pages} {104119} (\bibinfo {year} {2022})}\BibitemShut
  {NoStop}%
\bibitem [{\citenamefont {Debets}\ \emph {et~al.}(2023)\citenamefont {Debets},
  \citenamefont {L\"owen},\ and\ \citenamefont {Janssen}}]{Debets2023}%
  \BibitemOpen
  \bibfield  {author} {\bibinfo {author} {\bibfnamefont {V.~E.}\ \bibnamefont
  {Debets}}, \bibinfo {author} {\bibfnamefont {H.}~\bibnamefont {L\"owen}},\
  and\ \bibinfo {author} {\bibfnamefont {L.~M.~C.}\ \bibnamefont {Janssen}},\
  }\bibfield  {title} {\bibinfo {title} {Glassy dynamics in chiral fluids},\
  }\href {https://doi.org/10.1103/PhysRevLett.130.058201} {\bibfield  {journal}
  {\bibinfo  {journal} {Phys. Rev. Lett.}\ }\textbf {\bibinfo {volume} {130}},\
  \bibinfo {pages} {058201} (\bibinfo {year} {2023})}\BibitemShut {NoStop}%
\bibitem [{\citenamefont {Shee}\ \emph {et~al.}(2024)\citenamefont {Shee},
  \citenamefont {Henkes},\ and\ \citenamefont {Huepe}}]{Shee2024}%
  \BibitemOpen
  \bibfield  {author} {\bibinfo {author} {\bibfnamefont {A.}~\bibnamefont
  {Shee}}, \bibinfo {author} {\bibfnamefont {S.}~\bibnamefont {Henkes}},\ and\
  \bibinfo {author} {\bibfnamefont {C.}~\bibnamefont {Huepe}},\ }\bibfield
  {title} {\bibinfo {title} {{Emergent mesoscale correlations in active solids
  with noisy chiral dynamics}},\ }\href {https://doi.org/10.1039/D4SM00958D}
  {\bibfield  {journal} {\bibinfo  {journal} {Soft Matter}\ }\textbf {\bibinfo
  {volume} {20}},\ \bibinfo {pages} {7865} (\bibinfo {year}
  {2024})}\BibitemShut {NoStop}%
\bibitem [{\citenamefont {Marconi}\ and\ \citenamefont
  {Caprini}(2025)}]{Marconi2025}%
  \BibitemOpen
  \bibfield  {author} {\bibinfo {author} {\bibfnamefont {U.~M.~B.}\
  \bibnamefont {Marconi}}\ and\ \bibinfo {author} {\bibfnamefont
  {L.}~\bibnamefont {Caprini}},\ }\bibfield  {title} {\bibinfo {title}
  {Spontaneous generation of angular momentum in chiral active crystals},\
  }\href {https://doi.org/10.1039/D4SM01426J} {\bibfield  {journal} {\bibinfo
  {journal} {Soft Matter}\ }\textbf {\bibinfo {volume} {21}},\ \bibinfo {pages}
  {2586} (\bibinfo {year} {2025})}\BibitemShut {NoStop}%
\bibitem [{\citenamefont {Barman}(2025)}]{Barman2025}%
  \BibitemOpen
  \bibfield  {author} {\bibinfo {author} {\bibfnamefont {H.}~\bibnamefont
  {Barman}},\ }\href {https://doi.org/10.48550/arXiv.2510.15419} {\bibinfo
  {title} {Confinement-induced delay in chiral active brownian particles}},\
  \bibinfo {howpublished} {arXiv preprint arXiv:2510.15419} (\bibinfo {year}
  {2025})\BibitemShut {NoStop}%
\bibitem [{\citenamefont {Hermans}\ and\ \citenamefont
  {Ullman}(1952)}]{Hermans1952}%
  \BibitemOpen
  \bibfield  {author} {\bibinfo {author} {\bibfnamefont {J.}~\bibnamefont
  {Hermans}}\ and\ \bibinfo {author} {\bibfnamefont {R.}~\bibnamefont
  {Ullman}},\ }\bibfield  {title} {\bibinfo {title} {The statistics of stiff
  chains, with applications to light scattering},\ }\href
  {https://doi.org/10.1016/S0031-8914(52)80231-9} {\bibfield  {journal}
  {\bibinfo  {journal} {Physica}\ }\textbf {\bibinfo {volume} {18}},\ \bibinfo
  {pages} {951} (\bibinfo {year} {1952})}\BibitemShut {NoStop}%
\bibitem [{\citenamefont {Daniels}(1952)}]{Daniels1952}%
  \BibitemOpen
  \bibfield  {author} {\bibinfo {author} {\bibfnamefont {H.}~\bibnamefont
  {Daniels}},\ }\bibfield  {title} {\bibinfo {title} {Xxi.—the statistical
  theory of stiff chains},\ }\href {https://doi.org/10.1017/S0080454100007160}
  {\bibfield  {journal} {\bibinfo  {journal} {Proceedings of the Royal Society
  of Edinburgh Section A: Mathematics}\ }\textbf {\bibinfo {volume} {63}},\
  \bibinfo {pages} {290} (\bibinfo {year} {1952})}\BibitemShut {NoStop}%
\bibitem [{\citenamefont {Shee}\ \emph {et~al.}(2020)\citenamefont {Shee},
  \citenamefont {Dhar},\ and\ \citenamefont {Chaudhuri}}]{Shee2020}%
  \BibitemOpen
  \bibfield  {author} {\bibinfo {author} {\bibfnamefont {A.}~\bibnamefont
  {Shee}}, \bibinfo {author} {\bibfnamefont {A.}~\bibnamefont {Dhar}},\ and\
  \bibinfo {author} {\bibfnamefont {D.}~\bibnamefont {Chaudhuri}},\ }\bibfield
  {title} {\bibinfo {title} {Active brownian particles: mapping to equilibrium
  polymers and exact computation of moments},\ }\href
  {https://doi.org/10.1039/D0SM00367K} {\bibfield  {journal} {\bibinfo
  {journal} {Soft Matter}\ }\textbf {\bibinfo {volume} {16}},\ \bibinfo {pages}
  {4776} (\bibinfo {year} {2020})}\BibitemShut {NoStop}%
\bibitem [{\citenamefont {Chaudhuri}\ and\ \citenamefont
  {Dhar}(2021)}]{Chaudhuri2021}%
  \BibitemOpen
  \bibfield  {author} {\bibinfo {author} {\bibfnamefont {D.}~\bibnamefont
  {Chaudhuri}}\ and\ \bibinfo {author} {\bibfnamefont {A.}~\bibnamefont
  {Dhar}},\ }\bibfield  {title} {\bibinfo {title} {Active brownian particle in
  harmonic trap: exact computation of moments, and re-entrant transition},\
  }\href {https://doi.org/10.1088/1742-5468/abd031} {\bibfield  {journal}
  {\bibinfo  {journal} {Journal of Statistical Mechanics: Theory and
  Experiment}\ }\textbf {\bibinfo {volume} {2021}},\ \bibinfo {pages} {013207}
  (\bibinfo {year} {2021})}\BibitemShut {NoStop}%
\bibitem [{\citenamefont {Patel}\ and\ \citenamefont
  {Chaudhuri}(2023)}]{Patel2023}%
  \BibitemOpen
  \bibfield  {author} {\bibinfo {author} {\bibfnamefont {M.}~\bibnamefont
  {Patel}}\ and\ \bibinfo {author} {\bibfnamefont {D.}~\bibnamefont
  {Chaudhuri}},\ }\bibfield  {title} {\bibinfo {title} {Exact moments and
  re-entrant transitions in the inertial dynamics of active brownian
  particles},\ }\href {https://doi.org/10.1088/1367-2630/ad1538} {\bibfield
  {journal} {\bibinfo  {journal} {New Journal of Physics}\ }\textbf {\bibinfo
  {volume} {25}},\ \bibinfo {pages} {123048} (\bibinfo {year}
  {2023})}\BibitemShut {NoStop}%
\bibitem [{\citenamefont {L{\"o}wen}(2020)}]{Lowen2020}%
  \BibitemOpen
  \bibfield  {author} {\bibinfo {author} {\bibfnamefont {H.}~\bibnamefont
  {L{\"o}wen}},\ }\bibfield  {title} {\bibinfo {title} {Inertial effects of
  self-propelled particles: From active brownian to active langevin motion},\
  }\href {https://doi.org/10.1063/1.5134455} {\bibfield  {journal} {\bibinfo
  {journal} {The Journal of Chemical Physics}\ }\textbf {\bibinfo {volume}
  {152}},\ \bibinfo {pages} {040901} (\bibinfo {year} {2020})}\BibitemShut
  {NoStop}%
\bibitem [{\citenamefont {It{\^o}}(1975)}]{Ito1975}%
  \BibitemOpen
  \bibfield  {author} {\bibinfo {author} {\bibfnamefont {K.}~\bibnamefont
  {It{\^o}}},\ }\bibfield  {title} {\bibinfo {title} {Stochastic calculus},\
  }in\ \href {https://doi.org/10.1007/BFb0013327} {\emph {\bibinfo {booktitle}
  {International Symposium on Mathematical Problems in Theoretical Physics}}},\
  \bibinfo {series} {Lecture Notes in Physics}, Vol.~\bibinfo {volume} {39},\
  \bibinfo {editor} {edited by\ \bibinfo {editor} {\bibfnamefont
  {H.}~\bibnamefont {Araki}}}\ (\bibinfo  {publisher} {Springer},\ \bibinfo
  {address} {Berlin, Heidelberg},\ \bibinfo {year} {1975})\ pp.\ \bibinfo
  {pages} {218--223}\BibitemShut {NoStop}%
\bibitem [{\citenamefont {van~den Berg}\ and\ \citenamefont
  {Lewis}(1985)}]{VandenBerg1985}%
  \BibitemOpen
  \bibfield  {author} {\bibinfo {author} {\bibfnamefont {M.}~\bibnamefont
  {van~den Berg}}\ and\ \bibinfo {author} {\bibfnamefont {J.~T.}\ \bibnamefont
  {Lewis}},\ }\bibfield  {title} {\bibinfo {title} {{Brownian Motion on a
  Hypersurface}},\ }\href {https://doi.org/10.1112/blms/17.2.144} {\bibfield
  {journal} {\bibinfo  {journal} {Bulletin of the London Mathematical Society}\
  }\textbf {\bibinfo {volume} {17}},\ \bibinfo {pages} {144} (\bibinfo {year}
  {1985})}\BibitemShut {NoStop}%
\bibitem [{\citenamefont {Raible}\ and\ \citenamefont
  {Engel}(2004)}]{Raible2004}%
  \BibitemOpen
  \bibfield  {author} {\bibinfo {author} {\bibfnamefont {M.}~\bibnamefont
  {Raible}}\ and\ \bibinfo {author} {\bibfnamefont {A.}~\bibnamefont {Engel}},\
  }\bibfield  {title} {\bibinfo {title} {{Langevin equation for the rotation of
  a magnetic particle}},\ }\href {https://doi.org/10.1002/aoc.757} {\bibfield
  {journal} {\bibinfo  {journal} {Applied Organometallic Chemistry}\ }\textbf
  {\bibinfo {volume} {18}},\ \bibinfo {pages} {536} (\bibinfo {year}
  {2004})}\BibitemShut {NoStop}%
\bibitem [{\citenamefont {Mijatovi{\'{c}}}\ \emph {et~al.}(2020)\citenamefont
  {Mijatovi{\'{c}}}, \citenamefont {Mramor},\ and\ \citenamefont {{Uribe
  Bravo}}}]{Mijatovic2020}%
  \BibitemOpen
  \bibfield  {author} {\bibinfo {author} {\bibfnamefont {A.}~\bibnamefont
  {Mijatovi{\'{c}}}}, \bibinfo {author} {\bibfnamefont {V.}~\bibnamefont
  {Mramor}},\ and\ \bibinfo {author} {\bibfnamefont {G.}~\bibnamefont {{Uribe
  Bravo}}},\ }\bibfield  {title} {\bibinfo {title} {{A note on the exact
  simulation of spherical Brownian motion}},\ }\href
  {https://doi.org/10.1016/j.spl.2020.108836} {\bibfield  {journal} {\bibinfo
  {journal} {Statistics \& Probability Letters}\ }\textbf {\bibinfo {volume}
  {165}},\ \bibinfo {pages} {108836} (\bibinfo {year} {2020})}\BibitemShut
  {NoStop}%
\bibitem [{Note1()}]{Note1}%
  \BibitemOpen
  \bibinfo {note} {The unit vectors ${\protect \bf \protect \hat u}_\theta =
  \partial \protect \hat {\protect \bm {u}}/\partial \theta $ and ${\protect
  \bf \protect \hat u}_\phi =(1/\sin \theta ) (\partial \protect \hat {\protect
  \bm {u}}/\partial \phi )$.}\BibitemShut {Stop}%
\bibitem [{\citenamefont {Zheng}\ \emph {et~al.}(2013)\citenamefont {Zheng},
  \citenamefont {ten Hagen}, \citenamefont {Kaiser}, \citenamefont {Wu},
  \citenamefont {Cui}, \citenamefont {Silber-Li},\ and\ \citenamefont
  {L\"owen}}]{Zheng2013}%
  \BibitemOpen
  \bibfield  {author} {\bibinfo {author} {\bibfnamefont {X.}~\bibnamefont
  {Zheng}}, \bibinfo {author} {\bibfnamefont {B.}~\bibnamefont {ten Hagen}},
  \bibinfo {author} {\bibfnamefont {A.}~\bibnamefont {Kaiser}}, \bibinfo
  {author} {\bibfnamefont {M.}~\bibnamefont {Wu}}, \bibinfo {author}
  {\bibfnamefont {H.}~\bibnamefont {Cui}}, \bibinfo {author} {\bibfnamefont
  {Z.}~\bibnamefont {Silber-Li}},\ and\ \bibinfo {author} {\bibfnamefont
  {H.}~\bibnamefont {L\"owen}},\ }\bibfield  {title} {\bibinfo {title}
  {Non-gaussian statistics for the motion of self-propelled janus particles:
  Experiment versus theory},\ }\href
  {https://doi.org/10.1103/PhysRevE.88.032304} {\bibfield  {journal} {\bibinfo
  {journal} {Phys. Rev. E}\ }\textbf {\bibinfo {volume} {88}},\ \bibinfo
  {pages} {032304} (\bibinfo {year} {2013})}\BibitemShut {NoStop}%
\end{thebibliography}%

\onecolumngrid
\appendix

\section{Derivation of exact analytic moments in two-dimensions}
\label{app-A}

\medskip

\noindent
\textbf{Moments generator equation:~}
The probability distribution function $P(\textbf{r},\hat{\textbf{u}},t)$ of the particle follows the Fokker-Planck equation
\begin{align}
\partial_t P &=D \nabla^2 P+ D_r\partial_\phi^2 P+\nabla\cdot[(\mu  \br- v_0\hat{\textbf{u}}) P]-\omega\partial_\phi P\,.
\label{FPE_dimensionless_2d}
\end{align}
The detailed derivation of the Fokker–Planck equation~\eqref{FPE_dimensionless_2d} corresponding to Eqs.~\eqref{eom1:2d}–\eqref{eom2:2d} in the absence of harmonic confinement ($\mu = 0$) is presented in Pattanayak {\it et al.}~\cite{Pattanayak2024}.
By performing a Laplace transform $\tilde{P}(\br,\bu,s) = \int_0^\infty dt e^{-st} P (\br,\bu,t)$, the Fokker-Planck equation can be expressed as
\begin{align}
-P(\textbf{r},\hat{\textbf{u}},0)+s\tilde{P}(\textbf{r},\hat{\textbf{u}},s)=D\nabla^2\tilde{P} +D_r\partial_\phi^2\tilde{P}+\mu \nabla\cdot(\br \tilde{P}) - v_0\bu\cdot\nabla\tilde{P} -\omega\partial_\phi \tilde{P}\,,
\label{LFPE2d}
\end{align}
where the initial condition at $t=0$ is set by $P(\br,\bu,0)=\delta(\br)\delta(\bu -\bu_0)$, without any loss of generality. Finally, this leads to the moments generator equation
\begin{align}
-\langle\psi\rangle_0+s\langle\psi\rangle_s=v_0\langle\hat{\textbf{u}}\cdot\nabla\psi\rangle_s-\mu \langle \br \cdot\nabla\psi\rangle_s+D\langle\nabla^2\psi\rangle_s+D_r\langle\partial_\phi^2\psi\rangle_s+\omega\langle\partial_\phi\psi\rangle_s~,
\label{ME2}
\end{align}
for the mean of an arbitrary dynamical variable $\psi$ defined as $\langle\psi\rangle_s = \int d\br d\bu \psi(\br,\bu)\tilde{P}(\br,\bu,s)$, where the initial condition $\langle \psi \rangle_0 = \int d\br d\bu\, \psi(\br,\bu)\, P(\br,\bu,0)$. 
We use Eq.~\eqref{ME2} to derive the dynamical moments exactly and analyze them to characterize the nonequilibrium dynamics and deviations from Gaussian behavior.

\subsection{Lower order moments}

\noindent
\textbf{Orientation autocorrelation:~}
The orientation autocorrelation was explicitly calculated in our previous article~\cite{Pattanayak2024}. Here, we present it again for completeness and to facilitate understanding of the subsequent calculations.
To begin, we consider observable $\psi=\bu$ in Eq.~\eqref{ME2}. Therefore, $\nabla^2 \psi=0 $, $\nabla \psi= 0$ and $\langle \psi\rangle_0=\hat{\textbf{u}}_0$. Considering $\bu= \hat{x} \cos{\phi}+\hat{y}\sin{\phi}$ we get $\partial_\phi^2\psi=-\psi$ $\partial_\phi \text{u}_x=-\text{u}_y $ and $\partial_\phi \text{u}_y=\text{u}_x $. The Laplace transformed expressions of the components of $\bu$, 
\begin{align}
\langle \text{u}_x \rangle_s&=\frac{(s+D_r)\text{u}_{x0}-\omega \text{u}_{y0}}{(s+D_r)^2+\omega^2}\,,
\\
\langle \text{u}_y \rangle_s&=\frac{(s+D_r)\text{u}_{y0}+\omega \text{u}_{x0}}{(s+D_r)^2+\omega^2}\,.
\end{align} 
Performing inverse Laplace transform on the above expressions, we get the components of orientation vector, $\bu(t)$ as a function of time as, 
\begin{align}
 \langle \text{u}_x (t)\rangle &=e^{- D_r t}\left(\text{u}_{x0}\cos(\omega t)-\text{u}_{y0}\sin(\omega t)\right)=e^{-D_r t} \cos (\omega t+\phi_0)\,,
\\
\langle \text{u}_y (t)\rangle &=e^{- D_r t}\left(\text{u}_{y0}\cos(\omega t)+\text{u}_{x0}\sin(\omega t)\right)=e^{-D_r t}  \sin (\omega t+\phi_0)\,.
\end{align} 
The initial orientation is defined by $\hat{\textbf{u}}_0=\hat{x}\cos{\phi_0}+\hat{y}\sin{\phi_0}$. Finally, the orientation autocorrelation $\langle\bu\cdot\bu_0\rangle$~\cite{Caprini2019, Otte2021, Caprini2023, Pattanayak2024, Shee2024}
\begin{align}
\langle\bu\cdot\bu_0\rangle=e^{- D_r t}\cos (\omega t)\,.
\end{align}
In dimensionless unit, $\langle\bu\cdot\bu_0\rangle=e^{- \tilde{t}}\cos (\Omega \tilde{t})$. Here, $\langle\hat{\textbf{u}}\cdot\hat{\textbf{u}}_0\rangle$ decays with time-constant $1$ and shows an oscillatory nature with time period $2\pi/\Omega$. This expression is similar to that obtained for a cABP without any trap, as expected, since the trap has no impact on the orientation dynamics~\cite{Caprini2019, Otte2021, Caprini2023, Pattanayak2024, Shee2024}.

\medskip

\noindent
\textbf{Mean displacement:~}
To calculate mean displacement, we substitute $\psi=\textbf{r}$ in Eq.~\eqref{ME2} and obtain mean displacement in Laplace space $\langle \textbf{r}\rangle_s=v_0 \bu/(s+\mu)$. Using the expressions of $\langle \text{u}_i\rangle_s$'s from the previous calculation and then performing inverse Laplace transform one can get components of displacement as follows
\begin{align}
\langle x\rangle(t)=&\frac{v_0  }{(D_r-\mu)^2+\omega ^2}\left[e^{-D_r  t} \left(-D_r \cos \left(\omega t +\phi _0\right)+\mu  \cos \left(\omega t+\phi _0\right)+\omega  \sin \left(\omega t +\phi _0\right)\right)\right.
\nonumber\\
&\left.+e^{-\mu t } \left(\cos \phi _0 \left(D_r-\mu \right)-\omega  \sin \phi _0\right)\right]\,,\label{eq:xavg_time_2d}\\
\langle y\rangle(t)=&\frac{v_0  }{(D_r-\mu)^2+\omega ^2}\left[e^{-D_r  t} \left(-D_r \sin \left(\omega t+\phi _0\right)+\mu  \sin \left(\omega t +\phi _0\right)-\omega  \cos \left(\omega t +\phi _0\right)\right)\right.
\nonumber\\&\left.+e^{-\mu t} \left(\sin \phi _0 \left(D_r-\mu \right)+\omega  \cos \phi _0\right)\right]\,.\label{eq:yavg_time_2d}
\end{align}
At long times $t \to \infty$, mean displacement $\langle x\rangle_{t\to\infty} = \langle y\rangle_{t\to\infty} = 0$. 
Moreover, In the absence of noise $D_r=0$, the above Eqs.~\eqref{eq:xavg_time_2d} and \eqref{eq:yavg_time_2d} simplifies to
\begin{align}
&\langle x\rangle(t)=\frac{v_0  }{\mu^2+\omega ^2}\left[\left(\mu  \cos \left(\omega t +\phi _0\right)+\omega  \sin \left(\omega t +\phi _0\right)\right)-e^{-\mu t} \left(\mu\cos \phi _0 +\omega  \sin \phi _0\right)\right]\,,\\
&\langle y\rangle(t)=\frac{v_0  }{\mu^2+\omega ^2}\left[\left(\mu  \sin \left(\omega t +\phi _0\right)-\omega  \cos \left(\omega t +\phi _0\right)\right)-e^{-\mu t} \left(\mu\sin \phi _0 -\omega  \cos \phi _0\right)\right]\,.
\end{align}
At long times $t \to \infty$, the noiseless mean displacement does not decay but instead traces a circular orbit of radius $v_0/\sqrt{\mu^2+\omega^2}$ at angular frequency $\omega$, reflecting
the balance between the torque-driven rotation of the propulsion direction and the restoring force of the trap.
For a constant $\phi_0$, we define,
\begin{align}
    \tilde{\rm r}_{\rm mean}(\ttt)= \sqrt{[\langle \tilde x\rangle(\ttt)]^2+[\langle \tilde y\rangle(\ttt)]^2} = \frac{\Pe}{\sqrt{(\beta-1)^2+\Omega^2}}\left[e^{-2 \beta \ttt}+e^{-2 \ttt}-2\,e^{-(\beta+1)\ttt}\cos{(\Omega \ttt)}\right]^{1/2}\,.
    \label{eq:mean_disp_2d_dimensionless}
\end{align}

The bimodal probability distribution functions, plotted in Figs. \ref{fig3}[f,h,j,l] (see main text).
The peaks of the bimodal distributions are located near $\tilde{\rm r}_{\rm mean}$.
\medskip

\noindent
\textbf{Position-orientation cross-correlation:~}
To quantify how the instantaneous orientation of an active particle correlates with its position, 
we consider the cross-correlation function $\langle \mathbf{r} \cdot \bu\rangle$. 
This quantity measures the degree of alignment between the propulsion direction and the displacement 
from the trap center. Starting from the moment-generating Eq.~\eqref{ME2} with the choice $\psi = \mathbf{r} \cdot \bu$, 
we obtain
\begin{align}
\langle \textbf{r}\cdot\bu\rangle_s=\frac{1}{(s+D_r+\mu)}\left[\frac{v_0}{s}+\omega\langle \partial_\phi(\textbf{r}\cdot\hat{\textbf{u}})\rangle_s\right]\,.   
\label{eq:ru_cross_corr_Laplace_2d}
\end{align}
Using $\psi = \partial_\phi(\mathbf{r} \cdot \bu) = -(x u_y - y u_x)$, 
and the corresponding operator relations, we find
\begin{align}
\nabla\partial_\phi(\textbf{r}\cdot\hat{\textbf{u}})&=\partial_\phi\nabla(\textbf{r}\cdot\hat{\textbf{u}})=\partial_\phi(\hat{\textbf{u}})~,
\nonumber\\
\hat{\textbf{u}}\cdot\nabla\partial_\phi(\textbf{r}\cdot\hat{\textbf{u}})&=0~,
\nonumber\\
\partial_\phi^2\partial_\phi\textbf{r}\cdot\bu&=-\partial_\phi(\textbf{r}\cdot \hat{\textbf{u}})~,
\nonumber\\
\partial_\phi\psi=\partial_\phi^2(\textbf{r}\cdot\hat{\textbf{u}})&=-(x\text{u}_x+y\text{u}_y)=-\textbf{r}\cdot\hat{\textbf{u}}~.\nonumber
\end{align}
Hence, from Eq.~\eqref{ME2}, we get
\begin{align}
\langle\partial_\phi(\textbf{r}\cdot\hat{\textbf{u}})\rangle_s=\frac{-\omega\langle\textbf{r}\cdot\hat{\textbf{u}}\rangle_s}{(s+D_r+\mu)}\,.
\end{align}
Substituting this into Eq.~\eqref{eq:ru_cross_corr_Laplace_2d} gives the position–orientation cross-correlation in Laplace space,
\begin{align}
\langle \textbf{r}\cdot\hat{\textbf{u}}\rangle_s=\frac{v_0(s+D_r+\mu)}{s[(s+D_r+\mu)^2+\omega^2]}\,. 
\label{eq:ur_cross_corr_Laplace_2d}
\end{align}
Taking the inverse Laplace transform gives the time-dependent correlation,
\begin{align}
\langle \textbf{r}\cdot\hat{\textbf{u}}\rangle (t)=\frac{v_0 \left[(D_r+\mu ) -(D_r+\mu ) e^{-t (D_r+\mu )} \cos ( \omega t)+\omega e^{-t (D_r+\mu )}  \sin ( \omega t)\right]}{(D_r+\mu )^2+\omega ^2}\,. 
\label{eq:ur_cross_corr_2d}
\end{align}
In dimensionless form, the cross-correlation becomes
\begin{align}
\langle \tbr\cdot\hat{\textbf{u}}\rangle (\ttt)=\frac{\Pe \left[(\beta+1) -(\beta+1) e^{-\ttt (\beta+1)} \cos ( \Omega \ttt)+\Omega e^{-\ttt (\beta+1)}  \sin ( \Omega \ttt)\right]}{(\beta+1 )^2+\Omega ^2}\,, 
\label{eq:ur_cross_corr_dimensionless_2d}
\end{align}
with its steady-state value
\begin{align}
\langle \tbr\cdot\hat{\textbf{u}}\rangle_{\rm st}=\frac{\Pe (\beta+1)}{(\beta+1)^2+\Omega ^2}\,. 
\end{align}
The cross-correlation with the perpendicular orientation component, 
$\bu^{\perp} = (-u_y, u_x)$, characterizes the extent of azimuthal coupling induced by torque. 
It is given by
\begin{align}
\langle\textbf{r}\cdot\hat{\textbf{u}}^{\perp}\rangle(t) &= \frac{v_0  \left(-\omega  +\left(D_r+\mu \right) e^{-t \left(D_r+\mu \right)}\sin ( \omega t)+\omega  e^{-t \left(D_r+\mu \right)} \cos ( \omega t)\right)}{\left(D_r+\mu \right)^2+\omega ^2} \,.
 \label{eq:uperpr_cross_corr_2d}
\end{align}
whose dimensionless form reads
\begin{align}
    \langle \tbr\cdot \hat{\textbf{u}}^{\perp}\rangle(\ttt) = \frac{\text{Pe} \left(-\Omega  +(\beta +1) e^{-(\beta +1) \ttt} \sin (\Omega \ttt)+\Omega  e^{-(\beta +1) \ttt} \cos (\Omega \ttt)\right)}{(\beta +1)^2+\Omega ^2} \,.
\label{eq:uperpr_cross_corr_dimensionless_2d}
\end{align}
In the steady state, this correlation becomes 
\begin{align}
  \langle \tbr\cdot \hat{\textbf{u}}^{\perp}\rangle_{\rm st} = -\frac{\text{Pe} ~\Omega }{(\beta +1)^2+\Omega ^2}\,.
  \label{eq:ruperp_st_2d}
\end{align}
The position–orientation correlation $\langle \tilde{\mathbf{r}} \cdot \bu\rangle$ 
quantifies how strongly the particle displacement aligns with its propulsion direction. 
It increases with activity, reflecting persistent self-propulsion, but decays over the timescale 
set by $(\beta+1)^{-1}$. 
The perpendicular correlation $\langle \tilde{\mathbf{r}} \cdot \bu^{\perp}\rangle$ 
arises solely from torque-induced rotational coupling.
Its steady-state magnitude increases with $\Omega$ at small chirality but decreases at large chirality(Eq.~\eqref{eq:ruperp_st_2d}).
At large $\Omega$, this term dominates, indicating that strong chiral rotation drives the particle 
to circulate around the trap center, reducing longitudinal alignment and producing a finite steady-state 
azimuthal offset in the position–orientation coupling.

\medskip

\noindent
\textbf{Mean-squared displacement (MSD):~} 
To calculate mean-squared displacement, we consider $\psi=\br^2$ in Eq.~\eqref{ME2} gives 

\begin{align}
\langle \br^2\rangle_s=\frac{1}{(s+2\mu)}\left[4D\langle 1\rangle_s+2v_0\langle \br\cdot\bu\rangle_s\right]\,.
\label{eq:r2avg_Laplace_2d}
\end{align}
where $\langle 1\rangle_s=1/s$ and position-orientation cross correlation $\langle \br\cdot\bu\rangle_s$ already calculated in Eq.~\eqref{eq:ur_cross_corr_Laplace_2d}.

Finally, substituting above equation into Eq.~\eqref{eq:r2avg_Laplace_2d} and taking its inverse Laplace transform, we get the mean-squared displacement (MSD), $\langle \br^2(t)\rangle$ as a function of time~\cite{Lowen2020, Debets2023, Shee2024, Barman2025} 
\begin{align}
\langle \mathbf{r}^2(t)\rangle 
&= \frac{2 D}{\mu}\left(1 - e^{-2\mu t}\right)
\nonumber\\
&\quad + \frac{v_0^2}{\big[(D_r+\mu)^2+\omega^2\big]\big[(D_r-\mu)^2+\omega^2\big]}
\Bigg[
\frac{(D_r+\mu)\big((D_r-\mu)^2+\omega^2\big)}{\mu}
-\frac{(D_r-\mu)\big((D_r+\mu)^2+\omega^2\big)e^{-2\mu t}}{\mu} \nonumber\\
&\quad
-2 e^{-(D_r+\mu)t}\Big[\big(\mu^2 - D_r^2 + \omega^2\big)\cos(\omega t)
+ 2D_r\omega \sin(\omega t)\Big]
\Bigg]\,.
\end{align}
The dimensionless form of the MSD presented in the main text in Eq.~\eqref{eq:msd_2d_dimensionless}.

\medskip

\noindent
\textit{Limiting cases:~} 
Now we proceed to calculate the two limiting cases of Eq.~\eqref{eq:msd_2d_dimensionless}: first by setting the trap stiffness $\beta$ very small, and second by setting chirality $\Omega$ very small.
Letting $\beta \to 0$, the MSD in Eq.~\eqref{eq:msd_2d_dimensionless} simplifies to free chiral active Brownian particle~\cite{Pattanayak2024}:
\begin{align}
\lim_{\beta \to 0} \langle \tilde{\mathbf{r}}^2 \rangle
&= 4\tilde{t}
+ \frac{2\mathrm{Pe}^2}{\Omega^2 + 1}\,\tilde{t}
+ \frac{2\mathrm{Pe}^2(\Omega^2 - 1)}{(\Omega^2 + 1)^2}  - \frac{2\mathrm{Pe}^2 e^{-\tilde{t}}
\left[(\Omega^2 - 1)\cos(\Omega \tilde{t})
+ 2\Omega \sin(\Omega \tilde{t})\right]}
{(\Omega^2 + 1)^2}\,.
\end{align}
Letting $\Omega \to 0$, the MSD in Eq.~\eqref{eq:msd_2d_dimensionless} simplifies to non-chiral active Brownian particle in a harmonic trap~\cite{Chaudhuri2021}:
\begin{align}
\lim_{\Omega \to 0} \langle \tilde{\mathbf{r}}^2\rangle
&= \frac{2}{\beta} + \frac{\mathrm{Pe}^2}{\beta(\beta+1)}
+ \frac{\left(\mathrm{Pe}^2 + 2 - 2\beta\right)e^{-2\beta\tilde{t}}}{\beta(\beta-1)}
- \frac{2\,\mathrm{Pe}^2 e^{-(\beta+1)\tilde{t}}}{\beta^2 - 1}\,.
\end{align}

\noindent
\textbf{Components of mean-squared displacement:~} 
Setting $\psi=x^2$, $y^2$ and $xy$ we obtain the expressions of Laplace transformed expressions of all the second-order correlation functions of the displacement vector as,
\begin{align*}
    &\langle  x^2\rangle_s  = \frac{2 D}{s (2 \mu +s)}  \\&+\frac{2v_0^2 \left(\omega ^2 \left(2 \mu -2 s \text{u}_{x0}^2+3 s+2 D_r\right)+(s+4 D_r) (\mu +s+D_r) \left(s \text{u}_{x0}^2+2 D_r\right)-s \text{u}_{x0} \text{u}_{y0} \omega  (2 \mu +3 s+6 D_r)\right)}{\left((s+4 D_r)^2+4 \omega ^2\right) \left((\mu +s+D_r)^2+\omega ^2\right)s (2 \mu +s)}\,,
    \\
    &\langle  y^2\rangle_s =\frac{2 D}{s (2 \mu +s)}
    \\&+\frac{2 v_0^2 \left(s \text{u}_{x0} \text{u}_{y0} \omega  (2 \mu +3 s+6 D_r)+\omega ^2 \left(2 \mu -2 s \text{u}_{y0}^2+3 s+2 D_r\right)+(s+4 D_r) (\mu +s+D_r) \left(s \text{u}_{y0}^2+2 D_r\right)\right)}{s (2 \mu +s) \left((s+4 D_r)^2+4 \omega ^2\right) \left((\mu +s+D_r)^2+\omega ^2\right)}\,,
    \\
    &\langle  x y\rangle_s = \frac{v_0^2 \left(\left(2 \text{u}_{x0}^2-1\right) \omega  (2 \mu +3 s+6 D_r)+2 (s+4 D_r) \text{u}_{x0} \text{u}_{y0} (\mu +s+D_r)-4 \text{u}_{x0} \text{u}_{y0} \omega ^2\right)}{(2 \mu +s) \left((s+4 D_r)^2+4 \omega ^2\right) \left((\mu +s+D_r)^2+\omega ^2\right)}\,.
\end{align*}
Averaging over initial orientations, we obtain $\langle x y\rangle_s=0$ and $\langle  x^2\rangle_s=\langle  y^2\rangle_s$. The time-dependent expressions of $\langle  x^2\rangle$ and $\langle  y^2\rangle$ in dimensionless unit,
\begin{align}
\langle \tilde{x}^2 \rangle = \langle \tilde{y}^2 \rangle
&= \frac{1}{\beta}\big(1 - e^{-2\beta \tilde{t}}\big) + \frac{\mathrm{Pe}^2}{2}\Bigg[
\frac{\beta+1}{\beta\big[(\beta+1)^2 + \Omega^2\big]}
+ \frac{(\beta-1)e^{-2\beta \tilde{t}}}{\beta\big[(\beta-1)^2 + \Omega^2\big]}
\nonumber\\
&
- \frac{2 e^{-(\beta+1)\tilde{t}}
\big[(\beta^2 + \Omega^2 - 1)\cos(\Omega \tilde{t})
+ 2\Omega \sin(\Omega \tilde{t})\big]}
{\big[(\beta-1)^2 + \Omega^2\big]\big[(\beta+1)^2 + \Omega^2\big]}
\Bigg]
\nonumber\\
&= \frac{\langle \tilde{\mathbf{r}}^2 \rangle}{2}\,,
\end{align}
which is expected in the case of isotropic mobility and for isotropic harmonic trap, where the motion of chiral active Brownian particle maintains spatial symmetry.

\begin{figure}[t]
\centering
\includegraphics[width=\linewidth]{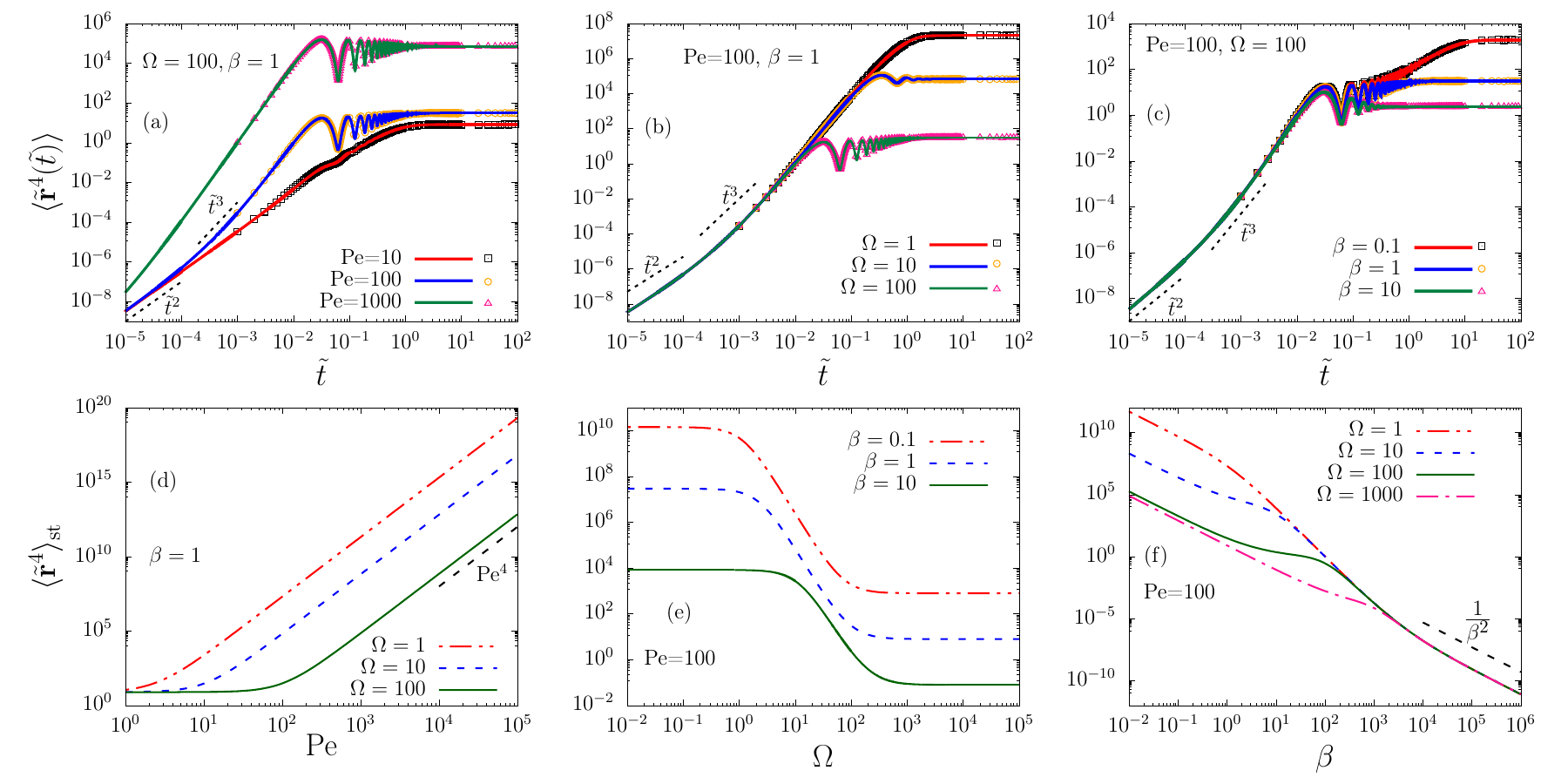}
\caption{
Fourth order moment of displacement of a chiral active Brownian particle (CABP) in a harmonic trap in two dimensions (2d) as a function of time for (a) $\text{Pe}=10, 100, 1000$ with $\Omega=100$ and $\beta=1$, (b) chirality $\Omega=1,10,100$ with activity $\text{Pe}=100$ and harmonic trap stiffness $\beta=1$, and (c) $\beta=0.1,1,10$ with $\Pe=100$ and $\Omega=100$.
Steady state fourth order moment of displacement as a function of (d) $\Pe$ for $\Omega=1,10,100$ with $\beta=1$, (e) $\Omega$ for $\beta=0.1,1,10$ with $\Pe=100$, and (f) $\beta$ for $\Omega=1,10,100,1000$ with $\Pe=100$.
The lines in (a), (b) and (c) are plots of Eq.~(\ref{eq:r4avg_2d_dimensionless}) and the points are from simulations. The lines in (d), (e), and (f) are plots of Eq.~(\ref{eq:r4avg_2d_st_dimensionless}).
}
\label{app_fig1}
\end{figure}

\begin{figure}[t]
\centering
\includegraphics[width=0.32\linewidth]{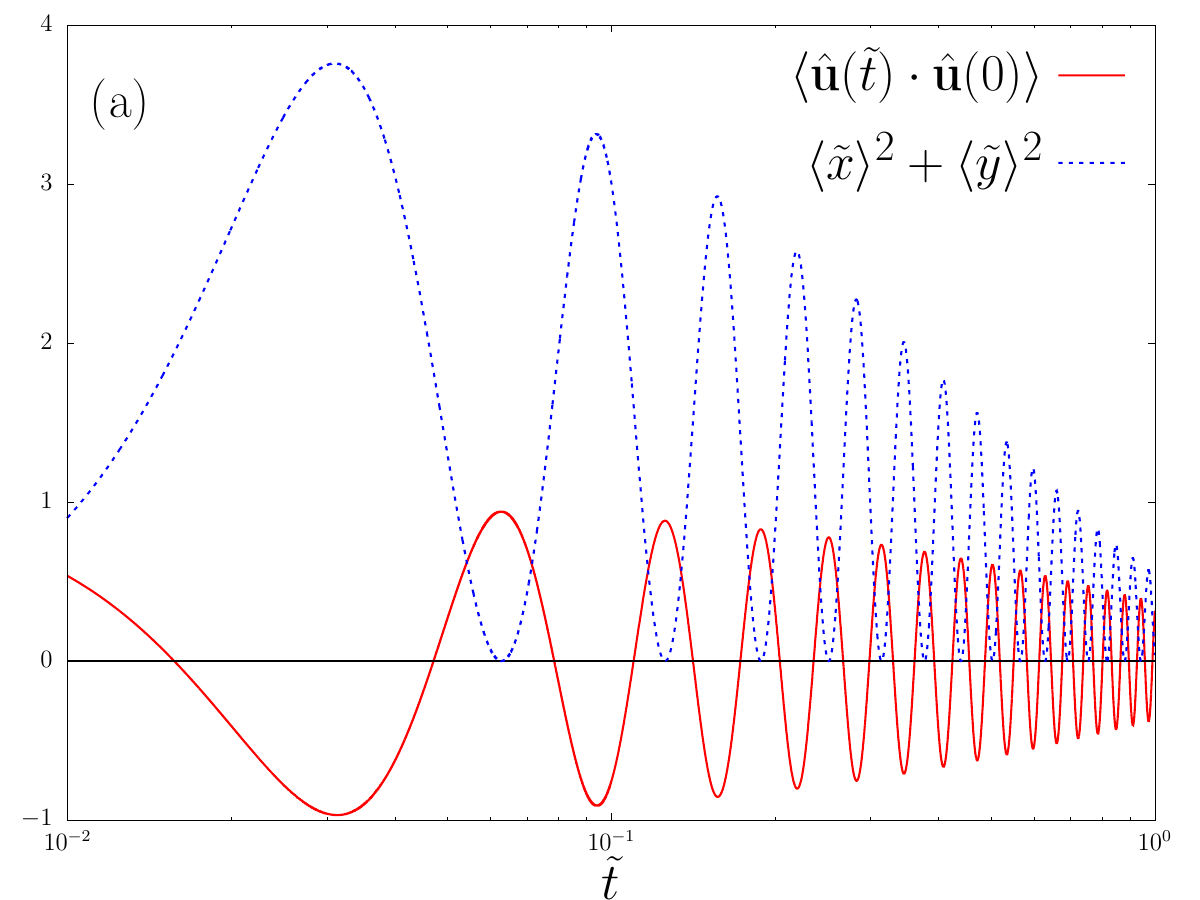}
\includegraphics[width=0.32\linewidth]{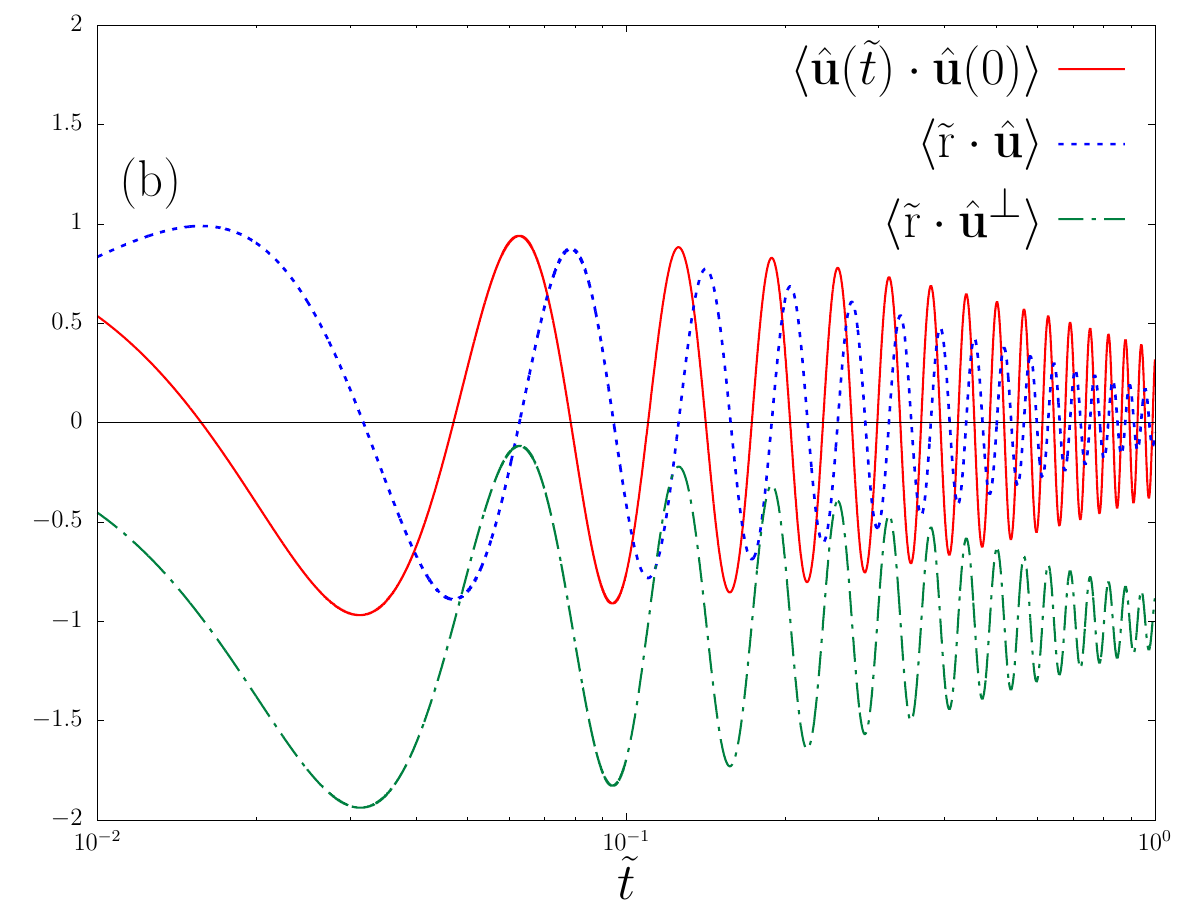}
\includegraphics[width=0.32\linewidth]{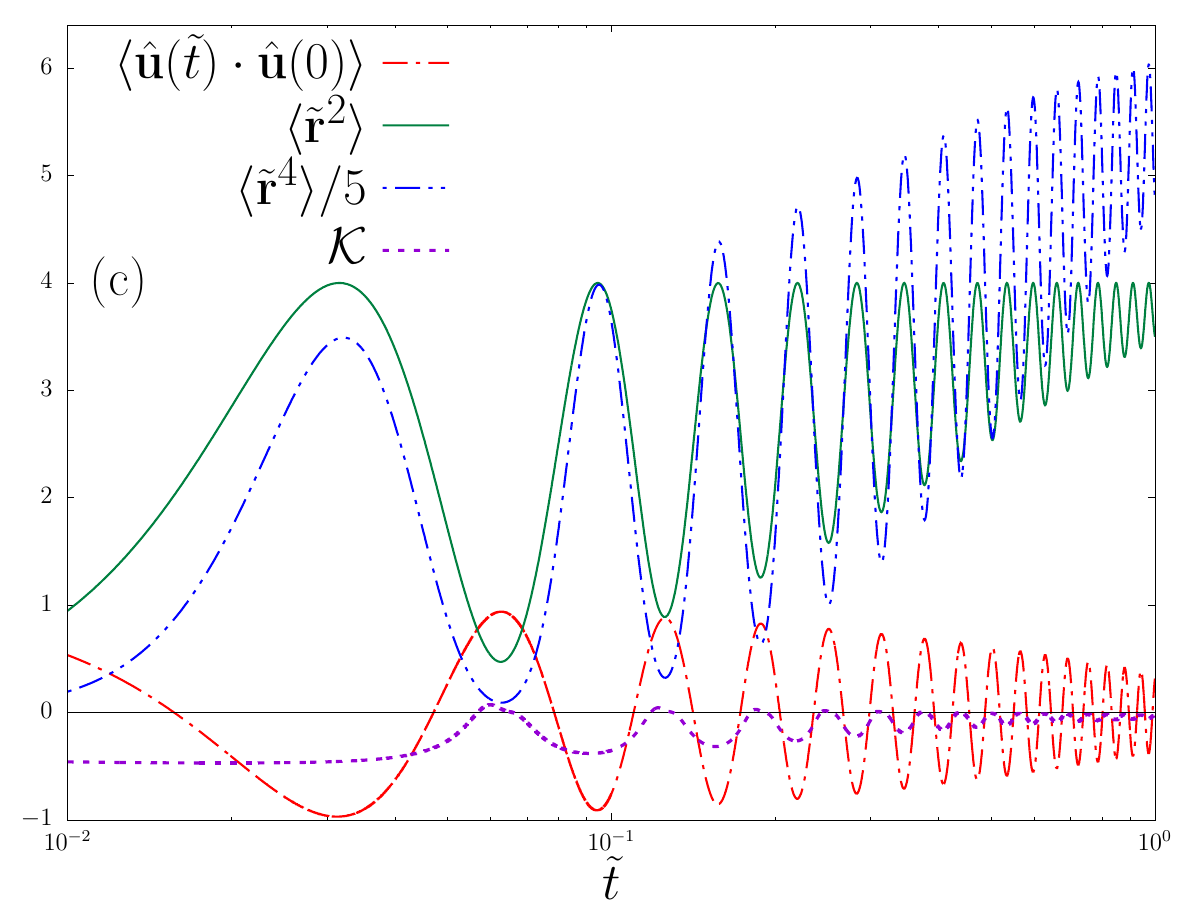}
\caption{
Oscillatory dynamics of a chiral active Brownian particle (CABP) confined in a two-dimensional harmonic trap, showing the time evolution of orientation autocorrelations $\langle \bu(\tilde{t}) \cdot \bu(0) \rangle=e^{- \tilde{t}}\cos (\Omega \tilde{t})$ (red line) and displacement moments, plotted as functions of the dimensionless time $\tilde{t}$.
(a) The quantity $\langle \tilde{x}\rangle^{2}  + \langle \tilde{y}\rangle^{2}$ (Eq.~\eqref{eq:mean_disp_2d_dimensionless}; blue line), showing damped oscillations that are out of phase with the orientation dynamics.
(b) The position–orientation cross-correlation $\langle \tilde{\mathbf{r}} \cdot \bu \rangle$ (Eq.~\eqref{eq:ur_cross_corr_dimensionless_2d}; blue line) exhibits oscillations that are out of phase, though not in exact opposite, while the perpendicular cross-correlation $\langle \tilde{\mathbf{r}} \cdot \bu_{\perp} \rangle$ (Eq.~\eqref{eq:uperpr_cross_corr_dimensionless_2d}; green line) oscillates in phase with the orientation dynamics.
(c) Mean-squared displacement(MSD) $\langle \tbr^{2} \rangle$ (Eq.~\eqref{eq:msd_2d_dimensionless} in main text; green line) and fourth-order moment of displacement $\langle \tbr^{4} \rangle / 5$ (Eq.~\eqref{eq:r4avg_2d_dimensionless}; blue line) exhibit out of phase oscillations, whereas excess kurtosis $\tilde{\mathcal{K}}$ (purple line) oscillates in phase with the underlying orientation dynamics.
The parameters used are activity $\mathrm{Pe}=100$, chirality $\Omega=100$, and trap stiffness $\beta=1$, which correspond to the same parameters used for excess kurtosis and orientation autocorrelation curves in Fig.~\ref{fig3}(d) and (e), respectively(see main text).
}
\label{app_fig2}
\end{figure}

\subsection{Fourth order moment of displacement}
To determine the higher-order statistics of the particle’s motion, we next compute the fourth order moment of the displacement. Substituting $\psi = \mathbf{r}^4$ into the moment-generating Eq.~\eqref{ME2} leads to
\begin{align}
\left\langle \br^{4}\right\rangle_s=\frac{1}{(s+4 \mu)}\left[16 D\left\langle \br^2\right\rangle_s+4 v_0\left\langle \br^2 \br\cdot \bu\right\rangle_{s}\right]\,.
\label{r4moment}
\end{align}
The second term in Eq.~\eqref{r4moment} can be obtained by evaluating the coupled moments involving 
$\langle \mathbf{r}^2\, \mathbf{r}\cdot\bu\rangle_s$. 
The detailed steps are outlined below.

\noindent
{\textit{Calculation of} $\left\langle \br^2 \br\cdot \bu\right\rangle_{s} $}

\begin{align}
&\left[s+D_r+3\mu+\frac{\omega^{2}}{s+D_r+3\mu}\right]\langle \br^2 \br\cdot \bu\rangle_s =8 D\left[\langle\br \cdot \bu\rangle_{s}+\frac{\omega\langle \partial_\phi(\br\cdot \bu)\rangle_{s}}{s+D_r+3\mu}\right]
\nonumber\\&+ v_0\left[2 \left\langle(\br \cdot \bu)^{2}\right\rangle_{s} 
+\frac{\omega\langle \partial_\phi(\br \cdot \bu)^{2}\rangle_{s}}{s+D_r+3\mu}+\langle \br^2\rangle_{s}\right]\,.
\end{align}

\noindent
{\textit{Calculation of} $\left\langle(\br \cdot \bu)^{2}\right\rangle_{s}$}

\begin{align}
&\left[s+4 D_r+2\mu+\frac{4 \omega^{2}}{s+4 D_r+2\mu}\right]\left\langle(\br\cdot \bu)^{2}\right\rangle_{s}= \frac{2 D}{s}+2 D_r\left\langle \br^2\right\rangle_{s}+2 v_0\langle\br\cdot \bu\rangle_{s} \nonumber\\
&+\frac{2 \omega v_0\left\langle \partial_\phi(\br\cdot \bu)\right\rangle_{s}}{s+4 D_r +2\mu} +2 \omega^{2} \frac{\left\langle \br^2\right\rangle_{s}}{s+4 D_r +2\mu}\,. 
\end{align}
Furthermore, the corresponding azimuthal derivative satisfies
\begin{align}
\left\langle \partial_\phi(\br\cdot \bu)^{2}\right\rangle_{s}  &= \frac{1}{\left(s+4 D_r+2 \mu\right)}\left[2 v_0\left\langle \partial_\phi(\br\cdot \bu)\right\rangle_{s}-4 \omega\left\langle(\br\cdot \bu)^{2}\right\rangle_{s} +2 \omega\left\langle \br^2\right\rangle_{s}\right]\,.
\end{align}

\newpage
\noindent
The fourth-order moment of displacement in dimensionless units,

    \begin{align}
    \begin{aligned}
        &\langle \tbr^4(\ttt) \rangle=-\frac{4 e^{-2 \beta  \ttt} \left(4 \left(\beta ^4+2 \beta ^2 \left(\Omega ^2-1\right)+\left(\Omega ^2+1\right)^2\right)-\left(\beta ^2-1\right) \text{Pe}^4+\text{Pe}^2 \left(-4 \beta ^2+4 \Omega ^2+4\right)\right)}{\beta ^2 \left(\beta ^4+2 \beta ^2 \left(\Omega ^2-1\right)+\left(\Omega ^2+1\right)^2\right)}
        \\& + e^{-4 \beta  \ttt} \left(\frac{8}{\beta ^2}+\frac{\text{Pe}^2 \left(-8 \beta +\frac{\text{Pe}^2 \left(9 \beta ^4-42 \beta ^3+67 \beta ^2+((\beta -4) \beta +2) \Omega ^2-42 \beta +8\right)}{\left((1-3 \beta )^2+\Omega ^2\right) \left((\beta -2)^2+\Omega ^2\right)}+8\right)}{\beta ^2 \left((\beta -1)^2+\Omega ^2\right)}\right)
        \\& +\frac{8}{\beta ^2}+\frac{\text{Pe}^2 \left(8 (\beta +1)+\frac{\text{Pe}^2 \left(9 \beta ^4+42 \beta ^3+67 \beta ^2+(\beta  (\beta +4)+2) \Omega ^2+42 \beta +8\right)}{\left((\beta +2)^2+\Omega ^2\right) \left((3 \beta +1)^2+\Omega ^2\right)}\right)}{\beta ^2 \left((\beta +1)^2+\Omega ^2\right)}
        \\&+ \text{Pe}^2 e^{-(3 \beta +1) \ttt}\cos{\Omega \ttt}  \left(\frac{2 (\beta +1) \left(8 \beta +(\beta -1) \text{Pe}^2\right)}{2 \beta ^3 \left((\beta +1)^2+\Omega ^2\right)}+\frac{(\beta -1) \left(16 \beta  (2 \beta -1)+(2 (\beta -2) \beta +1) \text{Pe}^2\right)}{2 \beta ^3 (2 \beta -1) \left((\beta -1)^2+\Omega ^2\right)}\right.
        \\&\left.-\frac{2 (\beta -3) \beta ^2 \text{Pe}^2}{2 \beta ^3 (2 \beta -1) \left((\beta -3)^2+\Omega ^2\right)}-\frac{(3 \beta +1) \text{Pe}^2}{2 \beta ^3 \left(4 \beta ^2-1\right) \left((3 \beta +1)^2+\Omega ^2\right)}-\frac{8 \beta ^4 \text{Pe}^2}{2 \beta ^3 \left(4 \beta ^2-1\right) \left(\beta ^2+\Omega ^2\right)}\right)
        \\& +\frac{4 \text{Pe}^2 \Omega e^{-(3 \beta +1) \ttt}\sin{\Omega \ttt} }{\beta  \left((\beta -3)^2+\Omega ^2\right) \left((\beta -1)^2+\Omega ^2\right) \left(\beta ^2+\Omega ^2\right) \left((\beta +1)^2+\Omega ^2\right) \left((3 \beta +1)^2+\Omega ^2\right)}\times
        \\&\Big(8 \left((\beta -3)^2+\Omega ^2\right) \left(\beta ^2+\Omega ^2\right) \left((3 \beta +1)^2+\Omega ^2\right)+\text{Pe}^2 \left(10 \left(-5 \beta ^3+8 \beta ^2+\beta -2\right) \Omega ^2+(12-5 \beta ) \Omega ^4\right.
        \\&\left.-\beta  (\beta  (\beta  (\beta  (45 \beta -116)+26)+68)+9)\right)\Big)   
        \\& + e^{-( \beta +1) \ttt}\sin{\Omega \ttt}\Bigg(-\frac{32 \text{Pe}^2 \Omega }{\beta  \left((\beta -1)^2+\Omega ^2\right) \left((\beta +1)^2+\Omega ^2\right)}
        \\&+\frac{4 \text{Pe}^4 \Omega  \left(10 \left(-5 \beta ^3-8 \beta ^2+\beta +2\right) \Omega ^2-\left((5 \beta +12) \Omega ^4\right)-\beta  (\beta  (\beta  (\beta  (45 \beta +116)+26)-68)+9)\right)}{\beta  \left((1-3 \beta )^2+\Omega ^2\right) \left((\beta -1)^2+\Omega ^2\right) \left(\beta ^2+\Omega ^2\right) \left((\beta +1)^2+\Omega ^2\right) \left((\beta +3)^2+\Omega ^2\right)}\Bigg)
        \\& +\text{Pe}^2e^{-( \beta +1) \ttt}\cos{\Omega \ttt} \left(-\frac{2 (\beta -1) \left(8 \beta +(\beta +1) \text{Pe}^2\right)}{2 \beta ^3 \left((\beta -1)^2+\Omega ^2\right)}-\frac{(\beta +1) \left(16 \beta  (2 \beta +1)+(2 \beta  (\beta +2)+1) \text{Pe}^2\right)}{2 \beta ^3 (2 \beta +1) \left((\beta +1)^2+\Omega ^2\right)}\right.
        \\&\left.+\frac{2 (\beta +3) \beta ^2 \text{Pe}^2}{2 \beta ^3 (2 \beta +1) \left((\beta +3)^2+\Omega ^2\right)}+\frac{(1-3 \beta ) \text{Pe}^2}{2 \beta ^3 \left(4 \beta ^2-1\right) \left((1-3 \beta )^2+\Omega ^2\right)}+\frac{8 \beta ^4 \text{Pe}^2}{2 \beta ^3 \left(4 \beta ^2-1\right) \left(\beta ^2+\Omega ^2\right)}\right)
        \\& +\frac{2 \text{Pe}^4 \left(\beta ^4+\beta ^2 \left(2 \Omega ^2-13\right)+\Omega ^4-37 \Omega ^2+36\right)e^{-2 (\beta +2) \ttt} \cos (2 \Omega \ttt )}{\left((\beta -3)^2+\Omega ^2\right) \left((\beta -2)^2+\Omega ^2\right) \left((\beta +2)^2+\Omega ^2\right) \left((\beta +3)^2+\Omega ^2\right)}
        \\&+ \frac{20 \text{Pe}^4 \Omega  \left(\beta ^2+\Omega ^2-6\right)e^{-2 (\beta +2) \ttt} \sin (2 \Omega \ttt )}{\left((\beta -3)^2+\Omega ^2\right) \left((\beta -2)^2+\Omega ^2\right) \left((\beta +2)^2+\Omega ^2\right) \left((\beta +3)^2+\Omega ^2\right)}\,.
        \label{eq:r4avg_2d_dimensionless}
        \end{aligned}
    \end{align}
We use Eq.~\eqref{eq:r4avg_2d_dimensionless} to compute the time-dependent excess kurtosis shown in Fig.~\ref{fig3} of the main text.
Small time expansion of fourth order moment as a function of time
\begin{align}
\lim_{\ttt \to 0} \langle \tbr^4\rangle(\ttt) &= 32 \ttt^2+16 \ttt^3 \left(\text{Pe}^2-4 \beta \right)+\ttt^4 \left(\frac{224 \beta ^2}{3}+\text{Pe}^4-\frac{16}{3} (6 \beta +1) \text{Pe}^2\right)\nonumber\\&- \ttt^5 \frac{2}{3}\bigg(96 \beta ^3+(3 \beta +1) \text{Pe}^4-2 \text{Pe}^2 \left(\beta  (27 \beta +8)-\Omega ^2+1\right)\bigg)+O\left(\ttt^6\right)\,.
\label{eq:r4avg_2d_small_time_dimensionless}
\end{align}
At very small times the dynamics is dominated by thermal fluctuations $\langle \tbr^4\rangle=32\ttt^2$, which crosses over to ballistic behavior $\langle \tbr^4\rangle\sim \ttt^3$ at $\ttt_{I}=2/(\Pe^2-4\beta)$.

\noindent
\textit{Steady state fourth order moment of displacement:~}
In the steady state, fourth order moment of displacement $\langle \tbr^4\rangle_{\rm st}= \lim_{\ttt \to \infty} \langle \tbr^4\rangle$~\cite{Pattanayak2025},
\begin{align}
\langle \tbr^4\rangle_{\rm st}
& =\frac{8}{\beta ^2}+\frac{8 (\beta +1) \text{Pe}^2}{\beta ^2 \left((\beta +1)^2+\Omega ^2\right)}\nonumber\\
&+\frac{\text{Pe}^4 \left(9 \beta ^4+42 \beta ^3+67 \beta ^2+(\beta  (\beta +4)+2) \Omega ^2+42 \beta +8\right)}{\beta ^2 \left((\beta +1)^2+\Omega ^2\right) \left((\beta +2)^2+\Omega ^2\right) \left((3 \beta +1)^2+\Omega ^2\right)}\,.
\label{eq:r4avg_2d_st_dimensionless}
\end{align}

Figure~\ref{app_fig1}(a)--(c) shows the time evolution and steady-state behaviour of the dimensionless fourth-order moment of displacement,
$\langle \tilde{\mathbf{r}}^{4}(\tilde{t})\rangle$, for a chiral active Brownian particle confined in a harmonic potential. The analytical results obtained from Eq.~\eqref{eq:r4avg_2d_dimensionless} (solid lines) show excellent agreement with simulation data (points) for all parameter sets. 
Similar to the mean-squared displacement, the dynamics exhibit a sequence of distinct regimes: an initial diffusive regime with $\langle \tilde{\mathbf{r}}^{4}\rangle\sim \tilde{t}^{2}$ dictated by thermal fluctuations; a ballistic growth regime, $\langle \tilde{\mathbf{r}}^{4}\rangle\sim \tilde{t}^{3}$, in which persistent propulsion dominates; an intermediate crossover regime where rotational decorrelation reduces the growth rate; and finally, saturation to a steady plateau determined by the trap stiffness and chiral rotation.  

At fixed $\Omega = 100$ and $\beta = 1$ [Fig.~\ref{app_fig1}(a)], increasing the activity ${\rm Pe}$ amplifies the magnitude and extends the duration of the ballistic regime, while also enhancing the prominence of the oscillatory features induced by chirality. 
For fixed ${\rm Pe}=100$ and $\beta=1$ [Fig.~\ref{app_fig1}(b)], increasing the chirality $\Omega$ introduces pronounced oscillatory modulations in the transient regime. As $\Omega$ grows, the overall amplitude of $\langle \tilde{\mathbf{r}}^{4}\rangle$ is progressively reduced, reflecting how rapid orientational precession averages out persistent propulsion and thereby suppresses large fluctuations.
Figure~\ref{app_fig1}(c) shows the effect of trap stiffness at fixed ${\rm Pe}=100$ and $\Omega=100$. Increasing $\beta$ damps the transient amplitude and speeds up relaxation to steady state. In the weak-trap limit ($\beta \ll 1$), the behavior shows diffusive behavior that of an unconfined chiral active particle saturates at relatively longer times. For large $\beta$, confinement dominates and strongly suppresses long time diffusive behavior, leading to a rapid relaxation towards steady-state distribution.

The steady-state fourth moment of displacement, plotted in Figs.~\ref{app_fig1}(d)--(f) of Eq.~\eqref{eq:r4avg_2d_st_dimensionless}, quantifies its dependence on activity ${\rm Pe}$, chirality $\Omega$, and trap strength $\beta$.
Figure~\ref{app_fig1}(d) highlights the passive Brownian limit, $\langle \tilde{\mathbf{r}}^{4}\rangle_{\rm st}^{\rm BP} \simeq 8/\beta^{2}$, at low activity ($\Pe \to 0$), and the strong quartic scaling $\langle \tilde{\mathbf{r}}^{4}\rangle_{\rm st} \propto \Pe^{4}$ that emerges at high activity ($\Pe \to \infty$).
Figure~\ref{app_fig1}(e) shows that $\langle \tilde{\mathbf{r}}^{4}\rangle_{\rm st}$ decreases monotonically with increasing chirality, transitioning from the non-chiral active Brownian limit, $\langle \tilde{\mathbf{r}}^{4}\rangle_{\rm st}^{\rm ABP}=\langle \tilde{\mathbf{r}}^{4}\rangle_{\rm st}(\Omega=0)$, in the low-chirality regime ($\Omega\to 0$), and saturating toward the passive Brownian value $\langle \tilde{\mathbf{r}}^{4}\rangle_{\rm st}^{\rm BP}$ in the high-chirality limit ($\Omega\to\infty$).
Figure~\ref{app_fig1}(f) demonstrates the characteristic inverse-square dependence on trap strength, $\langle \tilde{\mathbf{r}}^{4}\rangle_{\rm st} \sim 1/\beta^{2}$, reflecting how strong confinement suppresses spatial fluctuations across all orders.

\section{Derivation of exact analytic moments in three-dimensions}
\label{app-B}
\textbf{Moments generator equation:~}
The Fokker-Planck equation satisfied by the single particle probability distribution function, $P(\textbf{r},\hat{\textbf{u}},t)$,
\begin{align}
\partial_t P &=D\nabla^2 P+D_r\bR^2 P+\nabla\cdot[(\mu  \br- v_0\hat{\textbf{u}}) P]-\bm{\omega}\cdot\bR P\,.
\label{FPE_main}
\end{align}
Here $\bR\equiv \hat{\bm{u}}\times\nabla_\bu$ is equivalent to the gradient operator in the orientation vector space.
The detailed derivation of the Fokker–Planck equation~\eqref{FPE_main} corresponding to Eqs.~\eqref{eom1:3d}–\eqref{eom3:3d} in the absence of harmonic confinement ($\mu = 0$) is presented in Pattanayak {\it et al.}~\cite{Pattanayak2024}.
In the following sections, we have derived various moments associated with the motion of the particle.
By applying the Laplace transform to the Fokker-Planck equation,
\begin{align}
-P(\textbf{r},\hat{\textbf{u}},0)+s\tilde{P}(\textbf{r},\hat{\textbf{u}},s)=D\nabla^2\tilde{P} +D_r\bR^2\tilde{P}+\mu\nabla\cdot(\br \tilde{P}) - v_0\bu\cdot\nabla\tilde{P} -\bm{\omega}\cdot\bR \tilde{P}\,,
\label{LFPE3d}
\end{align}
where $\tilde{P}(\textbf{r},\hat{\textbf{u}},s)=\int_0^\infty dt e^{-st}P(\textbf{r},\hat{\textbf{u}},t)$ is the Laplace transformation of $P(\textbf{r},\hat{\textbf{u}},t)$ and $P(\textbf{r},\hat{\textbf{u}},0)=\delta^d(\textbf{r})\delta(\hat{\textbf{u}}-\hat{\textbf{u}}_0)$ is the initial probability distribution function.
Let us consider an arbitrary function $\psi=\psi(\textbf{r},\hat{\textbf{u}})$. If we multiply Eq.~\eqref{LFPE3d} with $\psi$ and integrate w.r.t $\textbf{r}$ and $\hat{\textbf{u}}$ over the whole space,
\begin{align}
-\psi_0+s\langle\psi\rangle_s=D\langle\nabla^2\psi\rangle_s+D_r\langle\bR^2\psi\rangle_s+v_0\langle\hat{\textbf{u}}\cdot\nabla\psi\rangle_s-\mu \langle \br \cdot\nabla\psi\rangle_s+\langle\boldsymbol{\Omega}\cdot\bR\psi\rangle_s\,,
\end{align}
where $\psi_0=\int d \textbf{r}\int d \hat{\textbf{u}} P(\textbf{r},\hat{\textbf{u}},0) \psi(\textbf{r},\hat{\textbf{u}})$ and $\langle\psi\rangle_s=\int d \textbf{r}\int d \hat{\textbf{u}} \tilde{P}(\textbf{r},\hat{\textbf{u}},s) \psi(\textbf{r},\hat{\textbf{u}})$
Let us consider the direction of the constant torque along the $z-$ axis, i.e. $\bm{\omega}=\omega \hat{z}$, which leads to a helical trajectory with a circular motion in the $x-y$ plane. The above equation for computing the moments further simplifies to,
\begin{align}
-\psi_0 +s\langle\psi\rangle_s =D\langle\nabla^2\psi\rangle_s+D_r\langle\bR^2\psi\rangle_s+v_0\langle\hat{\textbf{u}}\cdot\nabla\psi\rangle_s-\mu \langle \br \cdot\nabla\psi\rangle_s+\omega\langle\bR_z\psi\rangle_s\,.
\label{ME3}
\end{align}
Using the above equation, we compute the Laplace-transformed moments as outlined below. The main focus here is on the particle’s exact dynamics, supported by explicit analytical calculations.

\subsection{Lower order moments}

\noindent
\textbf{Orientation autocorrelation:~}
The explicit calculation of the orientation autocorrelation was already shown in our previous article~\cite{Pattanayak2024}. Here, we present it again for the sake of completeness.

First we take $\psi=\hat{\textbf{u}}$. Now, from the properties of $\bR$  operator, we can write $\mathcal{R}_i \text{u}_j=-\epsilon_{ijk}\text{u}_k$, $\bR^2 \text{u}_i=-(d-1) \text{u}_i$ where $d$ stands for dimension of the space.  $\nabla^2 \text{u}_i=0 $, $\nabla \psi= 0$ and $\psi_0=\hat{\textbf{u}}_0$. From Eq.~\eqref{ME3} for $d=3$,
\begin{align}
(s+2 D_r)\langle \text{u}_x \rangle_s +\omega \langle \text{u}_y \rangle_s &=\text{u}_{x0}\,,
\label{ux3d}
\\
(s+2 D_r)\langle \text{u}_y \rangle_s -\omega \langle \text{u}_x \rangle_s &=\text{u}_{y0}\,,
\label{uy3d}
\\
(s+2 D_r)\langle \text{u}_z \rangle_s  &=\text{u}_{z0}\,.
\label{uz3d}
\end{align}
Solving equations~$\eqref{ux3d}$ and $\eqref{uy3d}$ and rearranging Eq.~\eqref{uz3d}, we get
\begin{align}
\langle \text{u}_x \rangle_s&=\frac{(s+2 D_r)\text{u}_{x0}-\omega \text{u}_{y0}}{(s+2 D_r)^2+\omega^2}\,,
\\
\langle \text{u}_y \rangle_s&=\frac{(s+2D_r)\text{u}_{y0}+\omega \text{u}_{x0}}{(s+2 D_r)^2+\omega^2}\,,
\\
\langle \text{u}_z \rangle_s&=\frac{\text{u}_{z0}}{s+2D_r}\,.
\end{align} 
Performing inverse Laplace transform on the above expressions, we get the time evolution of the components of orientation vector, $\bu$ as,
\begin{align}
 \langle \text{u}_x (t)\rangle&=e^{-2 D_r t}\left(\text{u}_{x0}\cos(\omega t)-\text{u}_{y0}\sin(\omega t)\right)\,,
\\
\langle \text{u}_y (t)\rangle&=e^{-2 D_r t}\left(\text{u}_{y0}\cos(\omega t)+\text{u}_{x0}\sin(\omega t)\right)\,,
\\
\langle \text{u}_z (t)\rangle&=\text{u}_{z0}e^{-2 D_r t}\,.
\end{align} 
The above expressions are written in dimensionless units. 
The time correlation of the orientation vector $\hat{\textbf{u}}\cdot\hat{\textbf{u}}_0$ evolves as~\cite{Patel2023},
\begin{align}
\langle\hat{\textbf{u}}\cdot\hat{\textbf{u}}_0\rangle=e^{-2 D_r t}\Big[\cos \omega t+\text{u}_{z0}^2(1-\cos \omega  t)\Big]\,.
\end{align}
In dimensionless unit, $\langle\hat{\textbf{u}}\cdot\hat{\textbf{u}}_0\rangle=e^{-2 \tilde{t}}[\cos \Omega \tilde{t}+\text{u}_{z0}^2(1-\cos \Omega  \tilde{t})]$. In three dimensions, active particle’s orientation memory decays exponentially at rate $2D_r$ but also oscillates due to torque-induced precession, with the oscillation strength determined by how much of the initial direction lies perpendicular to the torque axis.

\medskip

\noindent
\textbf{Mean displacement:~}
Considering $\psi=\textbf{r}$ in equation \eqref{ME3}, we get $\langle \textbf{r}\rangle_s= v_0 \bu/(s+\mu)$. After using Expressions of $\langle \text{u}_i\rangle_s$ from the previous section and then performing its inverse Laplace transform one can get components of displacement in dimensionless units as follows
\begin{align}
    &\langle x(t)\rangle =\frac{-v_0 e^{-\mu t } \left[\left(\mu -2 D_r\right)\text{u}_{x0}+\omega  \text{u}_{y0}\right]}{\left(\mu -2 D_r\right)^2+\omega ^2}\nonumber\\
    & +\frac{v_0  e^{-2D_r t} \left[\sin (\omega t) \left(\omega  \text{u}_{x0}-\left(\mu -2 D_r\right) \text{u}_{y0} \right)+\cos (\omega t) \left( \left(\mu -2 D_r\right) \text{u}_{x0} +\omega  \text{u}_{y0}\right)\right]}{\left(\mu -2 D_r\right){}^2+\omega ^2}\,,
    \\
    &\langle y(t)\rangle =\frac{v_0  e^{- \mu t} \left(\omega  \text{u}_{x0}-\left(\mu -2 D_r\right) \text{u}_{y0}\right)}{\left(\mu -2 D_r\right){}^2+\omega ^2}\nonumber\\
    &+\frac{v_0  e^{-2 D_r  t} \left[\sin (\omega t) \left( \left(\mu -2 D_r\right) \text{u}_{x0} +\omega  \text{u}_{y0}\right)+\cos (\omega t) \left( \left(\mu -2 D_r\right) \text{u}_{y0} -\omega  \text{u}_{x0}\right)\right]}{\left(\mu -2 D_r\right){}^2+\omega ^2}\,,
    \\
    &\langle z(t)\rangle =-\frac{ v_0 \text{u}_{z0} \left(e^{-2 D_r t}-e^{  -\mu t}\right)}{2 D_r-\mu }\,.
\end{align}
These expressions reveal two competing dissipative scales: the trap relaxation rate $\mu$ and the rotational decorrelation rate $2D_r$. The $x$ and $y$ components display oscillatory relaxation reflecting chiral precession with frequency $\omega$, while the $z$–component decays monotonically since the torque does not act along that direction.

In the absence of rotational noise $D_r=0$,
\begin{align*}
\langle x(t)\rangle &= \frac{v_0}{\mu^2 + \omega^2} \left[ \cos(\omega t) \left(\mu \text{u}_{x0} + \omega \text{u}_{y0}\right) + \sin(\omega t) \left(\omega \text{u}_{x0} - \mu \text{u}_{y0}\right) - e^{-\mu t} \left(\mu \text{u}_{x0} + \omega \text{u}_{y0}\right) \right], \\
\langle y(t)\rangle &= \frac{v_0}{\mu^2 + \omega^2} \left[ \cos(\omega t) \left(\mu \text{u}_{y0} - \omega \text{u}_{x0}\right) + \sin(\omega t) \left(\mu \text{u}_{x0} + \omega \text{u}_{y0}\right) - e^{-\mu t} \left(\omega \text{u}_{x0} - \mu \text{u}_{y0}\right) \right], \\
\langle z(t)\rangle &= \frac{v_0 \text{u}_{z0} \left(1 - e^{-\mu t}\right)}{\mu}.
\end{align*}
In this deterministic limit, the particle traces a circular or helical trajectory, depending on the initial vertical component $u_{z0}$. The trap suppresses motion exponentially in time, but a steady oscillatory component remains in the transverse plane due to the torque.

At long times, by setting $t\to\infty$, we get
\begin{align*}
\langle x(t)\rangle &= \frac{v_0}{\mu^2 + \omega^2} \left[ \cos(\omega t) \left(\mu \text{u}_{x0} + \omega \text{u}_{y0}\right) + \sin(\omega t) \left(\omega \text{u}_{x0} - \mu \text{u}_{y0}\right) \right], \\
\langle y(t)\rangle &= \frac{v_0}{\mu^2 + \omega^2} \left[ \cos(\omega t) \left(\mu \text{u}_{y0} - \omega \text{u}_{x0}\right) + \sin(\omega t) \left(\mu \text{u}_{x0} + \omega \text{u}_{y0}\right) \right], \\
\langle z(t)\rangle &= \frac{v_0 \text{u}_{z0}}{\mu}.
\end{align*}
At long times, the motion in the $x$–$y$ plane approaches a periodic orbit determined by $\omega$ and the trap strength $\mu$, while the mean $z$–position saturates to a finite value. This reflects a balance between self–propulsion along the torque axis and confinement, giving a stationary helical orbit in the absence of noise.

\medskip

\noindent
\textbf{Position-orientation cross-correlation:~}
To calculate position-orientation cross-correlation, we consider $\psi = \textbf{r}\cdot\bu$ in Eq.~\eqref{ME3}, using $\nabla\psi=\bu$ and $\bR^2(\textbf{r}\cdot\hat{\textbf{u}})=-2\textbf{r}\cdot\hat{\textbf{u}}$, we get
\begin{align}
\langle \textbf{r}\cdot\hat{\textbf{u}}\rangle_s=\frac{1}{(s+2 D_r+\mu)}\left[\frac{v_0}{s}+\omega\langle \bR_z(\textbf{r}\cdot\hat{\textbf{u}})\rangle_s\right]\,.
\label{eq:ru_cross_corr_Laplace_3d}
\end{align}
Next we consider $\psi=\bR_z(\textbf{r}\cdot\hat{\textbf{u}})=-(x\text{u}_y-y\text{u}_x)$. 
\begin{align}
\nabla\bR_z(\textbf{r}\cdot\hat{\textbf{u}})&=\bR_z\nabla(\textbf{r}\cdot\hat{\textbf{u}})=\bR_z(\hat{\textbf{u}})\,,
\nonumber\\
\hat{\textbf{u}}\cdot\nabla\bR_z(\textbf{r}\cdot\hat{\textbf{u}})&=\hat{\textbf{u}}\cdot\bR_z\nabla(\textbf{r}\cdot\hat{\textbf{u}})=\hat{\textbf{u}}\cdot\bR_z\bu=0\,,
\nonumber\\
\bR^2\bR_z(\textbf{r}\cdot\hat{\textbf{u}})&=-2\bR_z(\textbf{r}\cdot \hat{\textbf{u}})\,,
\nonumber\\
\bR_z\bR_z(\textbf{r}\cdot\hat{\textbf{u}})&=\bR_z(-(x\text{u}_y-y\text{u}_x))=-(x\text{u}_x+y\text{u}_y)=-\textbf{r}\cdot\hat{\textbf{u}}+z\text{u}_z\,.\nonumber
\end{align}
Hence from Eq.~\eqref{ME3}, we get 
\begin{align}
\langle\bR_z(\textbf{r}\cdot\hat{\textbf{u}})\rangle_s=\frac{1}{(s+2 D_r+\mu)}\left[-\omega\langle\textbf{r}\cdot\hat{\textbf{u}}\rangle_s+\omega \langle z\text{u}_z\rangle_s\right]\,.
\end{align}
Putting this back into Eq.~\eqref{eq:ru_cross_corr_Laplace_3d}, we get 
\begin{align}
\langle \textbf{r}\cdot\hat{\textbf{u}}\rangle_s=\frac{v_0(s+2 D_r+\mu)+ s\, \omega^2 \langle z\text{u}_z\rangle_s}{s\left[(s+2 D_r+\mu)^2+\omega^2\right]}\,.
\label{eq:ru_cross_corr_Laplace_3d_final}
\end{align}
Following the same procedure and using the relation, $\bR^2\text{u}_i \text{u}_j=-6\text{u}_i\text{u}_j+2\delta_{ij}$, we calculate $\langle z\text{u}_z\rangle_s$ reads
\begin{align}
&\langle z\text{u}_z\rangle_s=\frac{v_0\langle \text{u}_z^2\rangle_s}{(s+2 D_r+\mu)}\,,
\end{align}
and $\langle \text{u}_z^2\rangle_s$
\begin{align}
&\langle \text{u}_z^2\rangle_s=\frac{2 D_r/s+\text{u}_{z0}^2}{s+6 D_r}\,.  
\end{align}
Finally, we obtain the position–orientation cross-correlation as a function of time by performing the inverse Laplace transform of Eq.~\eqref{eq:ru_cross_corr_Laplace_3d_final}
\begin{align}
\begin{aligned}
&\langle\textbf{r}\cdot\hat{\textbf{u}}\rangle (t) = \frac{v_0}{3}\Bigg(\frac{3 \left(2 D_r+\mu \right){}^2+\omega ^2}{\omega ^2 \left(2 D_r+\mu \right)+\left(2 D_r+\mu \right){}^3}+\frac{\omega ^2 \left(1-3 \text{u}_{z0}^2\right) e^{-6 t D_r}}{\left(4 D_r-\mu \right) \left(\left(\mu -4 D_r\right){}^2+\omega ^2\right)}
\\& -\frac{e^{-t \left(2 D_r+\mu \right)} \left(6 D_r \left(\text{u}_{z0}^2-1\right)+3 \mu  \text{u}_{z0}^2\right)}{\left(\mu -4 D_r\right) \left(2 D_r+\mu \right)}
\\& +\cos{\omega t}\frac{3 e^{-t \left(2 D_r+\mu \right)} \left(4 D_r \left(\mu ^2-\omega ^2 \text{u}_{z0}^2\right)+4 \mu  D_r^2 \left(1-3 \text{u}_{z0}^2\right)-16 D_r^3 \left(\text{u}_{z0}^2+1\right)+\mu  \left(\mu ^2+\omega ^2\right) \left(\text{u}_{z0}^2-1\right)\right)}{\left(4 \mu  D_r+4 D_r^2+\mu ^2+\omega ^2\right) \left(-8 \mu  D_r+16 D_r^2+\mu ^2+\omega ^2\right)}
\\& -\sin{\omega t}\frac{3 \omega  e^{-t \left(2 D_r+\mu \right)} \left(4 \mu  D_r \left(\text{u}_{z0}^2+1\right)+4 D_r^2 \left(\text{u}_{z0}^2-3\right)+\left(\mu ^2+\omega ^2\right) \left(\text{u}_{z0}^2-1\right)\right)}{\left(4 \mu  D_r+4 D_r^2+\mu ^2+\omega ^2\right) \left(-8 \mu  D_r+16 D_r^2+\mu ^2+\omega ^2\right)}\Bigg)\,.
\label{eq:ur_cross_corr_3d}
\end{aligned}
\end{align}
The quantity $\langle \mathbf{r}\cdot\bu\rangle$ measures how strongly the particle's displacement remains aligned with its propulsion direction.
In three dimensions, this correlation decays under the combined action of trap relaxation and rotational diffusion, while chirality introduces damped oscillations whose instantaneous frequency is set by the torque magnitude but whose amplitude and decay envelope are governed by the combined rate of orientational relaxation and trap confinement, as reflected in the full pole structure of Eq.~\eqref{eq:ur_cross_corr_3d}.
The presence of the additional term $\langle z u_z\rangle$ reflects the special role of the torque axis: motion parallel to the axis relaxes monotonically, whereas the transverse components precess, giving rise to oscillatory modulation.

In the steady state, Eq.~\eqref{eq:ur_cross_corr_3d} reads
\begin{align}
\langle\textbf{r}\cdot\hat{\textbf{u}}\rangle_{\rm st}=\frac{v_0}{3}\left[\frac{2 (\mu +2 D_r)}{(\mu +2 D_r)^2+\omega ^2}+\frac{1}{\mu +2 D_r}\right]\,.
\end{align}
At long times, the cross-correlation saturates to a finite, non-zero value,
indicating persistent alignment between position and orientation even in the presence of rotational noise.
In the limit $\omega\to 0$, the oscillatory contributions vanish, recovering the standard active Brownian limit.
Conversely, increasing chirality suppresses longitudinal alignment while generating a finite transverse component, consistent with helical trajectories in steady state.

The dimensionless form of position-orientation cross correlation,
\begin{align}
\langle \tbr\cdot\bu\rangle (\ttt) &=\frac{\text{Pe}}{3}\left[\frac{2 (\beta +2)}{(\beta +2)^2+\Omega ^2}+\frac{1}{\beta +2}+\frac{e^{-6 \ttt} \left(3 \text{u}_{z0}^2-1\right) \Omega ^2}{(\beta -4) \left((\beta -4)^2+\Omega ^2\right)}+\frac{e^{(\beta +2) (-\ttt)} \left(6-3 (\beta +2) \text{u}_{z0}^2\right)}{(\beta -4) (\beta +2)}\right]
\nonumber\\&
+\frac{\text{Pe}~e^{-(\beta +2) \ttt} \cos (\Omega \ttt) \left[-\beta  \left[(\beta -4) \beta +\Omega ^2-4\right]+(\beta -4) \left[(\beta +2)^2+\Omega ^2\right] \text{u}_{z0}^2-16\right]}{\left[(\beta -4)^2+\Omega ^2\right] \left[(\beta +2)^2+\Omega ^2\right]}
\nonumber\\& + \frac{\text{Pe} \Omega  e^{-(\beta +2)\ttt} \sin (\Omega \ttt) \left[(\beta -4) \beta -\left((\beta +2)^2+\Omega ^2\right) \text{u}_{z0}^2+\Omega ^2+12\right]}{\left[(\beta -4)^2+\Omega ^2\right] \left[(\beta +2)^2+\Omega ^2\right]}\,.
\label{eq:ur_cross_corr_dimensionless_3d}
\end{align}
In the steady state, Eq.~\eqref{eq:ur_cross_corr_dimensionless_3d} gives the dimensionless form of position-orientation cross-correlation:
\begin{align}
\langle\tbr\cdot\hat{\textbf{u}}\rangle_{\rm st}=\frac{\text{Pe}}{3}\left[\frac{2 (\beta +2)}{(\beta +2)^2+\Omega ^2}+\frac{1}{\beta +2}\right]\,.
\end{align}

Perpendicular position--orientation correlation:

\begin{align}
    \begin{aligned}
       & \langle\br\cdot\bu^{\perp}\rangle = \cos{\omega t}\frac{v_0 \omega  e^{-t \left(2 D_r+\mu \right)} \left(-4 D_r \left(\mu -3 D_r\right)-\text{u}_{z0}^2 \left(4 D_r \left(D_r+\mu \right)+\mu ^2+\omega ^2\right)+\mu ^2+\omega ^2\right)}{\left((\mu-4 D_r)^2+\omega ^2\right) \left((\mu+2 D_r)^2+\omega ^2\right)}
       \\& + \frac{v_0 e^{-t \left(2 D_r+\mu \right)} \sin{\omega t}}{\left((\mu-4 D_r)^2+\omega ^2\right) \left((\mu+2 D_r)^2+\omega ^2\right)}\bigg(-4 D_r \left(D_r \left(\mu -4 D_r\right)+\mu ^2\right)
       \\&-\left(\text{u}_{z0}^2 \left(\mu -4 D_r\right) \left(4 D_r \left(D_r+\mu \right)+\mu ^2+\omega ^2\right)\right)+\mu  \left(\mu ^2+\omega ^2\right)\bigg)
       \\&+\frac{v_0 \omega }{3}  \left(\frac{\left(3 \text{u}_{z0}^2-1\right) e^{-6 t D_r}}{-8 \mu  D_r+16 D_r^2+\mu ^2+\omega ^2}-\frac{2}{4 D_r \left(D_r+\mu \right)+\mu ^2+\omega ^2}\right) \,.
    \end{aligned}
\end{align}
The quantity $\langle \mathbf{r}\cdot \bu^{\perp}\rangle$ characterizes the coupling between the particle's position and the component of its self-propulsion direction perpendicular to both the torque axis and the instantaneous orientation, 
$\bu^{\perp} = -\hat{\mathbf{x}}\,u_y + \hat{\mathbf{y}}\,u_x$. This correlation captures the transverse displacement generated by chiral rotation. The full expression contains oscillatory $\cos(\omega t)$ and $\sin(\omega t)$ contributions, modulated by exponential damping $e^{-t(2D_r+\mu)}$, reflecting the competition between deterministic torque--driven precession and orientational decorrelation. The prefactors depend on the initial orientation through $u_{z0}$, demonstrating that the strength of transverse motion is maximal when the swimmer initially lies in the plane perpendicular to the torque axis, and vanishes when initially aligned with it. 
The final term, which includes $e^{-6D_r t}$, represents the decay of higher--order orientational correlations. Altogether, $\langle \mathbf{r}\cdot\bu^{\perp}\rangle$ reveals a damped helical drift orthogonal to the driving torque, which disappears in the limits $\omega\to 0$ or $D_r\to\infty$, and persists only when chirality breaks rotational symmetry.

In the long-time limit, the perpendicular position–orientation correlation converges to
\begin{align}
    \langle \br\cdot\bu^{\perp}\rangle_{\rm st} = \frac{-2 v_0 \omega }{3 \left(4 D_r \left(D_r+\mu \right)+\mu ^2+\omega ^2\right)}
\end{align}
This finite value reflects a persistent chiral drift transverse to the propulsion direction. 
The correlation is antisymmetric in $\omega$, indicating that the sign of the transverse alignment directly encodes the handedness of the torque. 
The magnitude decreases with increasing rotational noise $D_r$ or trap stiffness $\mu$, both of which suppress coherent precessional motion. 
In the limit $\omega\to 0$, the correlation vanishes, consistent with the disappearance of chirality-induced circulation. 
Thus, at steady state, confinement and rotational diffusion attenuate the helical bias, yet a residual transverse displacement persists whenever chirality breaks mirror symmetry.

\noindent\textit{Dimensionless form:}~
Expressing the perpendicular position--orientation correlation in terms of the scaled parameters 
\begin{align}
    \begin{aligned}
        & \langle \tbr\cdot\bu^{\perp}\rangle = e^{-(\beta +2) t}\cos{\Omega \ttt} \frac{\text{Pe} \Omega   \left((\beta -4) \beta -\left((\beta +2)^2+\Omega ^2\right) \text{u}_{z0}^2+\Omega ^2+12\right)}{\left((\beta -4)^2+\Omega ^2\right) \left((\beta +2)^2+\Omega ^2\right)}
        \\&+ e^{-(\beta +2) t}\sin{\Omega \ttt} ~\frac{\text{Pe}  \left(\beta  \left((\beta -4) \beta +\Omega ^2-4\right)-\left((\beta -4) \left((\beta +2)^2+\Omega ^2\right) \text{u}_{z0}^2\right)+16\right)}{\left((\beta -4)^2+\Omega ^2\right) \left((\beta +2)^2+\Omega ^2\right)}
        \\& \frac{\text{Pe} \Omega }{3}  \left(\frac{e^{-6 t} \left(3 \text{u}_{z0}^2-1\right)}{(\beta -4)^2+\Omega ^2}-\frac{2}{(\beta +2)^2+\Omega ^2}\right)
        \,.
    \end{aligned}
\end{align}

\noindent\textit{Steady state (dimensionless form):}~
In the long-time limit, the perpendicular position--orientation correlation saturates to
\begin{align}
 \langle \tbr\cdot\bu^{\perp}\rangle_{\rm st} = \frac{-2 \text{Pe} \Omega }{3 \left((\beta +2)^2+\Omega ^2\right)}\,. 
\end{align}
This steady value highlights the persistent transverse bias induced by chirality. 
The correlation is antisymmetric in $\Omega$, indicating that its sign directly reflects the handedness of rotation. 
Its magnitude decreases with increasing confinement strength $\beta$ or rotational noise (through the factor $(\beta+2)$), both of which diminish the coherence of the helical trajectory. 
In the achiral limit $\Omega \to 0$, the correlation vanishes, as expected for purely non-chiral active motion~\cite{Chaudhuri2021}. The weak trapping limit $\beta\to 0$ maximizes the steady transverse (helical) bias~\cite{Pattanayak2024}.
Thus, a finite transverse displacement persists at long times whenever a torque acts on an active particle in a trap, constituting a steady-state signature of chiral active dynamics.

\medskip

\noindent
\textbf{Mean-squared displacement (MSD):~} 
To calculate mean-squared displacement (MSD), we consider $\psi=\br^2$ in Eq.~\eqref{ME3}, which leads to 
\begin{align}
\langle \br^2\rangle_s=\frac{1}{s+2\mu}\left[6D\langle 1\rangle_s+2v_0\langle \textbf{r}\cdot\hat{\textbf{u}}\rangle_s\right]\,.
\label{eq:r2avg_Laplace_3d}
\end{align}
where $\langle 1\rangle_s=1/s$. The MSD depends on the position-orientation cross correlation $\langle \textbf{r}\cdot\hat{\textbf{u}}\rangle_s$ which we already calculated previously (Eq.~\eqref{eq:ru_cross_corr_Laplace_3d_final}).

Putting all these expressions back into the Eq.~\eqref{eq:r2avg_Laplace_3d} of $\langle \br^2\rangle_s$ and taking its inverse Laplace transform we obtain mean square displacement, $\langle \br^2(t)\rangle$ as a function of time,
\begin{align}
       &\langle \br^2(t)\rangle =  \frac{v_0^2 \omega ^2 e^{-6 D_r t} \left(3 \text{u}_{z0}^2-1\right)}{3(3 D_r-\mu ) (4 D_r-\mu ) \left(16 D_r^2-8 D_r \mu +\mu ^2+\omega ^2\right)}
       \nonumber
       \\&-\frac{2 v_0^2 e^{-t (2 D_r+\mu )} \left((2 D_r+\mu ) \text{u}_{z0}^2-2 D_r\right)}{(2 D_r-\mu ) (4 D_r-\mu ) (2 D_r+\mu )}
       \nonumber\\&+\frac{\frac{v_0^2 \left(12 D_r^2+12 D_r \mu +3 \mu ^2+\omega ^2\right)}{(2 D_r+\mu ) \left(4 D_r^2+4 D_r \mu +\mu ^2+\omega ^2\right)}+9 D}{3\mu }+ \frac{ e^{-2 \mu  t} \left(-\frac{v_0^2 \left(12 D_r^3-16 D_r^2 \mu +D_r \left(7 \mu ^2+\omega ^2\right)-\mu ^3-\mu  \omega ^2 \text{u}_{z0}^2\right)}{\left(6 D_r^2-5 D_r \mu +\mu ^2\right) \left(4 D_r^2-4 D_r \mu +\mu ^2+\omega ^2\right)}-3 D\right)}{\mu }
       \nonumber\\&+ \Bigg[\frac{ \left(-2 \omega ^2 \left(6 D_r^2-3 D_r \mu +\mu ^2\right)+\text{u}_{z0}^2 \left(8 D_r^2-6 D_r \mu +\mu ^2+\omega ^2\right) \left((2 D_r+\mu )^2+\omega ^2\right)\right)}{\left((\mu -4 D_r)^2+\omega ^2\right) \left((\mu -2 D_r)^2+\omega ^2\right) \left((2 D_r+\mu )^2+\omega ^2\right)}
       \nonumber\\&
       + \frac{\left((4 D_r-\mu ) (\mu -2 D_r)^2 (2 D_r+\mu )-\omega ^4\right)}{\left((\mu -4 D_r)^2+\omega ^2\right) \left((\mu -2 D_r)^2+\omega ^2\right) \left((2 D_r+\mu )^2+\omega ^2\right)}\Bigg]2 v_0^2 e^{-t (2 D_r+\mu )}\cos{\omega t}
       \nonumber\\&
       -\frac{4 D_r  \omega   \left(20 D_r^2-12 D_r \mu +\text{u}_{z0}^2 \left((2 D_r+\mu )^2+\omega ^2\right)+\mu ^2+\omega ^2\right)}{\left((\mu -4 D_r)^2+\omega ^2\right) \left((\mu -2 D_r)^2+\omega ^2\right) \left((2 D_r+\mu )^2+\omega ^2\right)}v_0^2e^{-t (2 D_r+\mu )}\sin \omega t \,.
       \label{msd_3d_eq}
\end{align}

\noindent
\textbf{Second-order displacement vector components moments:~}
Similar to the method described above, we use Eq.~\eqref{ME3} to obtain the coupled evolution equations for the second-order moments of the position vector components.

\noindent
\textit{Calculation of $\left\langle z ^2\right\rangle_s$}\\ 
\begin{align}
   & \left(s +2 \mu\right)\left\langle z ^2\right\rangle_s =2v_0 \langle z\text{u}_z\rangle_s+ 2 D/s\,,
   \\
   & \Big(s+2 D_r +\mu \Big) \langle z\text{u}_z\rangle_s= v_0 \langle \text{u}_z^2\rangle_s\,,
   \\
   & \Big(s+6 D_r \Big) \langle \text{u}_z^2\rangle_s= \text{u}_{z0}^2+ 2 D_r /s\,.
\end{align}
The final expression 
\begin{align}
\left\langle z ^2\right\rangle_s=\frac{2 v_0^2 \left(2 D_r+s\text{u}_{z0}^2\right)}{s(s+2\mu)(6 D_r+s) (\mu +2 D_r+s)}+\frac{2 D}{s (s+2\mu)}\,.
\label{eq_3d_zsq_Laplace}
\end{align}
Averaging over all possible initial orientations and taking inverse Laplace transform the time-dependent expression of $\langle z^2\rangle$,
\begin{align}
\langle z^2(t)\rangle
&= \frac{D}{\mu}\left(1 - e^{-2\mu t}\right)
+ \frac{v_0^2}{3\mu\left(4D_r^2 - \mu^2\right)}
\Big[\,2D_r\left(1 - e^{-2\mu t}\right)
+ \mu\left(2e^{-(2D_r+\mu)t} - 1 - e^{-2\mu t}\right)\Big]\,.
\end{align}
In dimensionless form:
\begin{align}
\langle \tilde z^2(\tilde t)\rangle
&= \frac{1}{\beta}\left(1 - e^{-2\beta \tilde t}\right)
+ \frac{\mathrm{Pe}^2}{3\beta\left(4-\beta^2\right)}
\Big[\,2\left(1 - e^{-2\beta \tilde t}\right)
+ \beta\left(2e^{-(2+\beta)\tilde t} - 1 - e^{-2\beta \tilde t}\right)\Big] \,,
\end{align}
The expression for $\langle \tilde{z}^2(\tilde{t})\rangle$ contains both passive and active contributions. 
The passive part represents thermal diffusion within the harmonic trap and corresponds to the equilibrium Brownian variance scaled by $1/\beta$. 
The active part, proportional to ${\rm Pe}^2$, arises from self-propelled motion and its coupling to rotational relaxation and confinement. 
Increasing $\beta$ suppresses the amplitude of active fluctuations by limiting the particle’s persistence length, while larger ${\rm Pe}$ enhances the contribution from active propulsion. 
In the long-time limit, the dimensionless mean-squared displacement along the torque axis saturates to
\begin{align}
\langle \tilde{z}^2\rangle_{\rm st} = \frac{1}{\beta}+\frac{{\rm Pe}^2}{3\beta(\beta+2)}\,,
\end{align}
The first term, $1/\beta$, represents the passive thermal contribution corresponding to the equilibrium variance of a Brownian particle confined in a harmonic trap. 
The second term, proportional to ${\rm Pe}^2$, arises from active propulsion and reflects the persistent motion of the particle along the torque axis before being randomized by rotational diffusion. 
This active contribution decreases with increasing $\beta$, as stronger confinement limits the persistence of directed motion. 
In the limit ${\rm Pe}\to 0$, the result correctly reduces to the equilibrium Brownian value, while for weak confinement ($\beta\to0$) it diverges, recovering the unconfined active motion regime. 
Hence, $\langle \tilde{z}^2\rangle_{\rm st}$ quantifies how activity, rotational relaxation, and confinement jointly determine the steady-state amplitude of fluctuations along the torque axis.

\noindent
\textit{Calculation of $\left\langle x ^2\right\rangle_s$}
\begin{align}
   & (s+2\mu) \left\langle x ^2\right\rangle_s =2v_0 \langle x \text{u}_x\rangle_s+ 2 D/s\,,\\
   & \Big(s+2 D_r +\mu\Big) \langle x\text{u}_x\rangle_s= v_0 \langle \text{u}_x^2\rangle_s+\omega \langle \mathcal{R}_z x \text{u}_x \rangle_s=v_0 \langle \text{u}_x^2\rangle_s-\omega \langle  x \text{u}_y \rangle_s\,.
   \end{align}
This leads to
   \begin{equation}
     \Big(s+2 D_r +\mu\Big) \langle x\text{u}_x\rangle_s +\omega \langle  x \text{u}_y \rangle_s= v_0 \langle \text{u}_x^2\rangle_s\,.
   \label{xux}  
   \end{equation}
Similarly $\psi= x \text{u}_y$ gives,
\begin{equation}
    \Big(s+2 D_r+\mu \Big) \langle x\text{u}_y\rangle_s -\omega \langle  x \text{u}_x \rangle_s= v_0 \langle \text{u}_x \text{u}_y\rangle_s\,.
\label{xuy}
\end{equation}
Solving equations~\eqref{xux} and \eqref{xuy} simultaneously,
\begin{align}
    \langle  x \text{u}_x \rangle_s\Big[\Big(s+2 D_r+\mu \Big)^2+\omega ^2\Big]&= v_0 \Big[\Big(s+2 D_r+\mu \Big)\langle \text{u}_x^2\rangle_s- \omega \langle \text{u}_x \text{u}_y\rangle_s\Big]\,,
    \\
    \langle  x \text{u}_y \rangle_s\Big[\Big(s+2D_r+\mu\Big)^2+\omega ^2\Big]&= v_0 \Big[\Big(s+2 D_r+\mu \Big)\langle \text{u}_x \text{u}_y\rangle_s+ \omega \langle \text{u}_x^2\rangle_s\Big]\,.
\end{align}
Now, for $\langle \text{u}_x^2\rangle_s$ and $\langle \text{u}_x \text{u}_y\rangle_s$,
\begin{align}
\nonumber
&\Big(s+6 D_r \Big) \langle \text{u}_x^2\rangle_s= \text{u}_{x0}^2+ 2 D_r /s- 2\omega \langle \text{u}_x \text{u}_y\rangle_s\,,
\\
\Rightarrow & \Big(s+6 D_r \Big) \langle \text{u}_x^2\rangle_s+2\omega \langle \text{u}_x \text{u}_y\rangle_s=\text{u}_{x0}^2+ 2 D_r /s\,. 
\\
\nonumber & \textrm{Considering $\psi= \text{u}_x \text{u}_y$},
\\ \nonumber
   & \Big(s+6 D_r \Big) \langle \text{u}_x \text{u}_y\rangle_s= \text{u}_{x0}\text{u}_{y0}+\omega \langle \text{u}_x^2\rangle_s -\omega \langle \text{u}_y^2\rangle_s=\text{u}_{x0}\text{u}_{y0}+2 \omega \langle \text{u}_x^2\rangle_s-\omega/s +\omega \langle \text{u}_z^2\rangle_s\,,
   \\
   \Rightarrow &  \Big(s+6 D_r \Big) \langle \text{u}_x \text{u}_y\rangle_s- 2 \omega \langle \text{u}_x^2\rangle_s=\text{u}_{x0}\text{u}_{y0}-\omega/s +\omega \langle \text{u}_z^2\rangle_s\,.
   \\ \nonumber
   &\textrm{The expressions of $\langle \text{u}_x^2\rangle_s$ and $\langle \text{u}_x \text{u}_y\rangle_s$,}
   \\
   & \Big[\Big(s+6 D_r \Big)^2+4 \omega ^2\Big] \langle \text{u}_x^2\rangle_s = \Big(s+6 D_r \Big)\Big(\text{u}_{x0}^2+ 2 D_r /s\Big)-2 \omega\Big(\text{u}_{x0}\text{u}_{y0}-\omega/s +\omega \langle \text{u}_z^2\rangle_s\Big)\,,
   \\&
   \Big[\Big(s+6 D_r \Big)^2+4 \omega ^2\Big] \langle \text{u}_x \text{u}_y\rangle_s = \Big(s+6 D_r \Big)\Big(\text{u}_{x0}\text{u}_{y0}-\omega/s +\omega \langle \text{u}_z^2\rangle_s\Big)+2 \omega\Big(\text{u}_{x0}^2+ 2 D_r /s\Big)\,.
\end{align}
Here, we used the normalization condition $\text{u}_x^2+\text{u}_y^2+\text{u}_z^2=1 $, and in Laplace space, it follows that  $\langle \text{u}_y^2\rangle_s =1/s-\langle \text{u}_x^2\rangle_s-\langle \text{u}_z^2\rangle_s$. The final expression of $\langle x^2\rangle_s$,
\begin{align}
    \langle x^2\rangle_s &= \frac{2 v_0^2 \left((\mu +2 D_r+s) \left((6 D_r+s)^2 \left(2 D_r+s \text{u}_{x0}^2\right)-2 s \omega  \text{u}_{x0} \text{u}_{y0} (6 D_r+s)+2 \omega ^2 \left(4 D_r-s \text{u}_{z0}^2+s\right)\right)\right)}{(s (6 D_r+s)) \left(\left((6 D_r+s)^2+4 \omega ^2\right) \left((\mu +2 D_r+s)^2+\omega^2\right)\right)}
    \\&+\frac{2 v_0^2 \left(-\omega  \left(\text{u}_{x0} \text{u}_{y0} (6 D_r+s)+\omega  \left(2 \text{u}_{x0}^2+\text{u}_{z0}^2-1\right)\right)\right)}{\left((6 D_r+s)^2+4 \omega ^2\right) \left((\mu +2 D_r+s)^2+\omega ^2\right)}+\frac{2 D}{s (2 \mu +s)}\,.
    \label{eq_3d_xsq_Laplace}
\end{align}
 Averaging over initial orientations and performing inverse Laplace transform,
\begin{align}
    \langle x^2\rangle(t) &= \frac{D \left(1- e^{-2 \mu  t}\right)}{ \mu } + \frac{v_0^2 \left(\frac{2 D_r+\mu }{(2 D_r+\mu )^2+\omega ^2}+\frac{(\mu -2 D_r) e^{-2 \mu  t}}{(\mu -2 D_r)^2+\omega ^2}\right)}{3 \mu }\nonumber\\ & +\frac{2 v_0^2 e^{-t (2 D_r+\mu )} \left(\left(4 D_r^2-\mu ^2-\omega ^2\right) \cos ( \omega t)-4 D_r \omega  \sin (\omega t)\right)}{3 \left((\mu -2 D_r)^2+\omega ^2\right) \left((2 D_r+\mu )^2+\omega ^2\right)}\,.
\end{align}
In dimensionless form:
\begin{align}
\langle \tilde{x}^{2}\rangle(\tilde{t})
&= \frac{1}{\beta}\left(1-e^{-2\beta \tilde{t}}\right) + \frac{\mathrm{Pe}^{2}}{3\beta}
\left[
\frac{\beta+2}{(\beta+2)^{2}+\Omega^{2}}
+\frac{(\beta-2)e^{-2\beta \tilde{t}}}{(\beta-2)^{2}+\Omega^{2}}
\right] \nonumber\\
&\quad
+\frac{2\,\mathrm{Pe}^{2}}{3}\,
\frac{e^{-(\beta+2)\tilde{t}}
\left[
(4-\beta^{2}-\Omega^{2})\cos(\Omega \tilde{t})
-4\Omega \sin(\Omega \tilde{t})
\right]}
{\left[(\beta-2)^{2}+\Omega^{2}\right]
\left[(\beta+2)^{2}+\Omega^{2}\right]} \,.
\end{align}

At long times ($\tilde t \to \infty$), the exponentials vanish and the steady state is
\begin{align}
\langle \tilde x^2\rangle_{\mathrm{st}}
= \frac{1}{\beta} + \frac{\mathrm{Pe}^2(\beta+2)}{3\beta[(\beta+2)^2+\Omega^2]}\,,
\end{align}
shows that confinement($\beta$) limits both thermal and active fluctuations, while torque($\Omega$) suppresses the active contribution by averaging out directed motion. For small torque or weak trapping, activity enhances the steady-state spread, whereas strong torque or confinement localizes the particle near the trap center.

\noindent
\textit{Calculation of $\left\langle y ^2\right\rangle_s$}\\ 
Using $r^2=x^2+y^2+z^2$, we can write $\left\langle y ^2\right\rangle_s=\left\langle r ^2\right\rangle_s-\left\langle x ^2\right\rangle_s-\left\langle z ^2\right\rangle_s$, gives
\begin{align}
\langle y^2\rangle_s &=  \frac{2 \omega ^2 v_0^2 \left(16 D_r^2+8 \mu  D_r+2 D_r s \left(6 \text{u}_{x0}^2+\text{u}_{z0}^2+3\right)+s \left(2 \mu +2 s \text{u}_{x0}^2-s \text{u}_{z0}^2+s-2 \mu  \text{u}_{z0}^2\right)\right)}{s (6 D_r+s) (2 \mu +s) \left((6 D_r+s)^2+4 \omega ^2\right) \left((\mu +2 D_r+s)^2+\omega ^2\right)}\nonumber\\
&+\frac{2 \omega  \text{u}_{x0} \text{u}_{y0} v_0^2 (2 \beta +10 D_r+3 s)}{(2 \beta +s) \left((6 D_r+s)^2+4 \omega ^2\right) \left((\mu +2 D_r+s)^2+\omega ^2\right)}\nonumber\\
&+\frac{2 v_0^2 (6 D_r+s) (\mu +2 D_r+s) \left(2 D_r+s \text{u}_{y0}^2\right)}{s (2 \mu +s) \left((6 D_r+s)^2+4 \omega ^2\right) \left((\mu +2 D_r+s)^2+\omega ^2\right)}+\frac{2 D}{s (2 \mu +s)}\,.
\label{eq_3d_ysq_Laplace}
\end{align}
Taking inverse Laplace transform and averaging over initial orientations,
\begin{align}
        \langle y^2\rangle(t) &= \frac{v_0^2 \left(\frac{2 D_r+\mu }{(2 D_r+\mu )^2+\omega ^2}+\frac{(\mu -2 D_r) e^{-2 \mu  t}}{(\mu -2 D_r)^2+\omega ^2}\right)}{3 \mu }+\frac{D \left(3-3 e^{-2 \mu  t}\right)}{3 \mu }
    \nonumber\\ & +\frac{2 v_0^2 e^{-t (2 D_r+\mu )} \left(\left(4 D_r^2-\mu ^2-\omega ^2\right) \cos ( \omega t)-4 D_r \omega  \sin (\omega t)\right)}{3 \left((\mu -2 D_r)^2+\omega ^2\right) \left((2 D_r+\mu )^2+\omega ^2\right)}\,.
\end{align}
In dimensionless form:
\begin{align}
\langle \tilde y^{2}\rangle(\tilde t)
&= \frac{1}{\beta}\left(1-e^{-2\beta \tilde t}\right) + \frac{\mathrm{Pe}^{2}}{3\beta}\left[
\frac{\beta+2}{(\beta+2)^{2}+\Omega^{2}}
+\frac{(\beta-2)\,e^{-2\beta \tilde t}}{(\beta-2)^{2}+\Omega^{2}}
\right] \nonumber\\
&\quad + \frac{2\,\mathrm{Pe}^{2}}{3}\,
\frac{e^{-(\beta+2)\tilde t}\left[\left(4-\beta^{2}-\Omega^{2}\right)\cos(\Omega \tilde t)
-4\Omega \sin(\Omega \tilde t)\right]}
{\left[(\beta-2)^{2}+\Omega^{2}\right]\left[(\beta+2)^{2}+\Omega^{2}\right]}\,.
\end{align}
As torque acts along $z$-axis, the dynamics in the $x-y$ plane remain radially symmetric, hence $\langle \tilde y^{2}\rangle(\tilde t)=\langle \tilde x^{2}\rangle(\tilde t)$. At long times ($\tilde t \to \infty$), the exponential terms vanish and the steady state is
\begin{align}
\langle \tilde y^{2}\rangle_{\mathrm{st}}
= \frac{1}{\beta} + \frac{\mathrm{Pe}^{2}(\beta+2)}{3\beta\big[(\beta+2)^{2}+\Omega^{2}\big]}\,.
\end{align}

\noindent
\textit{Calculation of $\left\langle xy\right\rangle_s$} 
\begin{align}
    &(s+2 \mu) \langle xy\rangle_s = v_0 \Big[\langle x \text{u}_y\rangle_s +\langle y\text{u}_x\rangle_s\Big]\,.
\end{align}
The expression of $\langle x \text{u}_y\rangle_s$ is already found in the previous section. $y \text{u}_x$ satisfies the following pair of equations,
\begin{align}
    &\Big(s+\mu+2 D_r \Big) \langle y\text{u}_x\rangle_s +\omega \langle  y \text{u}_y \rangle_s= v_0 \langle \text{u}_x \text{u}_y\rangle_s\,,
    \\
    &
    \Big(s+\mu+2 D_r\Big) \langle y\text{u}_y\rangle_s -\omega \langle  y\text{u}_x \rangle_s= v_0 \langle  \text{u}_y^2\rangle_s\,, 
    \\
    \nonumber
    & \textrm{ $\langle y\text{u}_x\rangle_s$ satisfies,}
    \\
    &\langle  y \text{u}_x \rangle_s\Big[\Big(s+2 D_r +\mu\Big)^2+\omega ^2\Big]= v_0 \Big[\Big(s+2D_r+\mu\Big)\langle \text{u}_x \text{u}_y\rangle_s- \omega \langle \text{u}_y ^2\rangle_s\Big]\,.
\end{align}

The final expression of  $\langle xy\rangle_s$,
\begin{align}
    \langle xy\rangle_s= \frac{v_0^2 \left(\omega  \left(2 \text{u}_{x0}^2+\text{u}_{z0}^2-1\right) (2 \mu +10 D_r+3 s)+2 \text{u}_{x0} \text{u}_{y0} (6 D_r+s) (\mu +2 D_r+s)-4 \omega^2 \text{u}_{x0} \text{u}_{y0}\right)}{(2 \mu +s) \left((6 D_r+s)^2+4 \omega ^2\right) \left((\mu +2 D_r+s)^2+\omega^2\right)}\,.
\end{align}

Averaging over all possible initial conditions, we get $\langle xy\rangle_s=0$.
Since the applied torque acts along the $z$-axis, the dynamics of the particle remain 
rotationally symmetric in the $x$--$y$ plane. As a result, cross correlations between 
orthogonal directions, such as $\langle xy\rangle$, vanish after averaging over all possible 
initial orientations. Although transient correlations $\langle xy\rangle(t)$ may arise for specific 
initial conditions—reflecting the instantaneous phase difference between $x$ and $y$ oscillations 
induced by torque—the isotropy of the $x$--$y$ plane ensures that these correlations cancel out 
in the ensemble average. Therefore, $\langle xy\rangle_{\mathrm{st}}=0$ signifies that the steady state 
preserves rotational symmetry about the torque axis, with no preferred azimuthal direction.

\noindent
\textit{Calculation of $\left\langle xz\right\rangle_s$} 
\begin{align*}
    &(s+2\mu) \langle xz\rangle_s = v_0 \Big(\langle x \text{u}_z\rangle_s +\langle z\text{u}_x\rangle_s\Big)\,,
    \\
    &
    \Big(s+2 D_r+\mu\Big) \langle z\text{u}_x\rangle_s +\omega \langle  z \text{u}_y \rangle_s= v_0 \langle \text{u}_x \text{u}_z\rangle_s\,,
    \\
    &
    \Big(s+ 2 D_r+\mu \Big) \langle z\text{u}_y\rangle_s -\omega \langle  z\text{u}_x \rangle_s= v_0 \langle  \text{u}_y \text{u}_z\rangle_s\,. 
    \end{align*} 
Solving the above equations for $\langle  z \text{u}_x \rangle_s$ and $\langle  z \text{u}_y \rangle_s$,
    \begin{align*}
    &\langle  z \text{u}_x \rangle_s\Big[\Big(s+ 2 D_r +\mu\Big)^2+\omega ^2\Big]= v_0 \Big[\Big(s+ 2 D_r+\mu \Big)\langle \text{u}_x \text{u}_z\rangle_s- \omega \langle \text{u}_z \text{u}_y \rangle_s\Big]\,,
     \\
    &\langle  z \text{u}_y \rangle_s\Big[\Big(s+2 D_r+\mu \Big)^2+\omega ^2\Big]= v_0 \Big[\Big(s+2 D_r+\mu\Big)\langle \text{u}_y \text{u}_z\rangle_s+ \omega \langle \text{u}_z \text{u}_x \rangle_s\Big]\,,
    \\
    &\Big(s+6 D_r \Big) \langle \text{u}_x \text{u}_z\rangle_s+  \omega \langle \text{u}_z \text{u}_y\rangle_s=\text{u}_{x0}\text{u}_{z0}\,,
    \\
    &\Big(s+6 D_r \Big) \langle \text{u}_y \text{u}_z\rangle_s-  \omega \langle \text{u}_z \text{u}_y\rangle_s=\text{u}_{y0}\text{u}_{z0}\,.
\end{align*}
If we average over all possible initial orientations, $\langle \text{u}_{x0}\text{u}_{z0}\rangle=0=\langle \text{u}_{y0}\text{u}_{z0}\rangle$, which gives $\langle \text{u}_x \text{u}_z\rangle_s=0=\langle \text{u}_y \text{u}_z\rangle_s$ resulting in $\langle  z \text{u}_x \rangle_s=0=\langle  z \text{u}_y \rangle_s$.

\noindent
$\langle x \text{u}_z\rangle_s$ satisfies,
\begin{align}
    \Big(s+2 D_r+\mu \Big) \langle x \text{u}_z\rangle_s = v_0 \langle \text{u}_x \text{u}_z\rangle_s=0\,.
\end{align}
Hence, $\langle xz\rangle_s=0$ if averaged over initial orientations.
Since the torque acts along the $z$-axis, the particle’s motion in the $x$--$y$ plane 
is decoupled from that along $z$. Consequently, the cross correlations between in-plane 
($x$ or $y$) and out-of-plane ($z$) components vanish after averaging over all possible 
initial orientations. While individual trajectories may show finite $\langle xz\rangle(t)$ 
due to specific initial alignment of the propulsion direction, ensemble averaging restores 
symmetry about the torque axis, leading to $\langle xz\rangle_{\mathrm{st}}=0$. 
This reflects the absence of any preferred tilt or coupling between transverse and axial 
fluctuations in the steady state.

\noindent
\textit{Calculation of $\left\langle yz\right\rangle_s$}\\ 
The cross-correlation between the $y$ and $z$ components satisfies
\begin{align}
    &(s+2\mu) \langle yz\rangle_s = v_0 \Big(\langle y \text{u}_z\rangle_s +\langle z \text{u}_y\rangle_s\Big)\,.
\end{align}
Here, $\langle y u_z\rangle_s$ obeys
\begin{align}
    \Big(s+2 D_r +\mu\Big) \langle y \text{u}_z\rangle_s = v_0 \langle \text{u}_y \text{u}_z\rangle_s=0\,.
\end{align}
The corresponding expression for $\langle z u_y\rangle_s$ has been obtained in the previous section. 
Averaging over all possible initial orientations gives $\langle yz\rangle_s = 0$.
The vanishing of $\langle yz\rangle_s$ has the same physical interpretation as that of $\langle xz\rangle_s$.

The mean-square displacement (MSD) of a three-dimensional chiral active Brownian particle (ABP) naturally decomposes into two independent contributions: one parallel to the torque axis, $\langle z^2(t) \rangle$, and the other in the plane perpendicular to the torque, $\langle \mathbf{r}_\perp^2(t) \rangle = \langle x^2(t) \rangle + \langle y^2(t) \rangle$.
These components are obtained by performing the inverse Laplace transform of the corresponding expressions given in Eq.~\eqref{eq_3d_zsq_Laplace}, Eq.~\eqref{eq_3d_xsq_Laplace}, and Eq.~\eqref{eq_3d_ysq_Laplace}.
The resulting exact expressions for the parallel and perpendicular MSD components, expressed in dimensionless form, are presented in the main text (see Eq.~\eqref{eq:msd_3d_rperp} and Eq.~\eqref{eq:msd_3d_z}).

\begin{figure}[t]
\centering
\includegraphics[width=\linewidth]{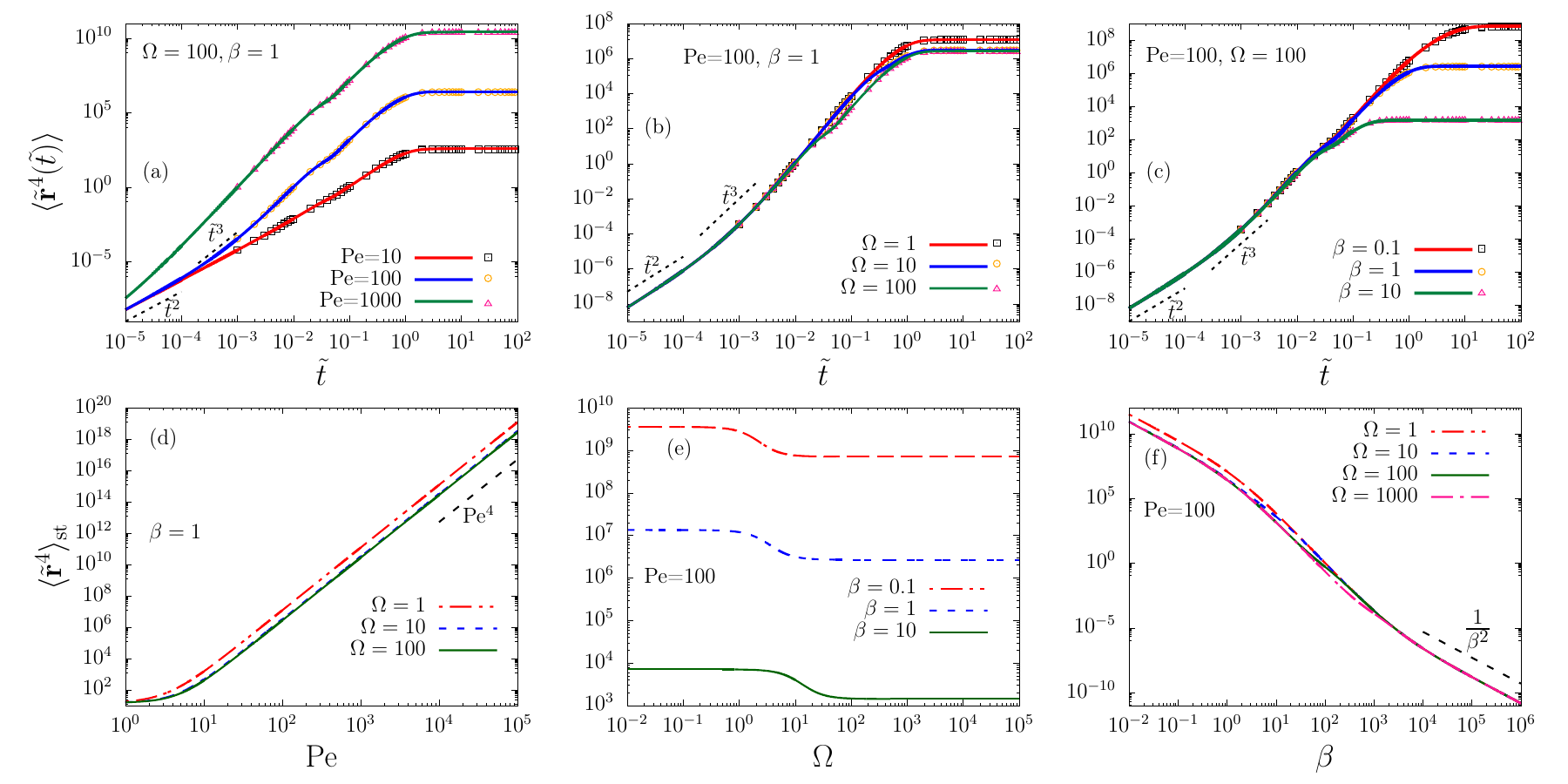}
\caption{
Fourth order moment of displacement of an active Brownian particle under torque in a harmonic trap in three dimensions (3d) as a function of time for (a) $\text{Pe}=10, 100, 1000$ with $\Omega=100$ and $\beta=1$, (b) torque $\Omega=1,10,100$ with activity $\text{Pe}=100$ and harmonic trap stiffness $\beta=1$, and (c) $\beta=0.1,1,10$ with $\Pe=100$ and $\Omega=100$.
Steady-state fourth-order moment of displacement as a function of (d) $\Pe$ for $\Omega=1,10,100$ with $\beta=1$, (e) $\Omega$ for $\beta=0.1,1,10$ with $\Pe=100$, and (f) $\beta$ for $\Omega=1,10,100,1000$ with $\Pe=100$.
The lines in (a), (b) and (c) are plots of Eq.~(\ref{eq:r4_3d_dimensionless}) and the points are from simulations. The lines in (d), (e), and (f) are plots of Eq.~(\ref{eq:r4_3d_st_dimensionless}).
}
\label{app_fig3}
\end{figure}
\subsection{Fourth order moment of displacement}
To compute the fourth-order displacement moment, we substitute $\psi = \mathbf{r}^4$ into the 
moment-generating Eq.~\eqref{ME3}. This leads to the following relation:
\begin{align}
(s+4 \mu)\left\langle \br^{4}\right\rangle_s=20 D \left\langle \br^2\right\rangle_s+4 v_0\left\langle \br^2 \br\cdot \bu\right\rangle_{s}\,,
\label{r4moment3d}
\end{align}
where $\left\langle \br^2\right\rangle_s$ is already calculated in previous section. The second term can be evaluated following the steps outlined below. During this calculation, we encounter the operator $R_z \psi$ which can be determined using the same procedure employed in the previous section for obtaining $\langle \textbf{r}\cdot\hat{\textbf{u}}\rangle_s$.

\noindent
{\textit{Calculation of} $\left\langle \br^2 \br\cdot \bu\right\rangle_{s}$}

To determine the mixed moment $\langle \mathbf{r}^2 \mathbf{r} \cdot \bu\rangle_s$, 
we substitute $\psi = \mathbf{r}^2\, \mathbf{r} \cdot \bu$ into the moment-generating 
Eq.~\eqref{ME3}. The resulting equation reads:
\begin{align}
&\left[s+2 D_r+3\mu+\frac{\omega^{2}}{s+2 D_r+3\mu}\right]\langle \br^2 \br\cdot \bu\rangle_s =10 D \left[\langle\br \cdot \bu\rangle_{s}+\frac{\omega\langle R_{z}(\br\cdot \bu)\rangle_{s}}{s+2 D_r+3\mu}\right]
\nonumber\\&+ v_0\left[2 \left\langle(\br \cdot \bu)^{2}\right\rangle_{s} 
+\frac{\omega\langle R_{z}(\br \cdot \bu)^{2}\rangle_{s}}{s+2 D_r+3 \mu}+\langle \br^2\rangle_{s}\right]+\frac{\omega^{2}\langle \br^2 z \text{u}_{z}\rangle_{s}}{s+2 D_r+3\mu}\,.
\end{align}
For the mixed moment $\left\langle \mathbf{r}^2 z u_z \right\rangle_s$, 
we substitute $\psi = \mathbf{r}^2 z u_z$ into the moment-generating Eq.~\eqref{ME3}. The resulting relation reads:
\begin{align}
&\left[s+2 D_r+3\mu\right]\left\langle \br^2z\text{u}_{z}\right\rangle_{s}=  10  D\left\langle z \text{u}_{z}\right\rangle_{s}+2 v_0\left\langle\br \cdot \bu z\text{u}_{z}\right\rangle_{s}+v_0\left\langle \br^2 \text{u}_{z}^{2}\right\rangle_{s}\,.
\end{align}
For the correlation term $\left\langle \mathbf{r} \cdot \bu z u_z \right\rangle_s$, 
we set $\psi = \mathbf{r} \cdot \bu z u_z$ in the moment-generating Eq.~\eqref{ME3}. The resulting equation is
\begin{align}
&\left[s+6D_r +2\mu+\frac{\omega^{2}}{s+6 D_r+2\mu}\right]\left\langle\br \cdot \bu z \text{u}_{z}\right\rangle_{s}=2 D\left\langle \text{u}_{z}^{2}\right\rangle_{s}+2 D_r\left\langle z^{2}\right\rangle_{s}+v_0\left\langle z \text{u}_{z}\right\rangle_{s}\nonumber\\
&+\frac{\omega^{2}\left\langle z^{2} \text{u}_{z}^{2}\right\rangle_{s}}{s+6 D_r+2\mu}+v_0\left[\left\langle\br \cdot \bu \text{u}_{z}^{2}\right\rangle_{s}+\frac{\omega\left\langle R_{z}\left(\br \cdot \bu \text{u}_{z}^{2})\right\rangle_{s}\right.}{s+6 D_r+2\mu}\right]\,,
\end{align}
and
\begin{align}
&\left(s+6 D_r+2\mu\right)\left\langle R_{z}\left(\br\cdot \bu z \text{u}_{z}\right)\right\rangle_{s}=v_0\left\langle R_{z}\left(\br\cdot \bu \text{u}_{z}^{2}\right)\right\rangle_{s}  -\omega\left\langle\br\cdot \bu z \text{u}_{z}\right\rangle_{s}+\omega\left\langle z^{2} \text{u}_{z}^{2}\right\rangle_{s}\,.
\end{align}
For the correlation term $\left\langle\br \cdot \bu \text{u}_{z}^{2}\right\rangle_{s}$, 
we set $\psi = \br \cdot \bu \text{u}_{z}^{2}$ in the moment-generating Eq.~\eqref{ME3}. The resulting equation is
\begin{align}
&\left[s+12 D_r+\mu+\frac{\omega^{2}}{s+12 D_r+\mu}\right] \left\langle\br\cdot \bu \text{u}_{z}^{2}\right\rangle_{s}=  2 D_r\langle\br\cdot \bu\rangle_{s}+4 D_r\left\langle z \text{u}_{z}\right\rangle_{s}+v_0\left\langle \text{u}_{z}^{2}\right\rangle_{s} \nonumber\\
&+\frac{\omega^{2}\left\langle z \text{u}_{z}^{3}\right\rangle_{s}}{s+12D_r+\mu}+\frac{2 \omega D_r\left\langle R_{z}(\br\cdot \bu)\right\rangle_{s} }{s+12 D_r+\mu}\,,
\end{align}
and
\begin{align}
{\left[s+12 D_r+\mu\right]\left\langle R_{z}\left(\br\cdot \bu \text{u}_{z}^{2}\right)\right\rangle_{s} = 2 D_r\left\langle R_{z}(\br\cdot \bu)\right\rangle_{s} } &-\omega\left\langle(\br\cdot \bu) \text{u}_{z}^{2}\right\rangle_{S}
+\omega\left\langle  z \text{u}_{z}^{3}\right\rangle_{s}\,.
\end{align}
Now,
\begin{align}
\left[s+12D_r+\mu\right]\left\langle z \text{u}_{z}^{3}\right\rangle_{s}=6 D_r\left\langle z \text{u}_{z}\right\rangle_{s}+v_0\left\langle \text{u}_{z}^{4}\right\rangle_{s}\,,
\end{align}
with
$$
\nonumber
\left[s+20 D_r\right]\left\langle \text{u}_{z}^{4}\right\rangle_{s}=12 D_r\left\langle \text{u}_{z}^{2}\right\rangle_{s}+\text{u}_{z0}^4\,,
$$
\begin{align}
\left(s+6 D_r+2 \mu\right)\left\langle z^{2} \text{u}_{z}^{2}\right\rangle_{s}=2 D\left\langle \text{u}_{z}^{2}\right\rangle_{s}+2 D_r \left\langle z^{2}\right\rangle_{s}+2 v_0\left\langle z \text{u}_{z}^{3}\right\rangle_{s}\,,
\end{align}
\begin{align}
\left(s+6 D_r+2 \mu\right)\left\langle \br^2 \text{u}_{z}^{2}\right\rangle_{s}=2 D_r\left\langle \br^2\right\rangle_{s}+6 D \left\langle \text{u}_{z}^{2}\right\rangle_{s}+2 v_0\left\langle\br \cdot \bu \text{u}_{z}^{2}\right\rangle_{s}\,.
\end{align}

\noindent
{\textit{Calculation of} $\left\langle(\br \cdot \bu)^{2}\right\rangle_{s}$}

To evaluate the orientationally coupled displacement moment 
$\langle (\mathbf{r} \cdot \bu)^{2} \rangle_s$, 
we substitute $\psi = (\mathbf{r} \cdot \bu)^{2}$ into the moment-generating Eq.~\eqref{ME3}. 
The corresponding relation reads:
\begin{align}
&\left[s+6 D_r+ 2\mu+\frac{4 \omega^{2}}{s+6 D_r +2\mu}\right]\left\langle(\br\cdot \bu)^{2}\right\rangle_{s}= \frac{2 D}{s}+2 D_r\left\langle \br^2\right\rangle_{s}+2 v_0\langle\br\cdot \bu\rangle_{s} \nonumber\\
&+\frac{2 \omega v_0\left\langle R_{z}(\br\cdot \bu)\right\rangle_{s}}{s+6 D_r+2\mu} +6 \omega^{2} \frac{\left\langle\br \cdot \bu z \text{u}_{z}\right\rangle_{s}}{s+6 D_r +2\mu}+2 \omega^{2} \frac{\left\langle \br^2\right\rangle_{s}}{s+6 D_r+2\mu}- \frac{2 \omega^{2}\left\langle z^{2}\right\rangle_{s}}{s+6 D_r +2\mu} \nonumber\\
&-\frac{2 \omega^{2}\left\langle \br^2 \text{u}_{z}^{2}\right\rangle_{s}}{s+6 D_r +2\mu}\,,
\end{align}
again,
\begin{align}
&{\left(s+6 D_r +2\mu\right) \left\langle R_{z}(\br\cdot \bu)^{2}\right\rangle_{s} } =2 v_0\left\langle R_{z}(\br\cdot \bu)\right\rangle_{s}-4 \omega\left\langle(\br\cdot \bu)^{2}\right\rangle_{s} +6 \omega\left\langle\br\cdot \bu z \text{u}_{z}\right\rangle _{s}\nonumber\\ 
&+2 \omega\left\langle \br^2\right\rangle_{s}-2 \omega\left\langle z^{2}\right\rangle_{s} -2 \omega\left\langle \br^2 \text{u}_{z}^{2}\right\rangle_{s}\,. 
\end{align}
Combining all the above expressions gives the complete time-dependent form of the fourth order moment of displacement. Its dimensionless representation, averaged over all possible initial orientations, is expressed as
{\allowdisplaybreaks
\begin{align}
        &\langle \tbr^4 (\ttt)\rangle = \frac{8 \text{Pe}^4 \left(4 \left(\beta ^2+16\right) \Omega ^2+3 \left(\beta ^2-16\right)^2+3 \Omega ^4\right) e^{-2 (\beta +3) \ttt}}{45 \left(\beta ^4-25 \beta ^2+144\right) \left(\beta ^4+2 \beta ^2 \left(\Omega ^2-16\right)+\left(\Omega ^2+16\right)^2\right)}
        \nonumber\\
       & +\frac{\text{Pe}^4 \Omega ^4 \left(759 \beta ^4+\beta ^2 \left(147 \Omega ^2+27802\right)+9 \Omega ^4+1367 \Omega ^2+42644\right)}{15 \beta  (\beta +2) (\beta +3) (3 \beta +2) \left((\beta +2)^2+\Omega ^2\right) \left((\beta +3)^2+\Omega ^2\right) \left(4 (\beta +3)^2+\Omega ^2\right) \left((3 \beta +2)^2+\Omega ^2\right)}
       \nonumber\\
       &+ \frac{2 \Omega ^4 \left(-45 \left(\beta ^2-4\right)+\text{Pe}^4+20 \text{Pe}^2\right) e^{-2 \beta  \ttt}}{3 (\beta -2) \beta ^2 (\beta +2) \left((\beta -2)^2+\Omega ^2\right) \left((\beta +2)^2+\Omega ^2\right)}
       \nonumber\\&
       +\frac{e^{-2 \beta  \ttt} \left(30 \left(\beta ^2-4\right)^2 \left(\left(\text{Pe}^2+6\right)^2-9 \beta ^2\right)+20 \Omega ^2 \left(-27 \beta ^4+\left(\beta ^2+4\right) \text{Pe}^4+96 \text{Pe}^2+432\right)\right)}{9 (\beta -2) \beta ^2 (\beta +2) \left((\beta -2)^2+\Omega ^2\right) \left((\beta +2)^2+\Omega ^2\right)}
       \nonumber\\&
       +\frac{\text{Pe}^4 \Omega ^2 \left(\left(151 \beta ^2+175\right) \Omega ^4+\beta ^2 \left(1489 \beta ^2+9874\right) \Omega ^2+\beta ^3 \left(27919 \beta ^2+165736\right)\right)}{3 \beta ^2 (\beta +2) (\beta +3) (3 \beta +2) \left((\beta +2)^2+\Omega ^2\right) \left((\beta +3)^2+\Omega ^2\right) \left(4 (\beta +3)^2+\Omega ^2\right) \left((3 \beta +2)^2+\Omega ^2\right)}
       \nonumber\\&
       +\frac{10 \text{Pe}^2 \Omega ^2 \left(209993 \beta ^4+\left(349 \beta ^2+995\right) \Omega ^4+7 \beta ^2 \left(431 \beta ^2+4372\right) \Omega ^2+11 \Omega ^6\right)}{3 \beta  (\beta +2) (\beta +3) (3 \beta +2) \left((\beta +2)^2+\Omega ^2\right) \left((\beta +3)^2+\Omega ^2\right) \left(4 (\beta +3)^2+\Omega ^2\right) \left((3 \beta +2)^2+\Omega ^2\right)}
       \nonumber\\&
       + \frac{16 (\beta -4) \beta  \text{Pe}^2 \left(5 (\beta -4) (3 \beta +2)+(4-3 \beta ) \text{Pe}^2\right) e^{-(3 \beta +2) \ttt}}{3 (3 \beta +2) \left(\beta ^2-4\right) \left((\beta -4)^2+\Omega ^2\right) \left(4 \beta ^2+\Omega ^2\right)}
       \nonumber\\&
       +\frac{4 \text{Pe}^2 \Omega ^2 \left(25 (\beta -4) (3 \beta +2) (\beta  (5 \beta -8)+16)-(3 \beta -4) (\beta  (17 \beta -44)+32) \text{Pe}^2\right) e^{-(3 \beta +2) \ttt}}{15 (\beta -4) \beta  (3 \beta +2) \left(\beta ^2-4\right) \left((\beta -4)^2+\Omega ^2\right) \left(4 \beta ^2+\Omega ^2\right)}
       \nonumber\\&
       +\frac{4 \text{Pe}^2 \Omega ^4 \left(25 (\beta -4) (3 \beta +2)+(12-9 \beta ) \text{Pe}^2\right) e^{-(3 \beta +2) \ttt}}{15 (\beta -4) \beta  (3 \beta +2) \left(\beta ^2-4\right) \left((\beta -4)^2+\Omega ^2\right) \left(4 \beta ^2+\Omega ^2\right)}
       \nonumber\\&
       +\frac{4 \text{Pe}^2 \Omega ^2 \left(-25 (\beta +4) (3 \beta -2) (\beta  (5 \beta +8)+16)-\left((3 \beta +4) (\beta  (17 \beta +44)+32) \text{Pe}^2\right)\right) e^{-(\beta +2) \ttt}}{15 (\beta -2) \beta  (\beta +2) (\beta +4) (3 \beta -2) \left(4 \beta ^2+\Omega ^2\right) \left((\beta +4)^2+\Omega ^2\right)}
       \nonumber\\&-\frac{16 \beta  (\beta +4) \text{Pe}^2 \left(5 (\beta +4) (3 \beta -2)+(3 \beta +4) \text{Pe}^2\right) e^{-(\beta +2) \ttt}}{3 (\beta -2) (\beta +2) (3 \beta -2) \left(4 \beta ^2+\Omega ^2\right) \left((\beta +4)^2+\Omega ^2\right)}
       \nonumber\\&+\frac{4 \text{Pe}^2 \Omega ^4 \left(-25 (\beta +4) (3 \beta -2)-3 (3 \beta +4) \text{Pe}^2\right) e^{-(\beta +2) \ttt}}{15 (\beta -2) \beta  (\beta +2) (\beta +4) (3 \beta -2) \left(4 \beta ^2+\Omega ^2\right) \left((\beta +4)^2+\Omega ^2\right)}
       \nonumber\\&+\frac{15}{\beta ^2}+\frac{4 (\beta +2) (\beta +3)^3 (3 \beta +2) (3 \beta +5) \text{Pe}^4}{\beta ^2 \left((\beta +2)^2+\Omega ^2\right) \left((\beta +3)^2+\Omega ^2\right) \left(4 (\beta +3)^2+\Omega ^2\right) \left((3 \beta +2)^2+\Omega ^2\right)}
       \nonumber\\&+\frac{\text{Pe}^4 \left(\left(\beta  \left(111 \beta ^6+1571 \beta ^5+29777 \beta ^3+59040 \beta +33552\right)+7920\right) \Omega ^2+\Omega ^8+1004 \Omega ^4\right)}{\beta ^2 (\beta +2) (\beta +3) (3 \beta +2) \left((\beta +2)^2+\Omega ^2\right) \left((\beta +3)^2+\Omega ^2\right) \left(4 (\beta +3)^2+\Omega ^2\right) \left((3 \beta +2)^2+\Omega ^2\right)}
       \nonumber\\&+ \frac{40 (\beta +2) (\beta +3)^4 (3 \beta +2)^2 \text{Pe}^2}{\beta ^2 \left((\beta +2)^2+\Omega ^2\right) \left((\beta +3)^2+\Omega ^2\right) \left(4 (\beta +3)^2+\Omega ^2\right) \left((3 \beta +2)^2+\Omega ^2\right)}
       \nonumber\\&+ \frac{10 \left(\beta  \left(\beta  \left(\beta  \left((\beta  (183 \beta +2873)+19103) \beta ^3+154070 \beta +208264\right)+169344\right)+76464\right)+14688\right) \text{Pe}^2 \Omega ^2}{\beta ^2 (\beta +2) (\beta +3) (3 \beta +2) \left((\beta +2)^2+\Omega ^2\right) \left((\beta +3)^2+\Omega ^2\right) \left(4 (\beta +3)^2+\Omega ^2\right) \left((3 \beta +2)^2+\Omega ^2\right)}
       \nonumber\\&+ \frac{10 \text{Pe}^2 \left(\left(\beta ^2+2\right) \Omega ^8+7 \left(\beta  \left(13 \beta ^5+638 \beta ^3+1820 \beta +1196\right)+312\right) \Omega ^4+\left(17 \beta ^4+293 \beta ^2+122\right) \Omega ^6\right)}{\beta ^2 (\beta +2) (\beta +3) (3 \beta +2) \left((\beta +2)^2+\Omega ^2\right) \left((\beta +3)^2+\Omega ^2\right) \left(4 (\beta +3)^2+\Omega ^2\right) \left((3 \beta +2)^2+\Omega ^2\right)}
       \nonumber\\&+ \frac{15 e^{-4 \beta  t}}{\beta ^2}-\frac{10 \text{Pe}^2 \left(3 (\beta -2)^2+\Omega ^2\right) e^{-4 \beta  t}}{3 (\beta -2) \beta ^2 \left((\beta -2)^2+\Omega ^2\right)}
       \nonumber\\&+\text{Pe}^4 \Bigg(\frac{(\beta -3) (\beta  (\beta  (\beta  (3 \beta  (111 \beta -905)+8632)-13104)+9424)-2640) \Omega ^2 e^{-4 \beta  \ttt}}{3 (\beta -2) \beta ^2 (3 \beta -2) \left((2-3 \beta )^2+\Omega ^2\right) \left((\beta -3)^2+\Omega ^2\right) \left(4 (\beta -3)^2+\Omega ^2\right) \left((\beta -2)^2+\Omega ^2\right)}
       \nonumber\\&+\frac{4 (\beta -3)^3 (3 \beta -5) \left(3 \beta ^2-8 \beta +4\right)^2 e^{-4 \beta  \ttt}}{(\beta -2) \beta ^2 (3 \beta -2) \left((2-3 \beta )^2+\Omega ^2\right) \left((\beta -3)^2+\Omega ^2\right) \left(4 (\beta -3)^2+\Omega ^2\right) \left((\beta -2)^2+\Omega ^2\right)}
       \nonumber\\&+\frac{(\beta  (\beta  (\beta  (\beta  (759 \beta -7445)+27802)-49370)+42644)-15060) \Omega ^4 e^{-4 \beta  \ttt}}{15 (\beta -3) (\beta -2) \beta ^2 (3 \beta -2) \left((2-3 \beta )^2+\Omega ^2\right) \left((\beta -3)^2+\Omega ^2\right) \left(4 (\beta -3)^2+\Omega ^2\right) \left((\beta -2)^2+\Omega ^2\right)}
       \nonumber\\&+\frac{\left(3 (3 \beta -5) \Omega ^8+(\beta  (\beta  (147 \beta -755)+1367)-875) \Omega ^6\right) e^{-4 \beta  \ttt}}{15 (\beta -3) (\beta -2) \beta ^2 (3 \beta -2) \left((2-3 \beta )^2+\Omega ^2\right) \left((\beta -3)^2+\Omega ^2\right) \left(4 (\beta -3)^2+\Omega ^2\right) \left((\beta -2)^2+\Omega ^2\right)}\Bigg) 
       \nonumber\\&+\frac{16 \text{Pe}^4 \left(9 \left(\beta ^2+12\right) \Omega ^4+16 \left(\beta ^2-16\right)^2 \left(\beta ^2-9\right)+4 \left(6 \beta ^4-77 \beta ^2+592\right) \Omega ^2+\Omega ^6\right) e^{-2 (\beta +3) \ttt} \cos ( \Omega \ttt)}{15 (\beta -4) (\beta +4) \left((\beta -4)^2+\Omega ^2\right) \left(4 (\beta -3)^2+\Omega ^2\right) \left(4 (\beta +3)^2+\Omega ^2\right) \left((\beta +4)^2+\Omega ^2\right)}
       \nonumber\\&+\frac{64 \text{Pe}^4 \Omega  \left(28 \beta ^4+\beta ^2 \left(11 \Omega ^2-784\right)+\left(\Omega ^2+56\right) \left(\Omega ^2+96\right)\right) e^{-2 (\beta +3) \ttt} \sin ( \Omega\ttt )}{15 (\beta -4) (\beta +4) \left((\beta -4)^2+\Omega ^2\right) \left(4 (\beta -3)^2+\Omega ^2\right) \left(4 (\beta +3)^2+\Omega ^2\right) \left((\beta +4)^2+\Omega ^2\right)}
       \nonumber\\&+\sin(\Omega\ttt)e^{-(3 \beta +2) \ttt}\Bigg(\frac{160 \text{Pe}^2 \Omega  }{3 \beta  \left((\beta -2)^2+\Omega ^2\right) \left((\beta +2)^2+\Omega ^2\right)}
       \nonumber\\&-32 \text{Pe}^4 \Omega  \Big(\frac{(\beta  (\beta  (\beta  (63 \beta -214)-68)+312)+96) \beta ^2}{3 \left((\beta -4)^2+\Omega ^2\right) \left((\beta -2)^2+\Omega ^2\right) \left(\beta ^2+\Omega ^2\right) \left(4 \beta ^2+\Omega ^2\right) \left((\beta +2)^2+\Omega ^2\right) \left((3 \beta +2)^2+\Omega ^2\right)}
       \nonumber\\&+\frac{(\beta  (\beta  (\beta  (\beta  (485 \beta -2772)+3376)+3656)-3392)-768) \Omega ^2}{15 (\beta -4) \left((\beta -4)^2+\Omega ^2\right) \left((\beta -2)^2+\Omega ^2\right) \left(\beta ^2+\Omega ^2\right) \left(4 \beta ^2+\Omega ^2\right) \left((\beta +2)^2+\Omega ^2\right) \left((3 \beta +2)^2+\Omega ^2\right)}
       \nonumber\\&+\frac{(\beta  (25 \beta -32)+108) \Omega ^6+2 (\beta  (\beta  (\beta  (97 \beta -381)+420)+440)-112) \Omega ^4+\Omega ^8}{15 (\beta -4) \beta  \left((\beta -4)^2+\Omega ^2\right) \left((\beta -2)^2+\Omega ^2\right) \left(\beta ^2+\Omega ^2\right) \left(4 \beta ^2+\Omega ^2\right) \left((\beta +2)^2+\Omega ^2\right) \left((3 \beta +2)^2+\Omega ^2\right)}\Big) \Bigg)
       \nonumber\\&+\Bigg(\frac{32 (4-3 \beta ) (\beta -4) (\beta -2) \beta ^3 (\beta +2) (3 \beta +2) \text{Pe}^4 }{3 \left((\beta -4)^2+\Omega ^2\right) \left((\beta -2)^2+\Omega ^2\right) \left(\beta ^2+\Omega ^2\right) \left(4 \beta ^2+\Omega ^2\right) \left((\beta +2)^2+\Omega ^2\right) \left((3 \beta +2)^2+\Omega ^2\right)}
       \nonumber\\&+\frac{8 \beta  (\beta  (\beta  (26320-\beta  (\beta  (\beta  (461 \beta -3176)+2768)+19584))+13440)-5888) \text{Pe}^4 \Omega ^2}{15 (\beta -4) \left((\beta -4)^2+\Omega ^2\right) \left((\beta -2)^2+\Omega ^2\right) \left(\beta ^2+\Omega ^2\right) \left(4 \beta ^2+\Omega ^2\right) \left((\beta +2)^2+\Omega ^2\right) \left((3 \beta +2)^2+\Omega ^2\right)}
       \nonumber\\&+\frac{32 (\beta  (\beta  (2648-\beta  (\beta  (5 \beta  (20 \beta -99)+284)+2428))+1408)-64) \text{Pe}^4 \Omega ^4}{15 (\beta -4) \beta  \left((\beta -4)^2+\Omega ^2\right) \left((\beta -2)^2+\Omega ^2\right) \left(\beta ^2+\Omega ^2\right) \left(4 \beta ^2+\Omega ^2\right) \left((\beta +2)^2+\Omega ^2\right) \left((3 \beta +2)^2+\Omega ^2\right)}
       \nonumber\\&-\frac{8 \text{Pe}^4 \Omega ^6 \left(4 (5 (\beta -1) \beta +16) \Omega ^2+6 \beta  (\beta  (\beta  (23 \beta -64)+40)+152)+\Omega ^4-976\right)}{15 (\beta -4) \beta  \left((\beta -4)^2+\Omega ^2\right) \left((\beta -2)^2+\Omega ^2\right) \left(\beta ^2+\Omega ^2\right) \left(4 \beta ^2+\Omega ^2\right) \left((\beta +2)^2+\Omega ^2\right) \left((3 \beta +2)^2+\Omega ^2\right)}
       \nonumber\\&+\frac{40 \text{Pe}^2 \left(\beta ^2+\Omega ^2-4\right)}{3 \beta  \left((\beta -2)^2+\Omega ^2\right) \left((\beta +2)^2+\Omega ^2\right)}\Bigg)\cos(\Omega \ttt)e^{-(3 \beta +2) \ttt}
       \nonumber\\&+\sin(\Omega\ttt)e^{-(\beta +2) \ttt}\Bigg(-\frac{160 \text{Pe}^2 \Omega  }{3 \beta  \left((\beta -2)^2+\Omega ^2\right) \left((\beta +2)^2+\Omega ^2\right)}
       \nonumber\\&-\frac{32 \text{Pe}^4\beta ^2 (\beta  (\beta  (\beta  (63 \beta +214)-68)-312)+96) \Omega }{3 \left((2-3 \beta )^2+\Omega ^2\right) \left((\beta -2)^2+\Omega ^2\right) \left(\beta ^2+\Omega ^2\right) \left(4 \beta ^2+\Omega ^2\right) \left((\beta +2)^2+\Omega ^2\right) \left((\beta +4)^2+\Omega ^2\right)}
       \nonumber\\&-\frac{32 \text{Pe}^4(\beta  (\beta  (\beta  (\beta  (485 \beta +2772)+3376)-3656)-3392)+768) \Omega ^3}{15 (\beta +4) \left((2-3 \beta )^2+\Omega ^2\right) \left((\beta -2)^2+\Omega ^2\right) \left(\beta ^2+\Omega ^2\right) \left(4 \beta ^2+\Omega ^2\right) \left((\beta +2)^2+\Omega ^2\right) \left((\beta +4)^2+\Omega ^2\right)}
       \nonumber\\&-\frac{32 \text{Pe}^4\Omega ^5 \left((\beta  (25 \beta +32)+108) \Omega ^2+2 \beta  (\beta  (\beta  (97 \beta +381)+420)-440)+\Omega ^4-224\right) }{15 \beta  (\beta +4) \left((2-3 \beta )^2+\Omega ^2\right) \left((\beta -2)^2+\Omega ^2\right) \left(\beta ^2+\Omega ^2\right) \left(4 \beta ^2+\Omega ^2\right) \left((\beta +2)^2+\Omega ^2\right) \left((\beta +4)^2+\Omega ^2\right)}\Bigg)
       \nonumber\\&+\Bigg(-\frac{32 (\beta -2) \beta ^3 (\beta +2) (\beta +4) (3 \beta -2) (3 \beta +4)\text{Pe}^4}{3 \left((2-3 \beta )^2+\Omega ^2\right) \left((\beta -2)^2+\Omega ^2\right) \left(\beta ^2+\Omega ^2\right) \left(4 \beta ^2+\Omega ^2\right) \left((\beta +2)^2+\Omega ^2\right) \left((\beta +4)^2+\Omega ^2\right)}
       \nonumber\\& -\frac{8 \beta  (\beta  (\beta  (\beta  (\beta  (\beta  (461 \beta +3176)+2768)-19584)-26320)+13440)+5888) \text{Pe}^4 \Omega ^2}{15 (\beta +4) \left((2-3 \beta )^2+\Omega ^2\right) \left((\beta -2)^2+\Omega ^2\right) \left(\beta ^2+\Omega ^2\right) \left(4 \beta ^2+\Omega ^2\right) \left((\beta +2)^2+\Omega ^2\right) \left((\beta +4)^2+\Omega ^2\right)}
       \nonumber\\&-\frac{32 (\beta  (\beta  (\beta  (\beta  (5 \beta  (20 \beta +99)+284)-2428)-2648)+1408)+64) \text{Pe}^4 \Omega ^4}{15 \beta  (\beta +4) \left((2-3 \beta )^2+\Omega ^2\right) \left((\beta -2)^2+\Omega ^2\right) \left(\beta ^2+\Omega ^2\right) \left(4 \beta ^2+\Omega ^2\right) \left((\beta +2)^2+\Omega ^2\right) \left((\beta +4)^2+\Omega ^2\right)}
        \nonumber\\&-\frac{8 \text{Pe}^4 \Omega ^6 \left(4 (5 \beta  (\beta +1)+16) \Omega ^2+6 \beta  (\beta  (\beta  (23 \beta +64)+40)-152)+\Omega ^4-976\right)}{15 \beta  (\beta +4) \left((2-3 \beta )^2+\Omega ^2\right) \left((\beta -2)^2+\Omega ^2\right) \left(\beta ^2+\Omega ^2\right) \left(4 \beta ^2+\Omega ^2\right) \left((\beta +2)^2+\Omega ^2\right) \left((\beta +4)^2+\Omega ^2\right)}
        \nonumber\\& -\frac{40 \text{Pe}^2 \left(\beta ^2+\Omega ^2-4\right)}{3 \beta  \left((\beta -2)^2+\Omega ^2\right) \left((\beta +2)^2+\Omega ^2\right)}\Bigg)\cos(\Omega\ttt)e^{-(\beta +2) \ttt}
        \nonumber\\&+ \frac{16 \text{Pe}^4 \left(\beta ^4+\beta ^2 \left(2 \Omega ^2-25\right)+\Omega ^4-73 \Omega ^2+144\right) e^{-2 (\beta +3) \ttt} \cos (2 \Omega \ttt)}{\left((\beta -4)^2+\Omega ^2\right) \left((\beta -3)^2+\Omega ^2\right) \left((\beta +3)^2+\Omega ^2\right) \left((\beta +4)^2+\Omega ^2\right)}
        \nonumber\\& +\frac{224 \text{Pe}^4 \Omega  \left(\beta ^2+\Omega ^2-12\right) e^{-2 (\beta +3) \ttt} \sin (2  \Omega\ttt )}{\left((\beta -4)^2+\Omega ^2\right) \left((\beta -3)^2+\Omega ^2\right) \left((\beta +3)^2+\Omega ^2\right) \left((\beta +4)^2+\Omega ^2\right)}\,.
        \label{eq:r4_3d_dimensionless}
\end{align}}
We use Eq.~\eqref{eq:r4_3d_dimensionless} to compute the time-dependent excess kurtosis shown in Fig.~\ref{fig5} of the main text. 
In the short-time limit, the fourth order moment of the displacement can be expanded as a power series in $\tilde{t}$. 
Retaining terms up to $\mathcal{O}(\tilde{t}^4)$, we obtain
\begin{align}
&\lim_{\ttt \to 0}\langle\tbr^4\rangle = 60 \ttt^2+20 \ttt^3 \left(\text{Pe}^2-6 \beta \right)+\ttt^4 \left(140 \beta ^2+\text{Pe}^4-\frac{40}{3} (3 \beta +1) \text{Pe}^2\right)+\mathcal{O}\left(\ttt^5\right)\,.
\label{eq:r4avg_3d_small_time_dimensionless}
\end{align}
In the short-time regime ($\tilde{t} \ll 1$), the leading term $60\tilde{t}^2$ arises purely from translational diffusion and corresponds to Gaussian fluctuations. 
The $\tilde{t}^3$ term captures the onset of active self-propulsion, which dominates over confinement when $\mathrm{Pe}^2 > 6\beta$, 
thereby marking the first deviation from purely diffusive, Gaussian behavior at $\tilde{t} = 3/(\mathrm{Pe}^2 - 6\beta)$.

In the long-time limit, the fourth order moment of the displacement attains a steady-state value as $\langle \mathbf{r}^4 \rangle_{\mathrm{st}} = \lim_{\tilde{t} \to \infty} \langle \mathbf{r}^4 \rangle$~\cite{Pattanayak2025}, 
the dimensionless expression is given by
\begin{align}
&\langle\tbr^4\rangle_{\rm st}= \frac{15}{\beta ^2}+\frac{10 \text{Pe}^2 \left(3 (\beta +2)^2+\Omega ^2\right)}{3 \beta ^2 (\beta +2) \left((\beta +2)^2+\Omega ^2\right)} 
\nonumber\\&+\text{Pe}^4 \Bigg(\frac{4 (\beta +2) (\beta +3)^3 (3 \beta +2) (3 \beta +5)}{\beta ^2 \left((\beta +2)^2+\Omega ^2\right) \left((\beta +3)^2+\Omega ^2\right) \left(4 (\beta +3)^2+\Omega ^2\right) \left((3 \beta +2)^2+\Omega ^2\right)}
\nonumber\\&+\frac{(\beta +3) (\beta  (\beta  (\beta  (3 \beta  (111 \beta +905)+8632)+13104)+9424)+2640) \Omega ^2}{3 \beta ^2 (\beta +2) (3 \beta +2) \left((\beta +2)^2+\Omega ^2\right) \left((\beta +3)^2+\Omega ^2\right) \left(4 (\beta +3)^2+\Omega ^2\right) \left((3 \beta +2)^2+\Omega ^2\right)}
\nonumber\\&
+\frac{(\beta  (\beta  (\beta  (\beta  (759 \beta +7445)+27802)+49370)+42644)+15060) \Omega ^4}{15 \beta ^2 (\beta +2) (\beta +3) (3 \beta +2) \left((\beta +2)^2+\Omega ^2\right) \left((\beta +3)^2+\Omega ^2\right) \left(4 (\beta +3)^2+\Omega ^2\right) \left((3 \beta +2)^2+\Omega ^2\right)}
\nonumber\\&
+\frac{3 (3 \beta +5) \Omega ^8+(\beta  (\beta  (147 \beta +755)+1367)+875) \Omega ^6}{15 \beta ^2 (\beta +2) (\beta +3) (3 \beta +2) \left((\beta +2)^2+\Omega ^2\right) \left((\beta +3)^2+\Omega ^2\right) \left(4 (\beta +3)^2+\Omega ^2\right) \left((3 \beta +2)^2+\Omega ^2\right)}\Bigg)\,.
\label{eq:r4_3d_st_dimensionless}
\end{align}

Figure~\ref{app_fig3} shows the time evolution and steady-state behavior of the dimensionless fourth order moment of displacement, 
$\langle \tilde{\mathbf{r}}^4(\tilde{t})\rangle$, for an active Brownian particle under torque confined in a harmonic potential. 
The $\langle \tilde{\mathbf{r}}^4(\tilde{t})\rangle$ evolution reveals four distinct dynamical regimes that appear consistently across parameters: an initial diffusive regime($\langle \tilde{\mathbf{r}}^4\rangle \sim \tilde{t}^2$) dominated by translational noise, ballistic regime($\langle \tilde{\mathbf{r}}^4\rangle \sim \tilde{t}^3$) dominated by persistent propulsion, an intermediate crossover regime without oscillatory behavior in $\langle \tilde{\mathbf{r}}^4(\tilde{t})\rangle$ characterized by rotational decorrelation, and a final steady-state plateau where confinement balances torque activity and diffusion. 
Figure~\ref{app_fig3}(a)--(c) present the temporal evolution of the $\langle \tilde{\mathbf{r}}^4(\tilde{t})\rangle$ for varying Péclet number ${\rm Pe}$, torque $\Omega$, and trap strength $\beta$, respectively, while Figs.~\ref{app_fig3}(d)--(f) show the corresponding steady-state values $\langle \tilde{\mathbf{r}}^4\rangle_{\rm st}$ as functions of these parameters.

For fixed $\Omega=100$ and $\beta=1$ [Fig.~\ref{app_fig3}(a)], increasing ${\rm Pe}$ enhances both the amplitude and duration of the ballistic regime, as stronger self-propulsion drives persistent motion before confinement becomes dominant. 
The steady-state value grows approximately as $\langle \tilde{\mathbf{r}}^4\rangle_{\rm st} \sim {\rm Pe}^4/\beta^2$, consistent with the analytical expression in Eq.~\eqref{eq:r4_3d_dimensionless}. 
At fixed ${\rm Pe}=100$ and $\beta=1$ [Fig.~\ref{app_fig3}(b)], increasing torque $\Omega$ suppresses $\langle \tilde{\mathbf{r}}^4(\tilde{t})\rangle$ in the intermediate regime. 
Larger $\Omega$ values suppress the overall $\langle \tilde{\mathbf{r}}^4(\tilde{t})\rangle$ amplitude, since rapid precession averages out directed motion and confines the particle more tightly near the trap center. 
For fixed ${\rm Pe}=100$ and $\Omega=100$ [Fig.~\ref{app_fig3}(c)], stronger confinement ($\beta$) accelerates relaxation and reduces the saturation value. 
In the weak-trap limit ($\beta \ll 1$), the $\langle \tilde{\mathbf{r}}^4(\tilde{t})\rangle$ remains unbounded for long times, approaching the unconfined active Brownian regime.

The steady-state results in Figs.~\ref{app_fig3}(d)--(f) quantify these dependencies in Eq.~\eqref{eq:r4_3d_st_dimensionless}. 
Figure~\ref{app_fig3}~(d) confirms a quadratic scaling $\langle \tilde{\mathbf{r}}^4\rangle_{\rm st} \propto {\rm Pe}^4$, indicating that steady-state fluctuations are primarily set by the strength of active propulsion. 
Figure~\ref{app_fig3}~(e) shows that $\langle \tilde{\mathbf{r}}^4\rangle_{\rm st}$ decreases monotonically from a constant value corresponds to low torque limit($\Omega\to 0$) 
\begin{align}
\langle \tilde{\mathbf{r}}^{4} \rangle_{\mathrm{st}}
&\simeq \frac{15}{\beta^{2}}
+ \frac{10\,\mathrm{Pe}^{2}}{\beta^{2}(\beta+2)}
+ \frac{(3\beta+5)\,\mathrm{Pe}^{4}}{\beta^{2}(\beta+2)(\beta+3)(3\beta+2)}\,,
\end{align}
with increasing torque $\Omega$, approaching a constant
\begin{align}
\langle \tilde{\mathbf{r}}^{4} \rangle_{\mathrm{st}}
&\simeq \frac{15}{\beta^{2}}
+ \frac{10\,\mathrm{Pe}^{2}}{3\,\beta^{2}(\beta+2)}
+ \frac{(3\beta+5)\,\mathrm{Pe}^{4}}{5\,\beta^{2}(\beta+2)(\beta+3)(3\beta+2)}\,,
\end{align}
at large torque limit($\Omega \to \infty$) where fast rotational dynamics average out propulsion. 
Finally, Fig.~\ref{app_fig3}~(f) demonstrates an inverse dependence $\langle \tilde{\mathbf{r}}^4\rangle_{\rm st} \sim 1/\beta^2$, highlighting that confinement directly suppresses the spatial extent of active fluctuations. 
In the strong-trap regime, all curves collapse, showing that torque becomes irrelevant once the trap dominates the dynamics.

In summary, it illustrates the complete dynamical crossover from diffusive to ballistic to an intermediate crossover regime without oscillation to localized motion. 
The observed trends from the analytical predictions(Eq.~\eqref{eq:r4_3d_dimensionless}; lines) are in excellent agreement with simulation results(points; Figs.~\ref{app_fig3}(a)--(c)).

\end{document}